\documentclass[10pt,journal,compsoc]{IEEEtran}
\usepackage[T1]{fontenc}
\usepackage[latin9]{inputenc}
\usepackage{color}
\usepackage{array}
\usepackage{verbatim}
\usepackage{multirow}
\usepackage{amsmath}
\usepackage{amsthm}
\usepackage{graphicx}
\ifCLASSOPTIONcompsoc
  \usepackage[nocompress]{cite}
\else
  \usepackage{cite}
\fi
\ifCLASSINFOpdf
\else
\fi
\makeatletter

\providecommand{\tabularnewline}{\\}

\theoremstyle{plain}
\newtheorem{thm}{\protect\theoremname}
\theoremstyle{plain}
\newtheorem{lem}[thm]{\protect\lemmaname}
\theoremstyle{plain}
\newtheorem{cor}[thm]{\protect\corollaryname}

\date{}
\usepackage[margin=0.2cm,figurename=Fig.]{caption}
\usepackage[font=footnotesize]{subfig}

\author{Jin Xu, I-Hong Hou and
Natarajan Gautam
\thanks{Jin Xu is with School of Science and Engineering, the Chinese University of Hong Kong, Shenzhen, China;
Shenzhen Institute of Artificial Intelligence and Robotics for Society, Shenzhen, China; and University of Science and Technology of China,
Hefei, Anhui, China (Email: xujin@cuhk.edu.cn)}
\thanks{I-Hong Hou is with Department of Electrical and Computer Engineering,
Texas A\&M University, College Station, TX, USA (Email:ihou@tamu.edu)}
\thanks{Natarajan Gautam is with Department of Industrial and Systems Engineering,
Texas A\&M University, College Station, TX, USA (Email:gautam@tamu.edu)}}

\usepackage{cuted}

\allowdisplaybreaks

\usepackage{array}
\newcolumntype{P}[1]{>{\hspace{0pt}}p{#1}}

\@ifundefined{showcaptionsetup}{}{%
 \PassOptionsToPackage{caption=false}{subfig}}
\usepackage{subfig}
\makeatother

\usepackage{babel}
\providecommand{\corollaryname}{Corollary}
\providecommand{\lemmaname}{Lemma}
\providecommand{\theoremname}{Theorem}

\begin{document}

\title{Age of Information for Single Buffer Systems with Vacation Server }
\IEEEtitleabstractindextext{%
\begin{abstract}
In this research, we study the information freshness in M/G/1 queueing
system with a single buffer and the server taking multiple vacations. This system has
wide applications in communication systems. We aim to evaluate the
information freshness in this system with both i.i.d. and non-i.i.d.
vacations under three different scheduling policies, namely Conventional
Buffer System (CBS), Buffer Relaxation System (BRS), and Conventional
Buffer System with Preemption in Service (CBS-P). For the systems
with i.i.d. vacations, we derive the closed-form expressions of information
freshness metrics such as the expected Age of Information (AoI), the
expected Peak Age of Information (PAoI), and the variance of peak
age under each policy. For systems with non-i.i.d. vacations, we use
the polling system as an example and provide the closed-form expression
of its PAoI under each policy. We explore the conditions under which
one of these policies has advantages over the others for each information
freshness metric. We further perform numerical studies to validate
our results and develop insights.
\end{abstract}

}

\maketitle

\section{Introduction\label{sec:Introduction}}

{Information freshness has recently drawn the wide
attention of researchers due to its applications in many communication
settings \cite{kaul2012real,costa2016age}. In a communication system,
the data receiver (user) usually needs fresh information sampled at
the physical process for on-the-fly decision-making. Unlike long-established
queueing metrics such as throughput or waiting time, information freshness
measures how timely the user is informed about the physical process,
and a large information freshness would enable the user to react timely
to different changes in the physical process \cite{zhong2018two,huang2015optimizing,bedewy2019age}.
Therefore, guaranteeing the information freshness for users in communication
systems is of great importance. }

{This paper studies the information freshness in a
queueing system where a data source generates data packets and sends
them to a server over time. The server needs to process the packets
to extract useful information for the user. The server would take
vacations after processing a packet, and the server would resume working
if it finds a packet in the queue after returning from a vacation
period. Such a vacation server system where information freshness
is of interest is an abstraction of real-life communication systems,
and it can be found in many application scenarios.}

{One scenario is in a smart manufacturing system where
fresh data sampled at machines would be helpful for the decision-maker
(user) in estimating the Remaining Useful Life (RUL) \cite{song2019generic},
detecting defects of manufactured products \cite{cheng2008vision},
or making process controls \cite{yao2018constrained}. In such a system,
the energy-saving sleeping period that the server (computer or processor)
takes when it has no information to process can be regarded as a vacation
period \cite{guo2016delay}.}

{Another scenario of such a system is in underwater
sensor networks of the petroleum industry or aquatic environment monitoring,
where people need to obtain timely updates about underwater environment
status. A rechargeable autonomous underwater vehicle can be sent from
the surface to upload or collect data from the underwater node in
a periodic manner (see \cite{heidemann2012underwater,vasilescu2005data}).
This way of collecting data can avoid frequent battery replacement
resulted from acoustic transmissions (see \cite{mohapatra2012combined,heidemann2006research}),
and the period that the vehicle travels between the surface and underwater
node can be regarded as the server vacation.}

{The third scenario is in remote health monitoring,
where the health data is acquired by a wearable device from a patient
and transmitted to the healthcare provider over time (see \cite{doukas2011managing,majumder2017wearable}).
The most recent health status of the patient will be useful for tracking
the patient's health status, but the doctor at the health center cannot
wait for the update from a single patient all the time without performing
other duties. }

{Besides the examples mentioned above, the vacation
server systems also exist in sensor networks and computer-communication
systems where the server has additional tasks aside from processing
the primary data source of interest, such as priority queue systems
\cite{takagi1991priority,xu2019towards,kella1988priorities} and polling
systems \cite{boon2011applications}. Systems with server maintenance
(see \cite{fuhrmann1985stochastic}) and systems with on/off servers
(see \cite{maccio2015optimal}) can be regarded as vacation server
systems as well. }

{Although the vacation server system widely exists
in various applications, its performance on information freshness
has not been fully understood. In this work, we aim to answer the
following key research questions:}
\begin{itemize}
\item There are several widely discussed metrics to measure
information freshness, such as Age of Information (AoI) \cite{kaul2012real},
Peak Age of Information (PAoI) \cite{huang2015optimizing}, and variance
of peak age \cite{mankar2021spatial}. How do we evaluate these metrics
in the systems with server vacations?
\item How are these information freshness metrics determined
by packet arrival rate, service time, vacation time, and scheduling
policies?
\item For different scheduling policies applied in this
system, which policy performs the best in terms of each information
freshness metric? 
\end{itemize}
{Answering those research questions would provide
us with theoretical performance guarantees and guidelines for designing
scheduling policies in various communication systems and real-life
applications, thereby improving information freshness for users. Several
challenges exist in answering these questions, which are: 1) the information
freshness metrics in vacation server systems have not been fully studied,
and results for systems with no vacations cannot be applied in our
system; 2) when the packet processing time and vacation time are non-exponential,
the technics that rely on exponential assumptions (such as Continuous
Time Markov Chain analysis \cite{maatouk2018age} and Stochastic Hybrid
System analysis \cite{yates2019age}) cannot be applied; 3) when the
vacation time is non-i.i.d., the age metrics could be difficult to
solve.}

{To address the research questions and overcome the
challenges above, we focus our discussion on scheduling policies applied
to the single buffer system due to the benefit of having a single
buffer in improving information freshness \cite{kosta2019age,xu2019towards}.
In particular, we study three different scheduling policies in the
single buffer system, denoted as Conventional Buffer System (CBS),
Buffer Relaxation System (BRS), and Conventional Buffer System with
Preemption in Service (CBS-P), with detailed descriptions provided
in Section \ref{sec:System-Model}. We show that the information freshness
metrics under these policies can be decomposed into several computable
components. Using this decomposition approach, we further evaluate
the AoI, PAoI, and variance of peak age for systems with i.i.d. vacations,
and PAoI for systems with non-i.i.d. vacations, under each scheduling
policy. The main contributions of this paper are summarized as follows:}
\begin{itemize}
\item {We propose a novel analytical approach to derive
the closed-form expressions of AoI, PAoI, and variance of peak age
for CBS, BRS, and CBS-P in systems with i.i.d. vacations. For systems
with non-i.i.d. vacations, we derive the PAoI for polling systems
with Markovian polling schemes. We show that this analytical approach
can be potentially applied to evaluate the age metrics for systems
with other types of server vacations.}
\item {We provide the conditions under which one scheduling
policy has the advantage over the others. Specifically, we prove that
when vacation times are i.i.d., the PAoI in BRS is always no greater
than that in CBS, regardless of the vacation or service time distributions.
We also show that when the arrival rate is low, BRS can have a significant
advantage over CBS in minimizing AoI, PAoI, and variance of peak age,
which shows the advantage of allowing the buffer to be available all
the time in light-traffic systems.}
\item {We provide sufficient conditions under which CBS-P
has a smaller PAoI than CBS. Our analysis reveals the advantage of
having packet preemption in vacation server systems when the packet
processing time is Gamma distributed with small scale parameters.}
\item {Our results show that under some specific processing
time distributions, reducing vacation time does not always decrease
the AoI, due to the particular definition of AoI. }
\item {Our work reveals that for polling systems with multiple
queues, reducing the vacation time for a specific queue could reduce
the PAoI for this queue but significantly increase the PAoI for other
queues. So the cyclic routing scheme is recommended in polling systems
for minimizing the average PAoI across queues.}
\end{itemize}
{The rest of this paper is organized as follows: Section
\ref{sec:Related-Works} provides a summary of the literature. The
system model is then introduced in Section \ref{sec:System-Model}.
In Section \ref{sec:Age-of-Information}, we consider the cases where
the server takes i.i.d. vacations. In Section \ref{sec:AoI-Polling},
we consider the case with non-i.i.d. vacations and discuss the polling
system as an example of the non-i.i.d. vacation model. We perform
numerical studies, develop insights in Section \ref{sec:Numerical-Study},
and provide concluding remarks and insights for future work in Section
\ref{sec:Concluding-Remarks}.}

\section{Related Work\label{sec:Related-Works}}

{The queueing systems with server vacations have been
widely investigated due to their wide applications. Most of the early
studies focused on classic queueing metrics such as average waiting
time, queue length, throughput, and blocking probability in vacation
server systems, without considering information freshness. Fuhrmann
\cite{fuhrmann1984note} studied the sojourn time in M/G/1 system
with the server following the multiple vacation scheme. Lee studied
the queue length distribution for M/G/1/N queue with vacations in
\cite{lee1984m,lee1989m}. Kella and Yechiali \cite{kella1988priorities}
studied the moments for waiting time in M/G/1 system with server vacations
and customer priorities. Other models about server vacations can be
found in \cite{frey1997note,takagi1991analysis,hur2003analysis,artalejo2001m}.
As a concept developed recently, information freshness was not considered
in these early works. }

{Most of the studies about information freshness focused
on queueing systems without vacations. Kaul et al. \cite{kaul2012real}
provided the average AoI for M/M/1, M/D/1, and D/M/1 queues. Costa
et al. \cite{costa2016age} provided the average AoI and PAoI for
M/M/1/1, M/M/1/2, and M/M/1/2{*} queues (the asterisk means keeping
the most recent packet in the buffer). Najm and Telatar \cite{najm2018status}
considered M/G/1/1 queue with preemption. Zou et al. \cite{zou2019benefis}
discussed the waiting procedure in M/G/1/1 and M/G/1/2{*} systems.
Huang and Modiano \cite{huang2015optimizing} considered PAoI of multi-class
M/G/1 and M/G/1/1 queues. Kaul and Yates \cite{kaul2018age} considered
a model with priority queues for preempted packets with and without
waiting rooms. Other queueing systems with AoI consideration can be
found in \cite{yates2019age,maatouk2019age,kaul2018age,moltafet2019age,kam2018age,inoue2019general,soysal2018age,najm2016age,najm2017status,chen2016age}.
Scheduling policies for optimizing information freshness in discrete-time
queueing systems have been studied in \cite{kosta2019age,jiang2019timely,he2017optimal,hsu2017age,talak2019optimizing,kadota2018scheduling}.
However, all these studies only considered the systems without server
vacations. }

{There are very few papers discussing information
freshness in systems with server vacations. Maatouk et al. \cite{maatouk2018age}
considered a system where the server sleeps and wakes randomly following
exponential distributions. We will show later in our work that allowing
the service time to be non-exponential will lead to some counter-intuitive
results. Moreover, the Continuous Time Markov Chain analysis used
in \cite{maatouk2018age} cannot be applied to systems where vacation
and service times are generally distributed. Najm et al. \cite{najm2019content}
considered a system of two data streams with different priorities,
and discussed several service disciplines for the low priority stream.
Xu and Gautam \cite{xu2019towards} discussed the PAoI in M/G/1/2{*}
and M/G/1 priority queueing systems. However, the priority queue system
is a special type of vacation model. The analysis in \cite{najm2019content,xu2019towards}
cannot be applied in our vacation server system under different scheduling
policies. Tripathi et al. \cite{tripathi2019age1,tripathi2019age}
provided analysis for discrete-time FCFS Ber/G/1 vacation server queue.
However, Talak et al. \cite{talak2019optimizing} found that the AoI
and PAoI in discrete-time queues could be significantly different
from their counterparts in continuous-time systems. Therefore, there
is a need to investigate a continuous-time vacation server model.
Moreover, as pointed out in \cite{xu2019towards}, the single buffer
systems are usually more efficient than FCFS in guaranteeing information
freshness. It thus motivates us to consider the systems with a single
buffer. The technics to derive age-related metrics for FCFS discipline
systems cannot be applied to single buffer systems. }

{In summary, the information freshness metrics in
systems with server vacations have not been fully investigated. It
is still unclear which scheduling policy in the single buffer system
achieves the smallest AoI or PAoI. The methodologies used in previous
studies cannot be applied in our models to derive the information
freshness metrics. This paper aims to provide a general mathematical
framework to compute the information freshness in systems with server
vacations. We also hope to understand how to manage the single buffer
to guarantee information freshness for both independent and dependent
vacation cases. }

\section{System Model\label{sec:System-Model}}

To better describe the models we analyze in this paper, we first describe
the single-queue system with server vacations. In Section \ref{sec:AoI-Polling}
we will show that a polling system with multiple queues can be regarded
as a single-queue system with non-i.i.d. vacations. To avoid introducing
more notations at this stage, we leave the detailed description of
the polling system in Section \ref{sec:AoI-Polling}. 

We consider a single-server system, where a data source generates
data packets following a Poisson process with rate $\lambda$. The
data packets are sent to the server as soon as they are generated.
The processing time (service time) $H$ for each packet is i.i.d.
and generally distributed. Once a packet has been processed, the server
takes a vacation. If the server finds no packet waiting in the buffer
upon returning from a vacation, it takes another vacation. Otherwise,
it starts processing the packet. This type of vacation scheme is called
\emph{multiple vacation} scheme, and it has wide applications (see
\cite{niu2003vacation,gupta2006computing}). In Section \ref{sec:Age-of-Information},
we discuss the case where each vacation period is i.i.d., and we discuss
the case of non-i.i.d. vacations in Section \ref{sec:AoI-Polling}.We
suppose that the buffer at the server only holds the freshest data
packet, i.e, the new packet will replace the old one in the buffer,
if the buffer is available. The buffer availability is determined
by scheduling policies defined as follows:
\begin{itemize}
\item Conventional Buffer System (CBS) (see \cite{lee1996exact,takine1988exact,chung1994performance,mukherjee1990comments}):
In this system, the buffer becomes available only when the server
is on vacation. Packets that arrive when the server is processing
will be rejected. The vacation starts once a packet has been processed.
\item Buffer Relaxation System (BRS) (see \cite{takine1988exact,chung1994performance}):
The buffer becomes available as soon as the server starts serving.
After completing a packet, the server will start a vacation, regardless
of whether the buffer is empty or not. 
\item Conventional Buffer System with Preemption in Service (CBS-P): In
this system, new arrival during processing will preempt the packet
in service. The preempted packet will be discarded. The vacation starts
once the system becomes empty. 
\end{itemize}
When the vacation time becomes zero, CBS becomes M/G/1/1 system, BRS
collapses into M/G/1/2{*} system, and CBS-P reduces to M/G/1/1/preemptive
system. We will discuss these special cases in Subsection \ref{subsec:Discussions-for-Systems}.

We consider age-related metrics in these systems. The age at time
$t$, for a single queue system, is defined as $\Delta(t)=t-\max\{r_{\{l\}}:C_{\{l\}}\leq t\},$
where $C_{\{l\}}$ is the completion time of the $l^{th}$ packet
that completes processing at the server, and $r_{\{l\}}$ is the generation
time of this packet. Note that the preempted or discarded packets
are not indexed, and we refer to those packets that complete the service
(incur age drops) as \emph{informative packets}. The time-average
age is then defined as $\bar{\Delta}=\lim_{T\rightarrow\infty}\frac{1}{T}\int_{0}^{T}\Delta(t)dt$.
By assuming the system is ergodic, we have $\boldsymbol{E}[\Delta]=\lim_{t\rightarrow\infty}\boldsymbol{E}[\Delta(t)]=\bar{\Delta},$
and we use the term ``AoI'' to refer $\boldsymbol{E}[\Delta]$.
While AoI is a useful metric to measure data freshness, many researchers
also analyzed a metric called Peak Age of Information (PAoI) due to
its tractability \cite{costa2016age,huang2015optimizing,inoue2019general}.
We let the $l^{th}$ peak of $\Delta(t)$ be $A_{\{l\}}$, and the
time-average peak value is then given by $\bar{A}=\lim_{k\rightarrow\infty}\frac{1}{k}\sum_{l=1}^{k}A_{\{l\}}$.
Again, by assuming the ergodicity, we have the expected peak age value,
i.e., $\boldsymbol{E}[A]=\lim_{l\rightarrow\infty}\boldsymbol{E}[A_{\{l\}}]$,
to be equal to $\bar{A}$. In this paper, we use the term ``PAoI''
to refer $\boldsymbol{E}[A]$. Throughout this paper, we let $X^{*}(s)$
denote the Laplace--Stieltjes Transform (LST) of random variable
$X$, $X^{*(n)}(s)$ be the $n^{th}$ derivative of $X^{*}(s)$, and
$F_{X}(x)$ be the cumulative distribution function (CDF) for $X$. 

\section{Age of Information for Systems with i.i.d. Vacations \label{sec:Age-of-Information}}

\begin{figure}[h]
\begin{center}\includegraphics[scale=0.27]{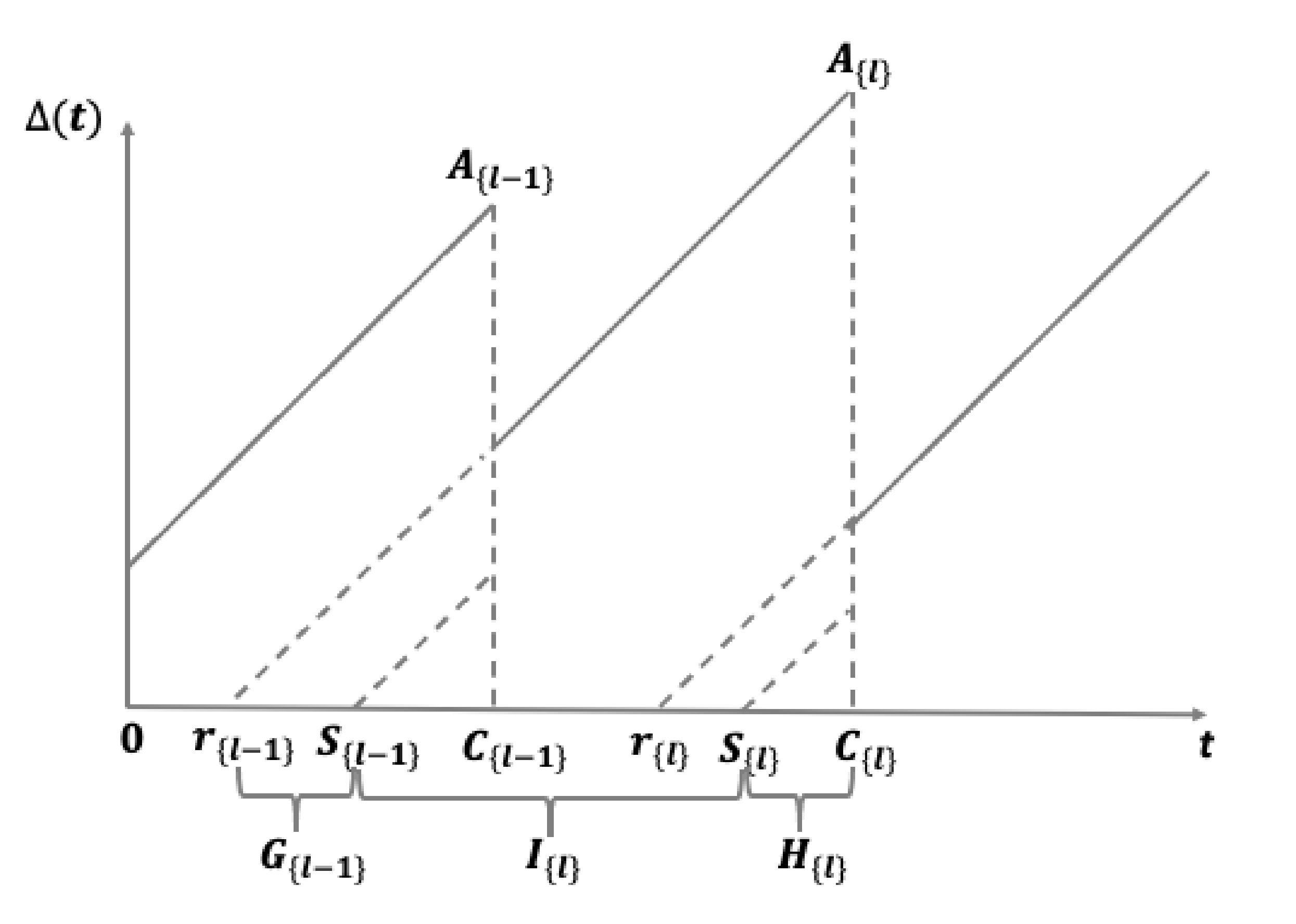}\end{center}

\caption{Age of Information Decomposition For Non-preemptive Service Systems.
The second age peak $A_{\{l\}}$ is decomposed into three components
$A_{\{l\}}=G_{\{l-1\}}+I_{\{l\}}+H_{\{l\}}$. The first component
$G_{\{l-1\}}$ is the waiting time of the $(l-1)^{th}$ served packet.
The second component $I_{\{l\}}$ is the time between the server starts
serving two packets. The third component $H_{\{l\}}$ is the service
time of the $l^{th}$ served packet. \label{fig:Age-of-Information}}
\end{figure}

We now consider three variations of the system, i.e., CBS, BRS, and
CBS-P, which we defined earlier in Section \ref{sec:System-Model},
in the scenarios where each vacation $V$ that the server takes is
i.i.d. with LST $V^{*}(s)$. The AoI, PAoI, and variance of peak age
of these three systems can be calculated by decomposing the peak age
into different computable components. In this section, we only discuss
the decomposition method that applies to non-preemptive service systems,
i.e., CBS and BRS. The decomposition approach for CBS-P shares a similar
idea but requires a different set of notations. So we leave the derivations
for CBS-P in Appendix \ref{sec:Proof-for-Theorem-1} of the supplementary
material. 

We now let $S_{\{l\}}$ be the time when the server starts processing
the $l^{th}$ informative packet, and this informative packet is completed
at time $C_{\{l\}}.$ From Fig. \ref{fig:Age-of-Information}, we
find that the $l^{th}$ age peak $A_{\{l\}}$ in CBS or BRS is the
time span from the completion time $C_{\{l\}}$ of the $l^{th}$ informative
packet, to the generation time (arrival time) of its previous informative
packet, i.e., $r_{\{l-1\}}$. This time span can then be divided into
three components: waiting time $G_{\{l-1\}}$ (in queue) of the $(l-1)^{th}$
informative packet, time span $I_{\{l\}}$ from the service starting
time $S_{\{l-1\}}$ of the $(l-1)^{th}$ informative packet till the
time when the $l^{th}$ informative packet starts service (which we
call \emph{regenerative cycle}), and service time $H_{\{l\}}$ of
the $l^{th}$ packet. These three components are mutually independent
for the following reasons. For both CBS and BRS, the regenerative
cycle $I_{\{l\}}$ contains processing time $H_{\{l-1\}}$ and a vacation
period. Since how long the vacation lasts only depends on the events
during $I_{\{l\}}$, and the processing time $H_{\{l-1\}}$ is independent
of $G_{\{l-1\}}$ and $H_{\{l\}}$, we have $I_{\{l\}}$ to be independent
of $G_{\{l-1\}}$ and $H_{\{l\}}$. It is also obvious that $H_{\{l\}}$
is independent of $G_{\{l-1\}}$. Therefore $G_{\{l-1\}}$, $I_{\{l\}}$
and $H_{\{l\}}$ are mutually independent. By letting $G$, $I$ and
$H$ denote the limit distribution of $G_{\{l\}},$ $I_{\{l\}}$ and
$H_{\{l\}}$, the PAoI can be given as 
\begin{eqnarray}
\boldsymbol{E}[A] & = & \lim_{l\rightarrow\infty}\boldsymbol{E}[A_{\{l\}}]\nonumber \\
 & = & \lim_{l\rightarrow\infty}(\boldsymbol{E}[G_{\{l-1\}}]+\boldsymbol{E}[I_{\{l\}}]+\boldsymbol{E}[H_{\{l\}}])\nonumber \\
 & = & \boldsymbol{E}[G]+\boldsymbol{E}[I]+\boldsymbol{E}[H],\label{eq:0}
\end{eqnarray}
and the system AoI can be given as 
\begin{eqnarray}
 &  & \boldsymbol{E}[\Delta]\nonumber \\
 & = & \lim_{l\rightarrow\infty}\frac{1}{2\boldsymbol{E}[C_{\{l\}}-C_{\{l-1\}}]}\bigg[\boldsymbol{E}[(G_{\{l-1\}}+I_{\{l\}}+H_{\{l\}})^{2}]\nonumber \\
 &  & -\boldsymbol{E}[(G_{\{l\}}+H_{\{l\}})^{2}]\bigg]\nonumber \\
 & = & \lim_{l\rightarrow\infty}\frac{1}{2\boldsymbol{E}[I_{\{l\}}+H_{\{l\}}-H_{\{l-1\}}]}\nonumber \\
 &  & \bigg[\boldsymbol{E}[(G_{\{l-1\}}+I_{\{l\}}+H_{\{l\}})^{2}]-\boldsymbol{E}[(G_{\{l\}}+H_{\{l\}})^{2}]\bigg]\nonumber \\
 & = & \frac{\boldsymbol{E}[I^{2}]}{2\boldsymbol{E}[I]}+\boldsymbol{E}[G]+\boldsymbol{E}[H].\label{eq:1}
\end{eqnarray}
Note that Equation (\ref{eq:0}) and (\ref{eq:1}) are for non-preemptive
service systems, i.e., CBS and BRS. In Appendix \ref{sec:Proof-for-Theorem}
of the supplementary material we show that the PAoI and AoI for CBS-P
can be calculated in a similar manner. Let the LST of $G,I$ and $H$
be $G^{*}(s),I^{*}(s)$ and $H^{*}(s)$, and then the LST of $A$
can be given as
\begin{eqnarray}
A^{*}(s) & = & G^{*}(s)I^{*}(s)H^{*}(s).\label{eq:2}
\end{eqnarray}
PAoI can be easily obtained by calculating the first moment of $A^{*}(s)$
at $s=0$. The variance of peak age can be used as a metric to measure
the age violations, and the variance of peak age can be given as 
\begin{eqnarray}
Var(A) & = & Var(G)+Var(I)+Var(H).\label{eq:1.1}
\end{eqnarray}
In this work, $H$ is a system parameter. We only need to obtain $G^{*}(s)$
and $I^{*}(s)$ to derive $\boldsymbol{E}[A]$, \textbf{$\boldsymbol{E}[\Delta]$
}and $Var(A)$. 

\begin{figure}
\includegraphics[scale=0.3]{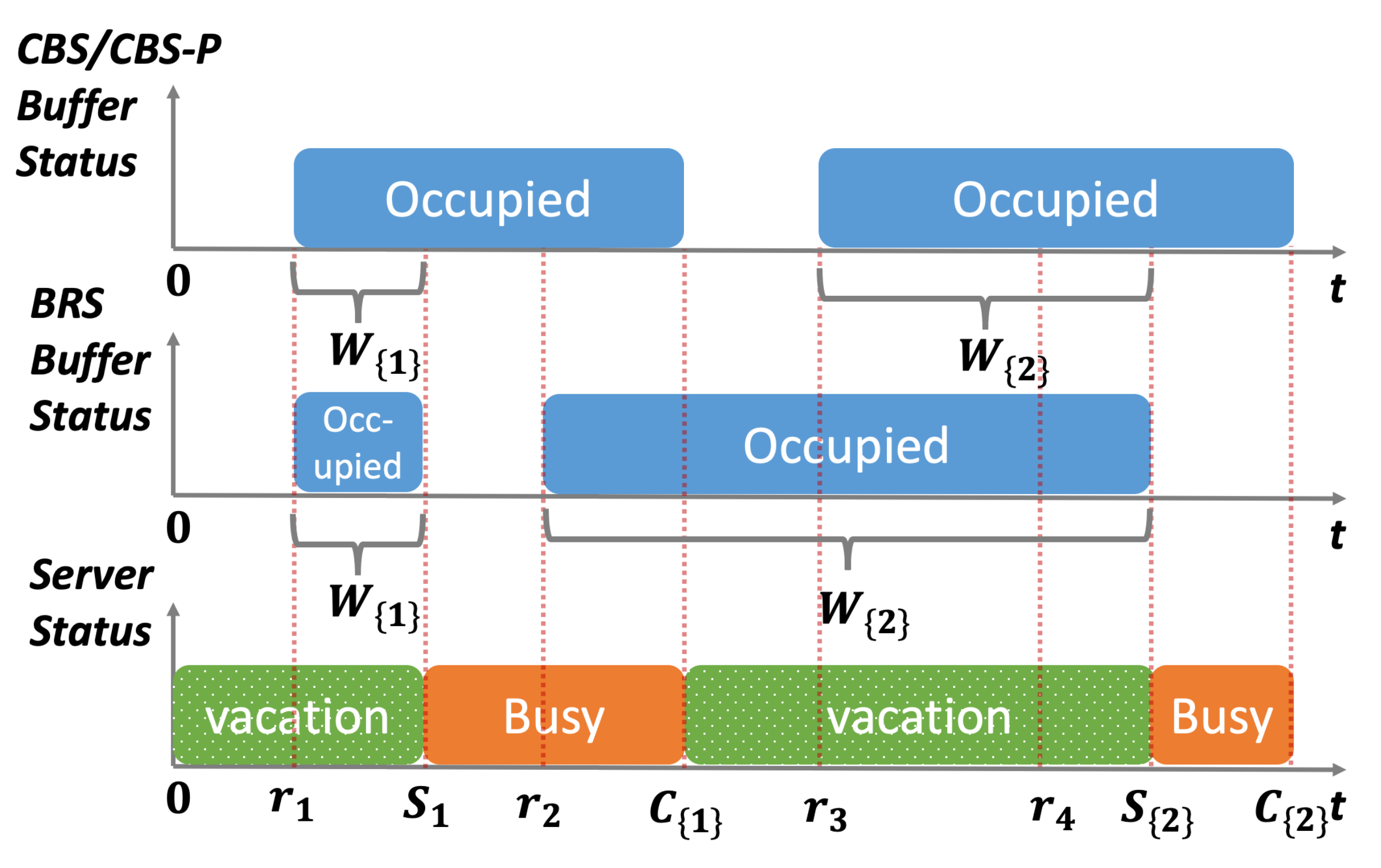}

\caption{Buffer Status, with $W_{\{l\}}$ being $W$ within the $l^{th}$ regenerative
cycle. \label{fig:Buffer-Status}}
\end{figure}

We now define a random variable that is useful in deriving $G^{*}(s)$.
Let $W$ be the period that starts from when the buffer becomes non-empty
within a regenerative cycle $I$, to the end of the regenerative cycle.
Note that $W$ in CBS and CBS-P only starts when the server is on
vacation (i.e., the system is empty). But $W$ in BRS could start
when the server is processing because the buffer becomes available
as soon as processing starts. Another way to understand $W$ is as
follows. We consider a dummy system for each of the three systems.
In each dummy system, packet replacement in the buffer is not allowed.
Each dummy system has the same vacation distribution as its original
counterpart, so that $W$ is equivalent to the packet waiting time
in the dummy system. A demonstrative graph about buffer status and
period $W$ is provided in Fig. \ref{fig:Buffer-Status}. The following
lemma reveals the relation between $G^{*}(s)$ and $W^{*}(s)$, which
will be useful for our derivations later.

\begin{lem}
\label{lem:1}For CBS, BRS, and CBS-P, it holds that $G^{*}(s)=\frac{\lambda}{\lambda+s}+\frac{s}{\lambda+s}W^{*}(\lambda+s)$.
\end{lem}
\begin{IEEEproof}
Since $G$ is the waiting time of the last packet that arrived during
$W$, from Campbell Theorem (P173, Theorem 5.14 in \cite{kulkarni2016modeling}),
we have that $\boldsymbol{P}(G\leq x|m(t)=m,W=t)=1-(\frac{t-x}{t})^{m}$
for $x\leq t$ and $m\geq1$. Since $W$ is the period during which
the buffer is occupied, if there is no arrival during $W,$ then $G=W$.
So that $\boldsymbol{E}[e^{-sG}|m(t)=0,W=t]=e^{-st},$ and we have
\begin{eqnarray*}
 &  & \boldsymbol{E}[e^{-sG}|W=t]\\
 & = & \int_{x=0}^{t}e^{-sx}\sum_{m=1}^{\infty}\frac{m(t-x)^{m-1}}{t^{m}}e^{-\lambda t}\frac{(\lambda t)^{m}}{m!}dx\\
 &  & +e^{-st}e^{-\lambda t}\\
 & = & \frac{\lambda}{\lambda+s}+\frac{s}{\lambda+s}e^{-(\lambda+s)t}.
\end{eqnarray*}
By unconditioning on $W=t$, the lemma can be proven.
\end{IEEEproof}
From Lemma \ref{lem:1} we can get
\begin{eqnarray}
\boldsymbol{E}[G] & = & \frac{1}{\lambda}(1-\boldsymbol{E}[e^{-\lambda W}]),\label{eq:1.2}
\end{eqnarray}
Lemma \ref{lem:1} implies that one can derive $G^{*}(s)$ by obtaining
the LST of waiting time in the dummy system. This result will be useful
in our derivation later. It was shown in \cite{xu2019towards} that
$\boldsymbol{E}[I]=\frac{1}{\lambda}+\boldsymbol{E}[W]$ for systems
with Poisson arrivals. However, this fact cannot be used to calculate
AoI or variance of peak age, as the time period during which the buffer
is empty (with expectation $\frac{1}{\lambda}$) and the time period
$W$ are not independent. As we will show later in this section, the
relation between $W^{*}(s)$ and $I^{*}(s)$ could be involved. Therefore,
to obtain the AoI, PAoI, and variance of peak age for CBS, BRS, and
CBS-P with i.i.d. vacations, we will need to derive $I^{*}(s)$ first,
and then characterize the relation between $W^{*}(s)$ and $I^{*}(s)$.

\subsection{Conventional Buffer System }

In this subsection, we will derive the information freshness metrics
for CBS. Recall that in CBS, the buffer will not be available until
the processing is completed, and the server will start a vacation
once the buffer becomes empty. We provide the results for CBS in the
following theorem. 
\begin{thm}
\label{thm:The-AoI-of-CBS}The AoI of CBS is 
\begin{eqnarray*}
 &  & \boldsymbol{E}[\Delta_{CBS}]\\
 & = & -\frac{1}{2(H^{*(1)}(0)+\frac{V^{*(1)}(0)}{1-V^{*}(\lambda)})}\bigg[H^{*(2)}(0)\\
 &  & +2H^{*(1)}(0)\frac{V^{*(1)}(0)}{1-V^{*}(\lambda)}+\frac{2V^{*(1)}(0)V^{*(1)}(\lambda)}{(1-V^{*}(\lambda))^{2}}\\
 &  & +\frac{V^{*(2)}(0)}{1-V^{*}(\lambda)}\bigg]+\frac{1}{\lambda}+\frac{V^{*(1)}(\lambda)}{1-V^{*}(\lambda)}-H^{*(1)}(0),
\end{eqnarray*}
the PAoI of CBS is 
\begin{eqnarray*}
\boldsymbol{E}[A_{CBS}] & = & \frac{1}{\lambda}+\frac{V^{*(1)}(\lambda)-V^{*(1)}(0)}{1-V^{*}(\lambda)}-2H^{*(1)}(0),
\end{eqnarray*}
and the variance of peak age of CBS is 
\begin{eqnarray*}
 &  & Var(A_{CBS})\\
 & = & \frac{V^{*(2)}(0)-V^{*(2)}(\lambda)}{1-V^{*}(\lambda)}-\left(\frac{V^{*(1)}(\lambda)-V^{*(1)}(0)}{1-V^{*}(\lambda)}\right)^{2}\\
 &  & +\frac{1}{\lambda^{2}}+2H^{*(2)}(0)-2\left(H^{*(1)}(0)\right)^{2}.
\end{eqnarray*}
\end{thm}
\begin{IEEEproof}
We first show that $I^{*}(s)=H^{*}(s)\frac{V^{*}(s)-V^{*}(s+\lambda)}{1-V^{*}(s+\lambda)}$
for CBS. Notice that the period $I$ starts once the server starts
processing, and ends when the server returns from a vacation and observes
a packet waiting in the buffer. Therefore $I^{*}(s)=\boldsymbol{E}[e^{-s(H+B)}]$,
where $B$ is the time period during which the server is continuously
on vacation within $I$. Note that $B$ may consist multiple vacations.
Let $B^{*}(s)$ be the LST of $B$. We let $V_{1}$ be the first vacation
taken during $B$, and $V_{\infty}$ be the last vacation. Let $m(V_{1})$
be the number of arrivals during vacation $V_{1}$. If $m(V_{1})\geq1$,
then $V_{1}=V_{\infty}=B$. Therefore, by conditioning on $V_{1}$
and $m(V_{1})$, we obtain $\boldsymbol{E}[e^{-sB}|V_{1}=v_{1},m(V_{1})\geq1]=e^{-sv_{1}}$
and $\boldsymbol{E}[e^{-sB}|V_{1}=v_{1},m(V_{1})=0]=e^{-sv_{1}}B^{*}(s)$.
Unconditioning on $m(v_{1})$, we have $\boldsymbol{E}[e^{-sB}|V_{1}=v_{1}]=e^{-sv_{1}}(1-e^{-\lambda v_{1}})+e^{-sv_{1}}B^{*}(s)e^{-\lambda v_{1}}$.
We then obtain $B^{*}(s)=\frac{V^{*}(s)-V^{*}(s+\lambda)}{1-V^{*}(s+\lambda)}$
by further unconditioning on $V_{1}=v_{1}$. Then $I^{*}(s)=H^{*}(s)\frac{V^{*}(s)-V^{*}(s+\lambda)}{1-V^{*}(s+\lambda)}$.

Now we derive $W^{*}(s)$. We notice a fact that $W$ only occurs
in $V_{\infty}$ within period $B$. So the number of arrivals during
the last vacation $m(V_{\infty})$ always satisfies $m(V_{\infty})\geq1$.
From Campbell's Theorem (P173, Theorem 5.14 in \cite{kulkarni2016modeling}),
it holds that $\boldsymbol{E}[e^{-sW}|m(t)=m,V_{\infty}=t]=\int_{0}^{t}e^{-sx}\frac{mx^{m-1}}{t^{m}}dx.$
Unconditioning on $m(t)=m$ and using the fact that $\boldsymbol{P}(m(t)=m|m(t)\geq1)=\frac{(\lambda t)^{m}}{m!}\frac{e^{-\lambda t}}{1-e^{-\lambda t}},$
we have 
\begin{eqnarray*}
 &  & \boldsymbol{E}[e^{-sW}|V_{\infty}=t,m(V_{\infty})\geq1]\\
 & = & \sum_{m=1}^{\infty}\int_{x=0}^{t}e^{-sx}\frac{mx^{m-1}}{t^{m}}\frac{e^{-\lambda t}}{1-e^{-\lambda t}}\frac{(\lambda t)^{m}}{m!}dx\\
 & = & \int_{x=0}^{t}e^{-sx}\frac{e^{-\lambda t}}{1-e^{-\lambda t}}\sum_{m=1}^{\infty}\frac{(\lambda x)^{m-1}}{(m-1)!}\lambda dx\\
 & = & \frac{e^{-\lambda t}-e^{-st}}{(s-\lambda)(1-e^{-\lambda t})}\lambda.
\end{eqnarray*}
Now we need to derive $\boldsymbol{P}(t<V_{\infty}\leq t+dt|m(V_{\infty})\geq1).$
From 
\begin{eqnarray*}
 &  & \boldsymbol{P}(V_{\infty}\leq x|m(V_{\infty})\geq1)\\
 &  & =\frac{\boldsymbol{P}(V_{\infty}\leq x,m(V_{\infty})\geq1)}{\boldsymbol{P}(m(V_{\infty})\geq1)}=\frac{\int_{0}^{x}(1-e^{-\lambda u})dF_{V}(u)}{\int_{0}^{\infty}(1-e^{-\lambda u})dF_{V}(u)}\\
 &  & =\frac{\int_{0}^{x}(1-e^{-\lambda u})dF_{V}(u)}{1-V^{*}(\lambda)},
\end{eqnarray*}
we have $\boldsymbol{P}(t<V_{\infty}\leq t+dt|m(V_{\infty})\geq1)=\frac{(1-e^{-\lambda t})dF_{V}(t)}{1-V^{*}(\lambda)}$.
Therefore 
\begin{eqnarray*}
 &  & \boldsymbol{E}[e^{-sW}|m(V_{\infty})\geq1]\\
 & = & \int_{0}^{\infty}\boldsymbol{E}[e^{-sW}|V_{\infty}=t,m(V_{\infty})\geq1]\frac{(1-e^{-\lambda t})dF_{V}(t)}{1-V^{*}(\lambda)}\\
 & = & \int_{0}^{\infty}\frac{e^{-\lambda t}-e^{-st}}{(s-\lambda)(1-e^{-\lambda t})}\lambda\frac{(1-e^{-\lambda t})dF_{V}(t)}{1-V^{*}(\lambda)}\\
 & = & \frac{V^{*}(\lambda)-V^{*}(s)}{(s-\lambda)(1-V^{*}(\lambda))}\lambda.
\end{eqnarray*}
Since $W$ only occurs in the last vacation, and the last vacation
always has $m(V_{\infty})\geq1$, we thus have $\boldsymbol{E}[e^{-sW}|m(V_{\infty})\geq1]=\boldsymbol{E}[e^{-sW}].$
We then have $W^{*}(s)=\frac{V^{*}(\lambda)-V^{*}(s)}{(s-\lambda)(1-V^{*}(\lambda))}\lambda$.
By Lemma \ref{lem:1}, it holds that $G^{*}(s)=\frac{\lambda}{\lambda+s}\frac{1-V^{*}(s+\lambda)}{1-V^{*}(\lambda)}$
and $\boldsymbol{E}[G]=\frac{1}{\lambda}+\frac{V^{*(1)}(\lambda)}{1-V^{*}(\lambda)}$.
By taking the second derivative of $G^{*}(s)$, we obtain 
\begin{eqnarray*}
G^{*(2)}(0) & = & \frac{2}{\lambda^{2}}+\frac{2}{\lambda}\frac{V^{*(1)}(\lambda)}{1-V^{*}(\lambda)}-\frac{V^{*(2)}(\lambda)}{1-V^{*}(\lambda)}.
\end{eqnarray*}
Using Equation (\ref{eq:1.1}) we can obtain $Var(A)$.
\end{IEEEproof}
In the proof of Theorem \ref{thm:The-AoI-of-CBS}, we used the fact
that $W$ only occurs in the last $V$ during period $B$ that comprises
a sequence of vacations. This fact also holds for BRS and CBS-P since
we assume the multiple vacation scheme for all three systems.

{We find from Theorem \ref{thm:The-AoI-of-CBS} that
$\boldsymbol{E}[A_{CBS}]$ is determined by $\boldsymbol{E}[H]$,
which means that different service time distributions may have the
same expression for $\boldsymbol{E}[A_{CBS}]$, as long as their mean
values are equivalent. However, $\boldsymbol{E}[\Delta_{CBS}]$ and
$Var(A_{CBS})$ are determined by both $\boldsymbol{E}[H]$ and $\boldsymbol{E}[H^{2}]$.
When fixing $\boldsymbol{E}[H]$, the processing time distribution
with a small $\boldsymbol{E}[H^{2}]$ (i.e., a small variance) will
reduce both $\boldsymbol{E}[A_{CBS}]$ and $Var(A_{CBS})$. We also
find that the distribution of vacation time $V$ uniquely determines
the expressions of $\boldsymbol{E}[\Delta_{CBS}],$ $\boldsymbol{E}[A_{CBS}]$,
and $Var(A_{CBS})$} {since these three metrics are
functions of the LST of $V$. }As the expressions for $\boldsymbol{E}[\Delta_{CBS}],$
$\boldsymbol{E}[A_{CBS}]$, and $Var(A_{CBS})$ in Theorem \ref{thm:The-AoI-of-CBS}
are involved, in the next corollary, we provide the results for CBS
with exponential service and vacation times.
\begin{cor}
\label{cor:If-the-vacation}For exponential vacation time with parameter
$v$ and exponential service time with parameter $\mu$, we have 
\begin{eqnarray*}
\boldsymbol{E}[\Delta_{CBS}] & = & \frac{1}{\lambda}+\frac{1}{v}-\frac{\lambda+v+\mu}{v\lambda+\mu\lambda+\mu v}+\frac{1}{v+\lambda}+\frac{2}{\mu},
\end{eqnarray*}
\begin{eqnarray*}
\boldsymbol{E}[A_{CBS}] & = & \frac{1}{\lambda}+\frac{1}{v}+\frac{1}{v+\lambda}+\frac{2}{\mu},
\end{eqnarray*}
and $Var(A_{CBS})=\frac{1}{(\lambda+v)^{2}}+\frac{1}{\lambda^{2}}+\frac{1}{v^{2}}+\frac{2}{\mu^{2}}.$
\end{cor}
\begin{IEEEproof}
When the vacation time is exponentially distributed, we have $I^{*}(s)=\frac{\mu v\lambda}{(\mu+s)(v+s)(\lambda+s)}$.
So from $\boldsymbol{E}[I]=\frac{v\lambda+\mu\lambda+\mu v}{\mu v\lambda}$
and $\boldsymbol{E}[I^{2}]=2\frac{(v\lambda+\mu\lambda+\mu v)^{2}}{\mu^{2}v^{2}\lambda^{2}}-2\frac{\lambda+v+\mu}{\mu v\lambda}$,
we have $Var(I)=\frac{1}{\mu^{2}}+\frac{1}{v^{2}}+\frac{1}{\lambda^{2}}.$
Also we know $\boldsymbol{E}[G]=\frac{1}{v+\lambda}$ and $\boldsymbol{E}[G^{2}]=\frac{2}{(v+\lambda)^{2}}.$
So that the results can be obtained from Equations (\ref{eq:0}) and
(\ref{eq:1}).
\end{IEEEproof}
{We find from Corollary \ref{cor:If-the-vacation}
that $\boldsymbol{E}[A_{CBS}]$ is an upper bound for $\boldsymbol{E}[\Delta_{CBS}]$
in this special case. One can easily verify from Corollary \ref{cor:If-the-vacation}
that $\boldsymbol{E}[\Delta_{CBS}]$, $\boldsymbol{E}[A_{CBS}]$,
and $Var(A_{CBS})$ are decreasing on $\lambda,$ $\mu$ and $v$,
which means that increasing the sampling, service, and vacation rates
can reduce $\boldsymbol{E}[\Delta_{CBS}]$, $\boldsymbol{E}[A_{CBS}]$
and $Var(A_{CBS})$ for CBS, when both service and vacation times
are exponential. We will show in Section \ref{sec:Numerical-Study}
that when the service times are not exponentially distributed, increasing
the vacation rate does not always reduce $\boldsymbol{E}[\Delta_{CBS}]$. }

\subsection{Buffer Relaxation System}

In BRS, the server will take a vacation after processing a packet,
and the packet arriving during processing will be processed only when
the vacation is over. This service discipline is also called ``gated''
in some literature about vacation server systems (see \cite{takagi1991analysis,takagi1988queuing,gautam2012analysis}).
This gated policy prevents the server from serving the buffer continuously
without taking vacations when the arrival rate is large, which is
helpful for systems where vacations have to be taken, such as the
priority queue systems \cite{xu2019towards} where vacations correspond
to ``serving the prioritized queues''. Also, as we will see in Section
\ref{subsec:System-Comparison}, BRS has the advantage over CBS in
terms of minimizing PAoI. We now provide the AoI and PAoI for BRS
in the following theorem.
\begin{thm}
\label{thm:The-AoI-of}The AoI of BRS is 
\begin{eqnarray*}
 &  & \boldsymbol{E}[\Delta_{BRS}]\\
 & = & \frac{-I^{*(2)}(0)}{2I^{*(1)}(0)}+V^{*(1)}(\lambda)H^{*}(\lambda)+V^{*}(\lambda)H^{*(1)}(\lambda)\\
 &  & +\frac{V^{*(1)}(\lambda)}{1-V^{*}(\lambda)}V^{*}(\lambda)H^{*}(\lambda)+\frac{1}{\lambda}-H^{*(1)}(0),
\end{eqnarray*}
where $I^{*}(s)=H^{*}(s)V^{*}(s)+H^{*}(\lambda+s)\frac{V^{*}(s+\lambda)(V^{*}(s)-1)}{1-V^{*}(s+\lambda)}.$
The PAoI of BRS is 
\begin{eqnarray*}
\boldsymbol{E}[A_{BRS}] & = & -2H^{*(1)}(0)-V^{*(1)}(0)+\frac{1}{\lambda}\\
 &  & +V^{*(1)}(\lambda)H^{*}(\lambda)+V^{*}(\lambda)H^{*(1)}(\lambda)\\
 &  & +\frac{H^{*}(\lambda)V^{*}(\lambda)}{1-V^{*}(\lambda)}(V^{*(1)}(\lambda)-V^{*(1)}(0)).
\end{eqnarray*}
\end{thm}
\begin{IEEEproof}
The proof relies on the renewal argument, which is similar to the
proof of Theorem \ref{thm:The-AoI-of-CBS}. The detail of the proof
is shown in Appendix \ref{sec:Proof-for-Theorem-4-1} of the supplementary
material.
\end{IEEEproof}
We can also obtain the variance of peak age for BRS, albeit its closed-form
expression is involved. To obtain the variance of peak age, we need
the LST of $G$, $I$, and $H$, as shown in Equation (\ref{eq:1.1}).
The LST of $I$ has been given in Theorem \ref{thm:The-AoI-of}, which
is 
\begin{eqnarray*}
I^{*}(s)=H^{*}(s)V^{*}(s)+H^{*}(\lambda+s)\frac{V^{*}(s+\lambda)(V^{*}(s)-1)}{1-V^{*}(s+\lambda)}.
\end{eqnarray*}
The expression of $W^{*}(s)$ is given in the proof of Theorem \ref{thm:The-AoI-of},
from which we can obtain 
\begin{eqnarray*}
G^{*}(s) & = & \frac{\lambda}{\lambda+s}\bigg[1+\frac{V^{*}(\lambda)H^{*}(\lambda)}{1-V^{*}(\lambda)}(1-V^{*}(\lambda+s))\\
 &  & -V^{*}(\lambda+s)H^{*}(\lambda+s)\bigg].
\end{eqnarray*}
{One can see from Theorem \ref{thm:The-AoI-of} that
both $\boldsymbol{E}[\Delta_{BRS}]$ and $\boldsymbol{E}[A_{BRS}]$
are uniquely determined by the distributions of vacation time $V$
and processing time $H$. We will show the numerical results for the
variance of peak age for BRS in Section \ref{sec:Numerical-Study}.
In the next corollary, we provide the results for BRS with exponential
service and exponential vacation times.}
\begin{cor}
\label{cor:For-exponential-vacation-1}For exponential vacation time
with parameter $v$ and exponential service time with parameter $\mu$,
we have 
\begin{eqnarray*}
 &  & \boldsymbol{E}[\Delta_{BRS}]\\
 & = & \bigg[\frac{1}{v^{2}}+\frac{1}{v\mu}+\frac{1}{\mu^{2}}+\frac{\mu}{\lambda v(\lambda+\mu)}+\frac{\mu}{\lambda^{2}(\lambda+\mu)}\\
 &  & +\frac{\mu}{\lambda(\lambda+\mu)^{2}}\bigg]\bigg/(\frac{1}{v}+\frac{1}{\mu}+\frac{\mu}{\lambda(\lambda+\mu)})\\
 &  & +\frac{1}{\lambda+v}+\frac{\lambda v}{(\lambda+\mu)^{2}(\lambda+v)}+\frac{1}{\mu}
\end{eqnarray*}
 and 
\begin{eqnarray*}
\boldsymbol{E}[A_{BRS}] & = & \frac{\mu^{2}-\mu v+\lambda\mu}{(\lambda+\mu)^{2}(\lambda+v)}+\frac{1}{v}+\frac{2}{\mu}+\frac{1}{\lambda}.
\end{eqnarray*}
\end{cor}
\begin{IEEEproof}
The results follow from Theorem \ref{thm:The-AoI-of} with $V^{*}(s)=\frac{v}{v+s}$
and $H^{*}(s)=\frac{\mu}{\mu+s}$.
\end{IEEEproof}
{One can easily verify that $\boldsymbol{E}[A_{BRS}]$
decreases on $\lambda$, $\mu$, and $v$ by taking the derivative.
It implies that increasing the generation, service, and vacation rates
can reduce PAoI in this special case. In Section \ref{sec:Numerical-Study},
we will show $\boldsymbol{E}[\Delta_{BRS}]$ does not always decrease
as the vacation rate $v$ increases when the service time is not exponential.}

\subsection{Conventional Buffer System with Preemption in Service}

Note that when allowing preemption in service, both CBS and BRS will
reduce to CBS-P. Unlike the non-preemptive service case, in CBS-P,
the age peak cannot be decomposed as shown in Equation (\ref{eq:0}),
simply because a packet that results in age peak may not have the
waiting time $G$ (as it may be a preemptive packet). A detailed decomposition
approach for CBS-P is given in Appendix \ref{sec:Proof-for-Theorem}
of the supplementary material. The AoI and PAoI are CBS-P is given
in the following theorem.
\begin{thm}
\label{thm:The-PAoI-for}The AoI for CBS-P is 
\begin{eqnarray*}
 &  & \boldsymbol{E}[\Delta_{CBS-P}]\\
 & = & \frac{1}{2(-\frac{V^{*(1)}(0)}{1-V^{*}(\lambda)}+\frac{1-H^{*}(\lambda)}{\lambda H^{*}(\lambda)})}\bigg\{\frac{V^{*(2)}(0)}{1-V^{*}(\lambda)}\\
 &  & +2\frac{V^{*(1)}(0)V^{*(1)}(\lambda)}{(1-V^{*}(\lambda))^{2}}-2\frac{V^{*(1)}(0)}{1-V^{*}(\lambda)}\frac{1-H^{*}(\lambda)}{\lambda H^{*}(\lambda)}\\
 &  & +\frac{2}{\lambda H^{*}(\lambda)^{2}}\big[\frac{1}{\lambda}-\frac{H^{*}(\lambda)}{\lambda}+H^{*(1)}(\lambda)\big]\bigg\}\\
 &  & -\frac{H^{*(1)}(\lambda)}{H^{*}(\lambda)}+H^{*}(\lambda)\big(\frac{1}{\lambda}+\frac{V^{*(1)}(\lambda)}{1-V^{*}(\lambda)}\big),
\end{eqnarray*}
 and the PAoI for CBS-P is 
\begin{eqnarray*}
\boldsymbol{E}[A_{CBS-P}] & = & \frac{1-H^{*}(\lambda)-\lambda H^{*(1)}(\lambda)+H^{*}(\lambda)^{2}}{\lambda H^{*}(\lambda)}\\
 &  & +\frac{H^{*}(\lambda)V^{*(1)}(\lambda)-V^{*(1)}(0)}{1-V^{*}(\lambda)}.
\end{eqnarray*}
 
\end{thm}
\begin{IEEEproof}
See Appendix \ref{sec:Proof-for-Theorem} of the supplementary material.
\end{IEEEproof}
{We also find from Theorem \ref{thm:The-PAoI-for}
that $\boldsymbol{E}[\Delta_{CBS-P}]$ and $\boldsymbol{E}[A_{CBS-P}]$
are uniquely determined by the distributions of vacation time $V$
and processing time $H$. In the next corollary, we provide the expressions
for $\boldsymbol{E}[\Delta_{CBS-P}]$ and $\boldsymbol{E}[A_{CBS-P}]$
when both vacation and processing time distributions are exponential.}
\begin{cor}
\label{cor:For-exponential-vacation}For exponential vacation time
with parameter $v$ and exponential service time with parameter $\mu$,
we have 
\begin{eqnarray*}
\boldsymbol{E}[A_{CBS-P}] & = & \frac{1}{\lambda}+\frac{1}{\mu}+\frac{1}{v}+\frac{\lambda+\mu+v}{(\lambda+\mu)(\lambda+v)}
\end{eqnarray*}
 and 
\begin{eqnarray*}
\boldsymbol{E}[\Delta_{CBS-P}] & = & \frac{1}{v}+\frac{1}{\lambda}+\frac{1}{\mu}-\frac{\mu+v+\lambda}{\lambda\mu+v\mu+\lambda v}\\
 &  & +\frac{v+\mu+\lambda}{(\mu+\lambda)(v+\lambda)}.
\end{eqnarray*}
\end{cor}
It can be observed from Corollary \ref{cor:For-exponential-vacation}
that when service and vacation times are both exponential, $\boldsymbol{E}[\Delta_{CBS-P}]$
is always upper bounded by $\boldsymbol{E}[A_{CBS-P}]$. One can also
verify that in this case, $\boldsymbol{E}[A_{CBS-P}]$ and $\boldsymbol{E}[\Delta_{CBS-P}]$
are decreasing on parameters $\lambda$, $\mu$, and $v$ by taking
the derivatives. The variance of peak age for CBS-P can also be obtained
by the decomposition approach given in Appendix \ref{sec:Proof-for-Theorem},
but its expression is involved. We will show it numerically in Section
\ref{sec:Numerical-Study}. 

\subsection{System Comparison \label{subsec:System-Comparison}}

\begin{figure}[h]
\begin{center}\includegraphics[scale=0.4]{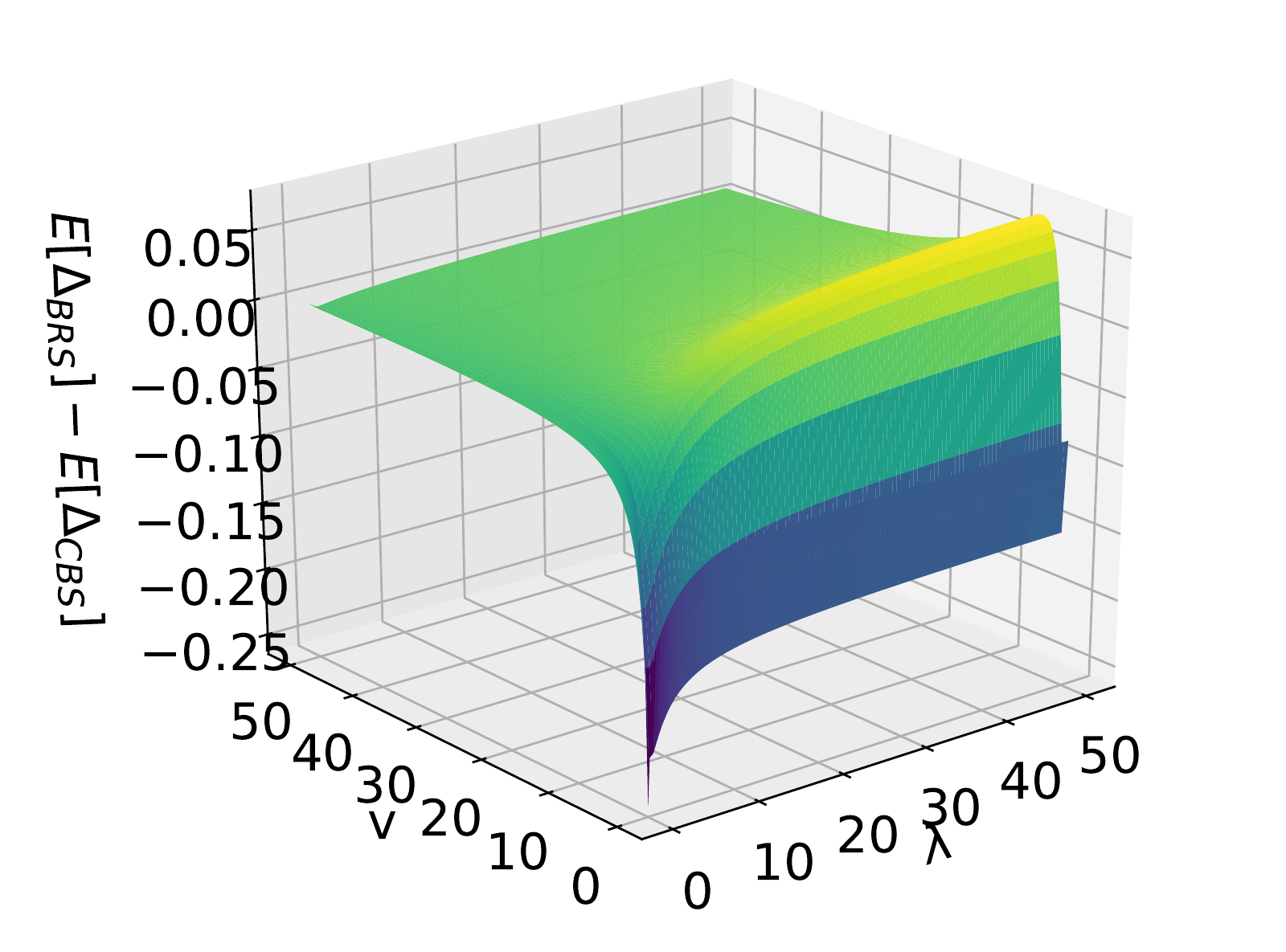}\end{center}\caption{AoI in CBS vs AoI in BRS, $H\sim exp(1),V\sim exp(v)$ \label{fig:AoI-of-CBS}}
\end{figure}

We mainly compare the AoI and PAoI under different policies in this
subsection. The expressions for variance of peak age for BRS and CBS-P
are involved, so that we will compare the variance of peak age numerically
in Section \ref{sec:Numerical-Study}. We first compare the PAoI for
CBS and BRS in the following theorem.
\begin{thm}
\label{thm:BRS-always-has} The PAoI in BRS is always no greater than
that in CBS, if the vacation times are i.i.d.
\end{thm}
\begin{IEEEproof}
See Appendix \ref{sec:Proof-for-Theorem-3} of the supplementary material.
\end{IEEEproof}
{Theorem \ref{thm:BRS-always-has} shows that allowing
the buffer to be available all the time, i.e., adopting BRS, can achieve
a smaller PAoI than CBS. However, as shown in Fig. \ref{fig:AoI-of-CBS},
BRS does not always have a smaller AoI than CBS, which implies that
a policy that reduces PAoI does not necessarily reduce AoI. One can
also find from Fig. \ref{fig:AoI-of-CBS} that CBS has a smaller AoI
than BRS when both $v$ and $\lambda$ are large, but the advantage
that CBS has over BRS is not significant. When both $v$ and $\lambda$
are small, $\boldsymbol{E}[\Delta_{BRS}]$ could be much smaller than
$\boldsymbol{E}[\Delta_{CBS}].$ This observation shows the advantage
of adopting BRS when the vacation time is large and the data generation
rate is low.}

\begin{figure}[h]
\begin{center}\subfloat[$H\sim Gamma(2,1)$]{\includegraphics[scale=0.33]{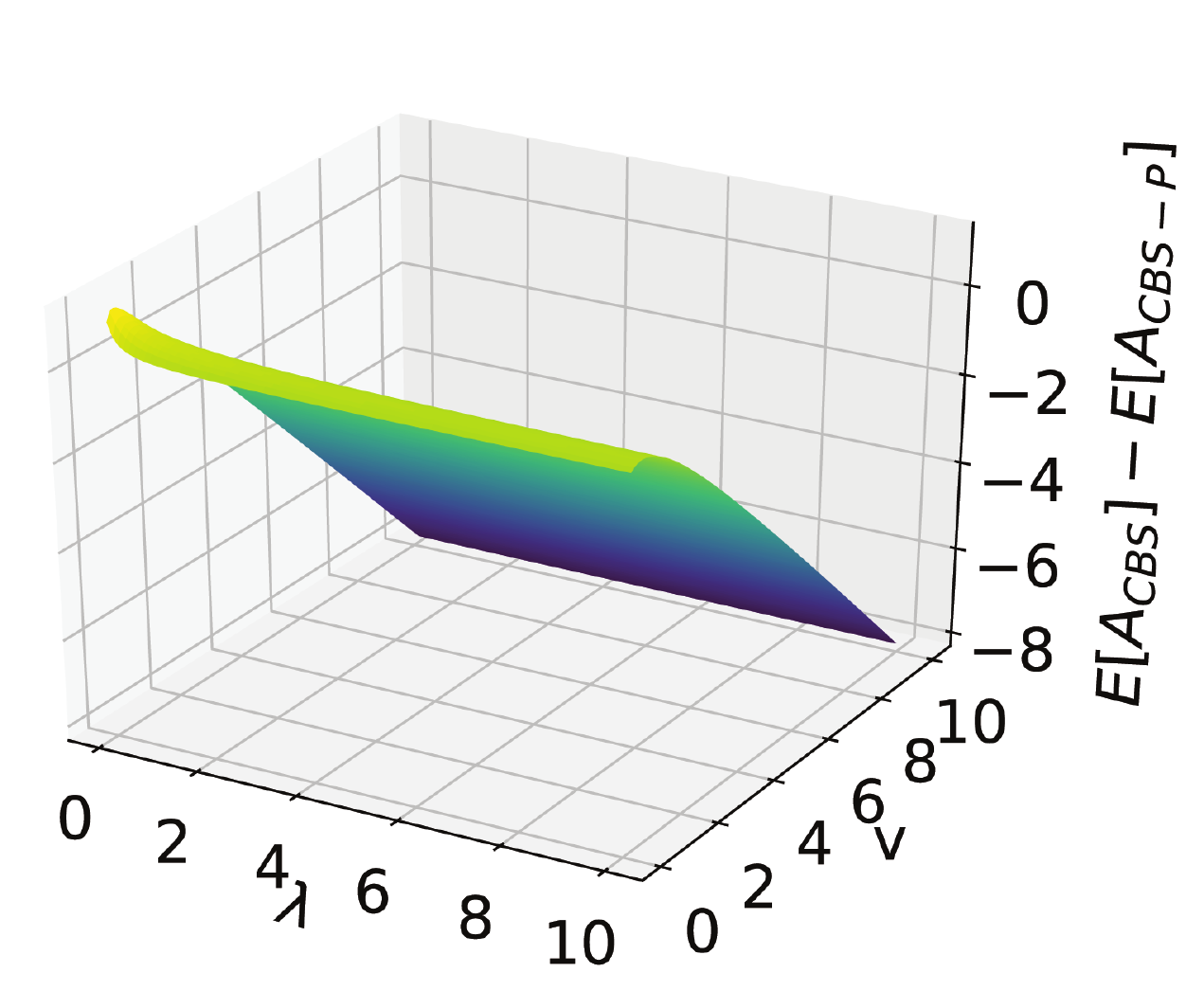}

}\subfloat[$H\sim Gamma(\frac{1}{2},1)$]{\includegraphics[scale=0.33]{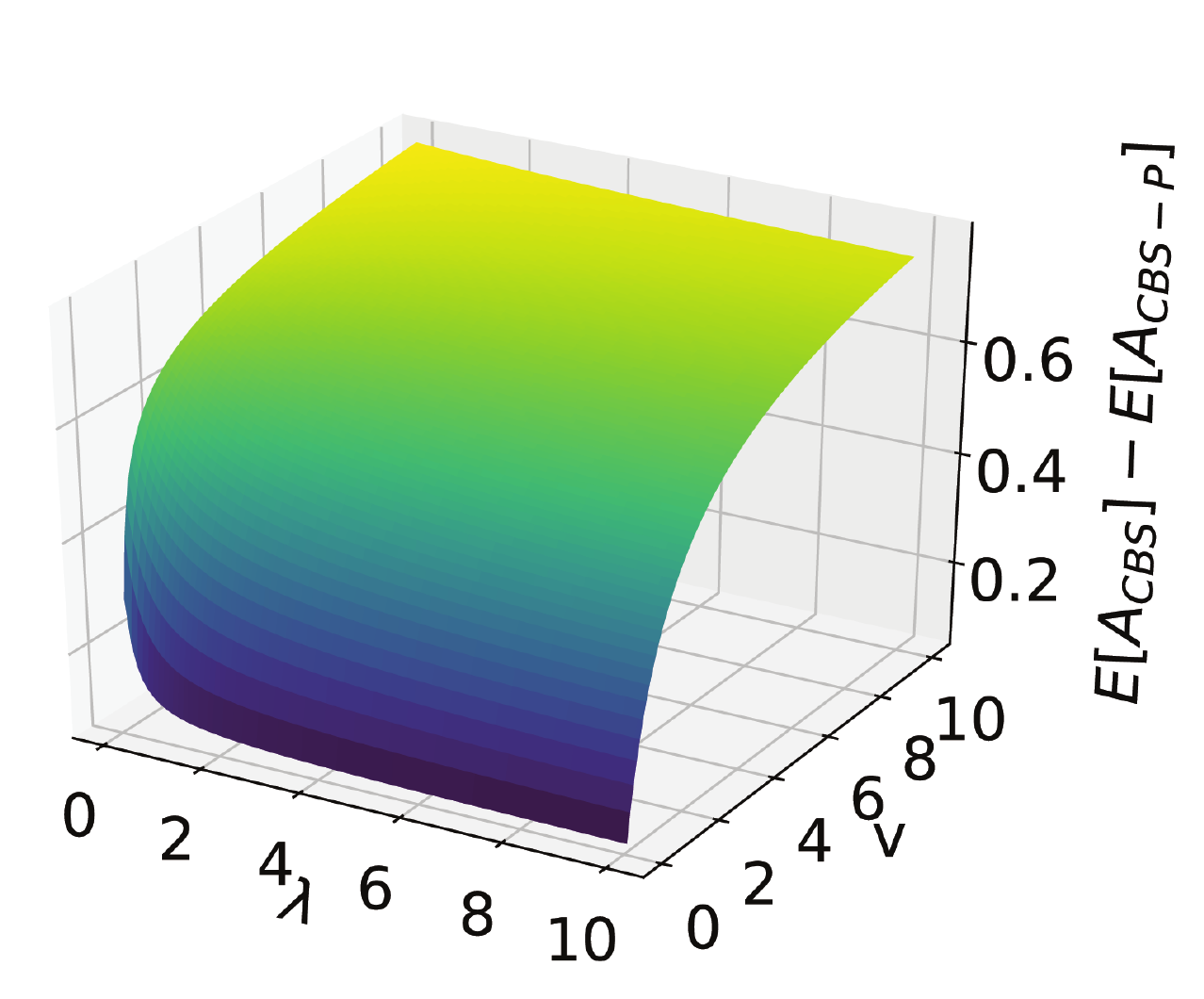}

}\end{center}

\caption{PAoI in CBS vs. PAoI in CBS-P. Service time is Gamma distributed.
Vacation time is exponentially distributed. \label{fig:PAoI-of-CBS}}

\end{figure}

We then compare CBS-P with CBS, and we have the following theorems.
\begin{thm}
\label{thm:If-the-service}If the service time is exponentially distributed,
then the AoI and PAoI in CBS-P are no greater than those in CBS, when
vacation times are i.i.d.
\end{thm}
\begin{IEEEproof}
See Appendix \ref{sec:Proof-for-Theorem-1} of the supplementary material.
\end{IEEEproof}
Note that Theorem \ref{thm:If-the-service} holds for systems with
vacation time being general and service time being exponential. It
does not always hold when the service time is non-exponential, as
shown numerically in Section \ref{sec:Numerical-Study}. In the following
theorem, we provide a sufficient condition under which CBS-P will
always have a PAoI no greater than CBS.
\begin{thm}
\label{thm:If-the-service-2}If the service time $H$ satisfies $\boldsymbol{E}[H]\geq\frac{1-H^{*}(s)}{sH^{*}(s)}$
for all $s>0$, then CBS-P always has a PAoI no greater than that
in CBS, when vacation times are i.i.d.
\end{thm}
\begin{IEEEproof}
See Appendix \ref{sec:Proof-for-Theorem-4} of the supplementary material.
\end{IEEEproof}
Theorem \ref{thm:If-the-service-2} provides a simple condition for
checking whether CBS-P has a smaller PAoI than CBS, and this sufficient
condition does not rely on the vacation time distribution. {We
now provide some examples of how Theorem \ref{thm:If-the-service-2}
can be applied. When the service time is exponential, we have $\frac{1-H^{*}(s)}{sH^{*}(s)}=\boldsymbol{E}[H].$
Then by Theorem \ref{thm:If-the-service-2} we can conclude that CBS-P
has a PAoI than no greater than that in CBS, which is the same as
our conclusion in Theorem \ref{thm:If-the-service}. We next give
an example where the processing time is Gamma distributed with parameters
$\alpha$ and $\beta$. Since the LST of Gamma distribution is given
by $H^{*}(s)=(1+\beta s)^{-\alpha}$, we have $\frac{1-H^{*}(s)}{sH^{*}(s)}=\frac{(1+\beta s)^{\alpha}-1}{s}.$
By Bernoulli's inequality we have that $(1+\beta s)^{\alpha}\geq1+\alpha\beta s$
when $\alpha\geq1,$ and $(1+\beta s)^{\alpha}<1+\alpha\beta s$ when
$\alpha<1$. From the fact that $\boldsymbol{E}[H]=\alpha\beta$,
we have $\frac{1-H^{*}(s)}{sH^{*}(s)}>\boldsymbol{E}[H]$ when $\alpha>1$,
and $\frac{1-H^{*}(s)}{sH^{*}(s)}\leq\boldsymbol{E}[H]$ when $\alpha\leq1$.
By Theorem \ref{thm:If-the-service-2}, CBS-P will have an advantage
over CBS when the service time is Gamma distributed with scale parameter
$\alpha\leq1$. For Gamma distributions with $\alpha\leq1$, the probability
density functions are more skewed than the exponential distribution.
Therefore, Theorem \ref{thm:If-the-service-2} implies that when service
time distribution is more skewed than the exponential distribution,
allowing preemption in processing would reduce PAoI. A numerical study
of this example is given in Fig. \ref{fig:PAoI-of-CBS}, from which
we find that when $\alpha=2,$ CBS-P does not always have a smaller
PAoI than CBS. When $\alpha=\frac{1}{2}$, CBS-P has a smaller PAoI
than CBS for all the positive values of $\lambda$ and $v$.}

Using the results of Theorems \ref{thm:The-AoI-of-CBS}, \ref{thm:The-AoI-of},
and \ref{thm:The-PAoI-for}, one can also derive other sufficient
and necessary conditions under which one policy performs better than
the others, by simply comparing the closed-form expressions. However,
those conditions may be complicated due to the closed-form expressions
for information freshness metrics being involved. 

\subsection{\label{subsec:Discussions-for-Systems}Discussions for Systems without
Server Vacation}

When the server takes no vacations (or takes vacation infinitely fast),
then CBS reduces to the M/G/1/1 non-preemptive system, BRS becomes
the M/G/1/2{*} system (the asterisk means that only the most recent
packet is kept in the buffer as defined in \cite{costa2016age,zou2019benefis}),
and CBS-P becomes M/G/1/1/preemptive system. Different variations
of these systems have been discussed in \cite{najm2018status,huang2015optimizing,zou2019benefis,costa2016age,kam2018age,inoue2019general}.
However, the variance of peak age in these single buffer systems has
not been studied. We here provide the variance of peak age for the
systems without server vacations as an extension of our discussion
about vacation server systems. With the decomposition approach that
we introduced earlier, we can provide the variance of peak age for
M/G/1/1, M/G/1/2{*}, and M/G/1/1/preemptive systems, as shown in Table
\ref{tab: information freshness metrics}. The detailed derivations
for Table \ref{tab: information freshness metrics} are provided in
Appendix \ref{sec:Derivations-for-Systems} of the supplementary material.
When the service time is exponentially distributed, we have Table
\ref{tab: information freshness metrics-1}.

\begin{table*}
\begin{center}%
\begin{tabular}{|>{\centering}p{2.7cm}|>{\centering}p{4.7cm}|>{\centering}p{3cm}|>{\centering}p{4.6cm}|}
\hline 
Systems & $\boldsymbol{E}[\Delta]$ & $\boldsymbol{E}[A]$ & $Var(A)$\tabularnewline
\hline 
\hline 
M/G/1/1 & $\big[\frac{2}{\lambda^{2}}-\frac{2}{\lambda}H^{*(1)}(0)+H^{*(2)}(0)\big]\big/\big[\frac{1}{\lambda}-H^{*(1)}(0)\big]-H^{*(1)}(0)$ & $\frac{1}{\lambda}-2H^{*(1)}(0)$ & $\frac{1}{\lambda^{2}}+2H^{*(2)}(0)-2\{H^{*(1)}(0)\}^{2}$\tabularnewline
\hline 
M/G/1/2{*} & $\big[\frac{1}{2}H^{*(2)}(0)+\frac{1}{\lambda^{2}}H^{*}(\lambda)-\frac{1}{\lambda}H^{*(1)}(\lambda)\big]\big/\big[-H^{*(1)}(0)+\frac{H^{*}(\lambda)}{\lambda}\big]+\frac{1}{\lambda}-\frac{1}{\lambda}H^{*}(\lambda)+H^{*(1)}(\lambda)-H^{*(1)}(0)$ & $-2H^{*(1)}(0)+\frac{1}{\lambda}+H^{*(1)}(\lambda)$ & $2H^{*(2)}(0)-2H^{*(1)}(0)+\frac{2H^{*}(\lambda)(1-H^{*}(\lambda))}{\lambda^{2}}+\frac{2H^{*}(\lambda)}{\lambda}[H^{*(1)}(0)+H^{*(1)}(\lambda)]+\frac{1}{\lambda^{2}}-H^{*(2)}(\lambda)-\frac{2}{\lambda}H^{*(1)}(\lambda)-H^{*(1)}(\lambda)^{2}$\tabularnewline
\hline 
M/G/1/1/Preemptive & $\frac{1}{\lambda H^{*}(\lambda)}$ & $\frac{-H^{*(1)}(\lambda)}{H^{*}(\lambda)}+\frac{1}{\lambda H^{*}(\lambda)}$ & $\frac{H^{*(2)}(\lambda)}{H^{*}(\lambda)}-\frac{\{H^{*(1)}(\lambda)\}^{2}}{H^{*}(\lambda)^{2}}+\frac{1}{\lambda^{2}H^{*}(\lambda)^{2}}+\frac{2H^{*(1)}(\lambda)}{\lambda H^{*}(\lambda)^{2}}$\tabularnewline
\hline 
\end{tabular}\end{center}

\caption{Information Freshness Metrics for Systems without Vacations \label{tab: information freshness metrics}}
\end{table*}

\begin{table*}
\begin{center}%
\begin{tabular}{|>{\centering}p{2.7cm}|>{\centering}p{4.6cm}|>{\centering}p{3.6cm}|>{\centering}p{3.6cm}|}
\hline 
Systems & $\boldsymbol{E}[\Delta]$ & $\boldsymbol{E}[A]$ & $Var(A)$\tabularnewline
\hline 
\hline 
M/M/1/1 & $\frac{1}{\lambda}+\frac{2}{\mu}-\frac{1}{\lambda+\mu}$ & $\frac{1}{\lambda}+\frac{2}{\mu}$ & $\frac{1}{\lambda^{2}}+\frac{2}{\mu^{2}}$\tabularnewline
\hline 
M/M/1/2{*} & $\frac{1}{\lambda}+\frac{2}{\mu}+\frac{\lambda}{(\lambda+\mu)^{2}}+\frac{1}{\lambda+\mu}-\frac{2(\lambda+\mu)}{\lambda^{2}+\lambda\mu+\mu^{2}}$ & $\frac{1}{\mu}+\frac{1}{\lambda}+\frac{\lambda}{(\mu+\lambda)^{2}}+\frac{\lambda}{\mu(\mu+\lambda)}$ & $\frac{1}{\lambda^{2}}+\frac{2}{\mu^{2}}-\frac{2\lambda^{2}+4\lambda\mu+3\mu^{2}}{(\lambda+\mu)^{4}}$\tabularnewline
\hline 
M/M/1/1/Preemptive & $\frac{1}{\mu}+\frac{1}{\lambda}$ & $\frac{1}{\mu+\lambda}+\frac{1}{\mu}+\frac{1}{\lambda}$ & $\frac{1}{(\lambda+\mu)^{2}}+\frac{1}{\lambda^{2}}+\frac{1}{\mu^{2}}$\tabularnewline
\hline 
\end{tabular}\end{center}

\caption{Information Freshness Metrics for Exponential Service Systems without
Vacations \label{tab: information freshness metrics-1}}
\end{table*}

The AoI and PAoI results for M/M/1/1 and M/M/1/2{*} systems in Table
\ref{tab: information freshness metrics-1} are the same as the ones
obtained in \cite{costa2016age}. The AoI and PAoI results for M/G/1/1/preemptive
system in Table \ref{tab: information freshness metrics} are the
same as the ones obtained in \cite{najm2018status}. The AoI and PAoI
results in Table \ref{tab: information freshness metrics-1} can also
be obtained from Corollaries \ref{cor:If-the-vacation}, \ref{cor:For-exponential-vacation-1}
and \ref{cor:For-exponential-vacation}, by letting $v\rightarrow\infty$.
These closed-form expressions enable us to evaluate the information
freshness in M/M/1/1, M/M/1/2{*}, and M/M/1/1/Preemptive systems.
Interestingly, no system always performs better or worse than the
other two systems in terms of all the three metrics: AoI, PAoI, and
variance of peak age. As shown in Table \ref{tab:Single-Queue-Systems},
although M/M/1/1/Preemptive has the smallest AoI among the three systems,
it does not have a smaller PAoI or variance of peak age than the other
two systems. M/M/1/1 turns out to perform worse than the other two
systems in terms of PAoI and variance of peak age, but its AoI is
not always greater than that in M/M/1/2{*} system. Note that Table
\ref{tab:Single-Queue-Systems} only compares the systems with exponential
service times. When service times are generally distributed, one can
easily verify that Theorems \ref{thm:BRS-always-has} and \ref{thm:If-the-service-2}
still hold for systems with no vacations. More numerical comparisons
are provided in Section \ref{sec:Numerical-Study}.

\begin{table}
\begin{center}%
\begin{tabular}{|c|>{\centering}p{1.4cm}|>{\centering}p{1.4cm}|>{\centering}p{1.4cm}|}
\hline 
Systems & $\boldsymbol{E}[\Delta]$ & $\boldsymbol{E}[A]$ & $Var(A)$\tabularnewline
\hline 
\hline 
M/M/1/1 & Could be smaller than M/M/1/2{*} & Largest & Largest\tabularnewline
\hline 
M/M/1/2{*} & Could be smaller than M/M/1/1 & Could be the smallest & Could be the smallest\tabularnewline
\hline 
M/M/1/1/Preemptive & Smallest & Could be the smallest & Could be the smallest\tabularnewline
\hline 
\end{tabular}\end{center}

\caption{Comparison for Systems without Vacations \label{tab:Single-Queue-Systems}}
\end{table}

\section{Peak Age of Information for Systems with Dependent Vacations \label{sec:AoI-Polling}}

We now extend our discussion to a more general case by allowing the
vacations to be non-i.i.d. Equations (\ref{eq:1}) and (\ref{eq:2})
may no longer hold in this case as $G$ and $I$ may not be independent.
However, we can still rely on Equation (\ref{eq:0}) to compute the
PAoI for each system. This section will discuss the approach for deriving
the exact solution for PAoI, and use PAoI to evaluate the information
freshness under each scheduling policy. In Section \ref{subsec:System-Comparison},
we showed that when vacation times are i.i.d., the PAoI in BRS is
always no greater than that in CBS, and the PAoI in CBS-P is always
no greater than that in CBS when the service time is exponential.
We aim to understand whether these results hold when vacation times
are non-i.i.d.

Because of the memoryless property of exponential inter-arrival times,
the component $\boldsymbol{E}[I]$ in Equation (\ref{eq:0}) satisfies
$\boldsymbol{E}[I]=\frac{1}{\lambda}+\boldsymbol{E}[W]$ for CBS and
BRS. By Equation (\ref{eq:1.2}) $\boldsymbol{E}[G]=\frac{1}{\lambda}(1-W^{*}(\lambda))$,
we can write the PAoI in CBS and BRS in terms of $\boldsymbol{E}[W]$
and $W^{*}(\lambda)$. Similarly, the PAoI in CBS-P can also be written
as a function of $\boldsymbol{E}[W]$ and $W^{*}(\lambda)$, with
detailed derivations in Appendix \ref{sec:Bounds-for-AoI} of the
supplementary material. Then we have Equation (\ref{eq:4.1}) in the
following:
\begin{equation}
\boldsymbol{E}[A]=\begin{cases}
-\frac{1}{\lambda}W^{*}(\lambda)+\frac{2}{\lambda}+\boldsymbol{E}[W]+2\boldsymbol{E}[H] & \mbox{for CBS,}\\
-\frac{1}{\lambda}W^{*}(\lambda)+\frac{2}{\lambda}+\boldsymbol{E}[W]+\boldsymbol{E}[H] & \mbox{for BRS, and}\\
-\frac{H^{*(1)}(\lambda)}{H^{*}(\lambda)}+H^{*}(\lambda)\frac{1}{\lambda}(1-W^{*}(\lambda))\\
+\boldsymbol{E}[W]+\frac{1}{\lambda H^{*}(\lambda)} & \mbox{for CBS-P.}
\end{cases}\label{eq:4.1}
\end{equation}

{As we mentioned in Section \ref{sec:Age-of-Information},
$W$ can be regarded as the waiting time of a packet in the dummy
system where packet replacement in the buffer is not allowed. Once
$W^{*}(s)$ is available, the closed-form expression of PAoI can be
obtained. Equation (\ref{eq:4.1}) does not require the vacation to
be i.i.d., so it can be applied to derive PAoI for general systems
with server vacations. One only needs to obtain the LST of packet
waiting time in the dummy system to calculate PAoI. In the remaining
part of this section, we will focus our discussion on the polling
system, as it is a system where the server takes non-i.i.d. vacations
(see \cite{kofman1993blocking}). We will show how to obtain $W^{*}(s)$
for polling systems, and then derive the PAoI for polling systems
based on Equation (\ref{eq:4.1}).}

\begin{figure}[h]
\hfill{}\includegraphics[scale=0.33]{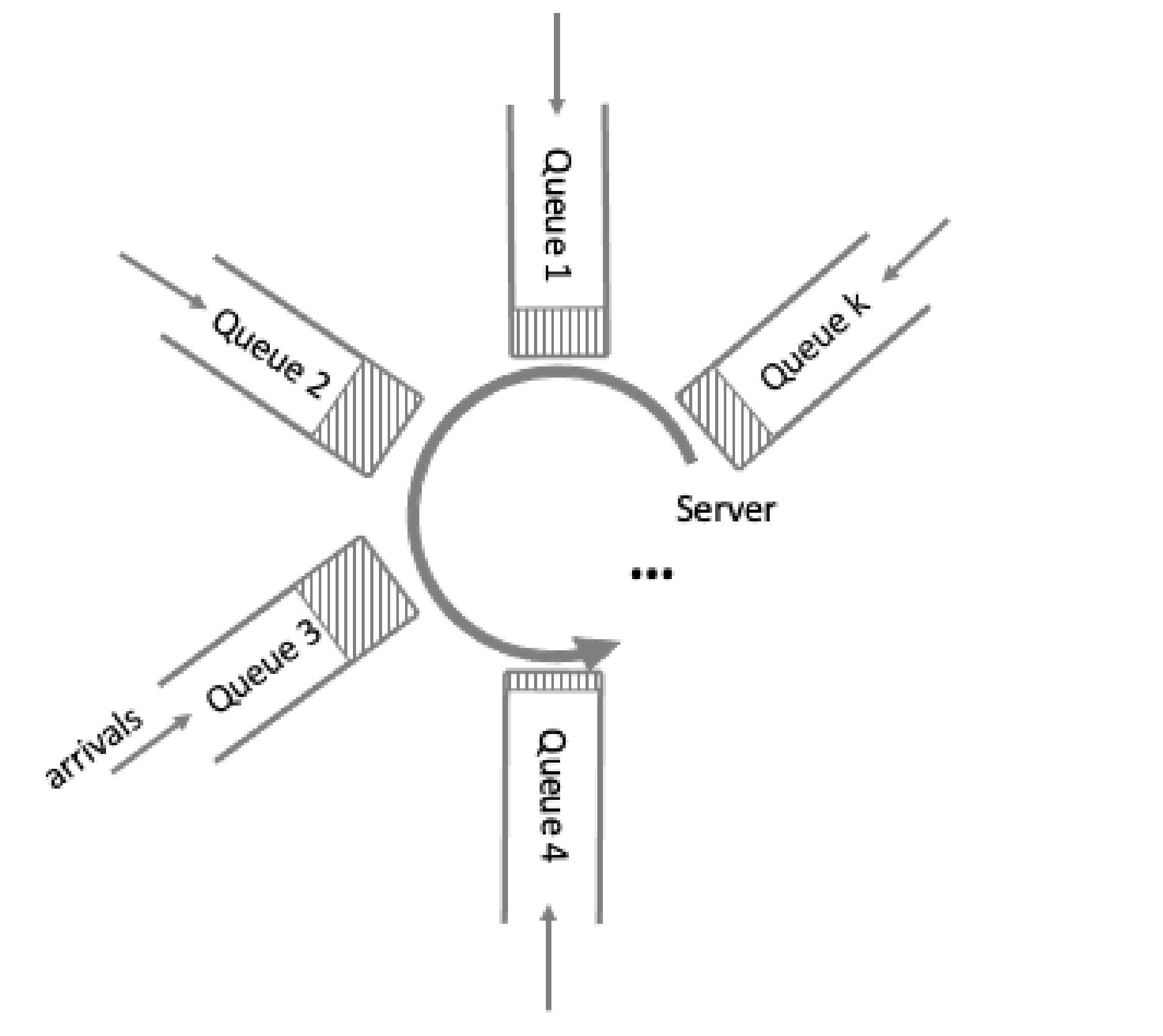}\hfill{}
\caption{A $k$-queue Polling System with Cyclic Polling Scheme \label{fig:A--queue-Polling}}
\end{figure}

A polling system is a queueing system that contains a single server
and $k$ classes of packets. Each packet class would have its own
queue, so there are $k$ queues in the system. The server serves packets
by switching between queues, and a switchover time is incurred when
the server switches from one queue to another. A demonstrative graph
of polling systems is provided in Fig. \ref{fig:A--queue-Polling}.
Polling systems have a wide application in communication networks
and other networks (see \cite{wierman2007scheduling,boon2011applications,2020arXiv200102530X}),
but the PAoI in polling systems has not been fully studied. Specifically,
suppose there are multiple data nodes in the underwater sensor network
example which we discussed in Section \ref{sec:Introduction} (also
see \cite{vasilescu2005data,heidemann2012underwater}). In this case,
we can model the underwater system as a polling system, where each
data node can be modeled as a queue/buffer, and the autonomous vehicle
can be regarded as the server that collects/processes data from each
node in a periodic manner. 

In this paper, we are interested in single buffer systems, so we assume
that each queue has a single buffer that can hold only one packet
at a time. Similar to our discussion in Section \ref{sec:Age-of-Information},
we assume that only the most recently arrived packet is kept in the
buffer, and we consider three variations of the polling system by
making different assumptions about the buffer availability and service
preemption. We still denote the polling systems under the three scheduling
policies as CBS, BRS, and CBS-P. In CBS, the buffer is not available
until the current packet completes its service. When the server is
busy processing, newly arrived packets in this queue will be rejected.
In BRS, the buffer becomes available once the service has started,
and the new arrival during the service time will be served in the
next polling instant. In CBS-P, the new arrival will preempt the packet
in service, and the preempted packet will be discarded. The server
will switch to the next queue when the service of a packet is complete.
In all these three systems, the server will start another switching
process immediately if it observes an empty queue. We assume that
the arrival process of packets in each queue $i$ follows a Poisson
process with rate $\lambda_{i}$, and the service time $H_{i}$ for
packets at each queue is i.i.d. with mean $h_{i}$ and LST $H_{i}^{*}(s).$
The switchover time $U_{ij}$ from queue $i$ to queue $j$ has mean
$u_{ij}$ and LST $U_{ij}^{*}(s)$. In the remaining part of this
section, we use the subscript $i$ to denote the parameter for queue
$i$ in the polling system. 

There are multiple widely used routing schemes that determine which
queue to switch to next for the server. Routing schemes include cyclic
\cite{mukherjee1990comments,takine1988exact,takagi2000analysis,ferguson1985exact},
random polling \cite{lee1996exact}, and Markovian polling \cite{chung1994performance,boxma1989waiting}.
In this work, we focus on the Markovian polling scheme since the random
polling and cyclic polling schemes are both special cases of the Markovian
polling scheme, as we will show later. In the Markovian polling scheme,
after serving queue $i$, the probability of serving queue $j$ next
is given by $p_{ij}$. Considering all the possible queue indices
before and after switching, we can characterize the switching process
by a discrete Markov chain with transition matrix $P=[p_{ij}].$ We
assume that $P$ is irreducible positive recurrent. 

For the cyclic polling scheme, the transition matrix is given by 
\begin{eqnarray*}
p_{ij} & = & \begin{cases}
1 & \mbox{if }j=i+1,\\
0 & \mbox{otherwise},
\end{cases}\mbox{for }i,j\in\{1,2,...,k\}.
\end{eqnarray*}
Two other polling schemes were discussed in \cite{chung1994performance}.
One is called load-oriented-policy (LOP), which is defined by the
transition matrix with $p_{ij}=\frac{\lambda_{j}}{\sum_{l=1}^{k}\lambda_{l}}$
for all $i$ and $j$. The other polling scheme is called symmetric
random polling, in which $p_{ij}=\frac{1}{k}$ for all $i$ and $j$.
We will show the performance of these schemes numerically in Section
\ref{sec:Numerical-Study}.

The service process for each queue in polling systems can be modeled
as a single server with multiple vacations: when the server polls
the queue, it serves the packet if the queue is not empty, and takes
a vacation (switches out and serves other queues) once the service
completes; if the queue is empty when polled, the server takes another
vacation. It is important to note that as pointed out by Kofman in
\cite{kofman1993blocking}, even when the cyclic polling scheme is
applied, the vacations that the server takes in a polling system are
non-i.i.d. Suppose $W_{i}$ is the packet waiting time in queue $i$
of the dummy system, with LST $W_{i}^{*}(s)$. Our methods for deriving
$W^{*}(s)$ for systems with i.i.d. vacations in Section \ref{sec:Age-of-Information}
cannot be applied here for deriving $W_{i}^{*}(s)$ in polling systems.

Chung et al. \cite{chung1994performance} provided the LST for waiting
time $W_{i}$ in the dummy systems of CBS and BRS i.e., without packet
replacement in the buffer. We can borrow the expressions of $W_{i}^{*}(s)$
for our system as whether there is preemption or not in the buffer
for CBS and BRS does not influence the vacation process. We now summarize
how $W_{i}^{*}(s)$ is obtained by Chung et al. \cite{chung1994performance}
and use it to derive the PAoI for queue $i$ (i.e., $\boldsymbol{E}[A_{i}]$).
The main idea in \cite{chung1994performance} of deriving $W_{i}^{*}(s)$
is to solve Equation (\ref{eq:8}), 
\begin{eqnarray}
 &  & F_{i}(z_{1},...,z_{k})\nonumber \\
 & = & \sum_{j=1}^{k}\frac{\pi_{j}}{\pi_{i}}p_{ji}\tilde{U}_{ij}^{*}\bigg\{(1-\tilde{H_{j}^{*}})F_{j}(z_{1},...,z_{k})_{z_{j}=0}\nonumber \\
 &  & +\tilde{H_{j}^{*}}F_{j}(z_{1},...,z_{k})_{z_{j}=1}\bigg\}\mbox{ for }i=1,...,k,\label{eq:8}
\end{eqnarray}
where $F_{i}(z_{1},...,z_{k})$ is a probability generating function
with $F_{i}(1,...,1)=1$, $(\pi_{1},...,\pi_{k})$ is the stationary
distribution of the transition matrix $P$, $\tilde{U}_{ij}^{*}=U_{ij}^{*}(\sum_{l=1}^{k}\lambda_{l}(1-z_{l}))$,
and $\tilde{H_{j}^{*}}$ is given in Equation (\ref{eq:10.1}) with
$\tilde{\lambda}_{j}(z)=\sum_{l=1,l\neq j}^{k}\lambda_{l}(1-z_{l}).$

\begin{equation}
\tilde{H_{j}^{*}}=\begin{cases}
H_{j}^{*}(\tilde{\lambda}_{j}(z)) & \mbox{for CBS,}\\
H_{j}^{*}(\sum_{l=1}^{k}\lambda_{l}(1-z_{l})) & \mbox{for BRS, and}\\
\frac{H_{j}^{*}(\tilde{\lambda}_{j}(z)+\lambda_{j})}{\frac{\tilde{\lambda}_{j}(z)}{\tilde{\lambda}_{j}(z)+\lambda_{j}}+\frac{\lambda_{j}}{\tilde{\lambda}_{j}(z)+\lambda_{j}}H_{j}^{*}(\tilde{\lambda}_{j}(z)+\lambda_{j})} & \mbox{for CBS-P.}
\end{cases}\label{eq:10.1}
\end{equation}

Chung et al. \cite{chung1994performance} only showed that Equation
(\ref{eq:8}) holds for CBS and BRS. However, we show that Equation
(\ref{eq:8}) also holds for CBS-P, with $\tilde{H_{j}^{*}}$ given
in Equation (\ref{eq:10.1}). The analysis is as follows. In CBS-P,
the server switches out from queue $j$ only when one packet has been
completely served. If we regard the period during which the server
is continuously serving packets as the service time for ``one packet'',
then we can also regard CBS-P as CBS. The only difference is in the
distribution of completing one packet. In CBS, completing one packet
in queue $j$ takes $H_{j}$ amount of time. While in CBS-P, completing
one packet in queue $j$ takes time $L_{j}$ with LST 
\begin{eqnarray}
L_{j}^{*}(s) & = & \frac{H_{j}^{*}(s+\lambda_{j})}{\frac{s}{s+\lambda_{j}}+\frac{\lambda_{j}}{s+\lambda_{j}}H_{j}^{*}(s+\lambda_{j})}.\label{eq:10.2}
\end{eqnarray}
A detailed derivation of Equation (\ref{eq:10.2}) can be found in
Appendix \ref{sec:Proof-for-Theorem} of the supplementary material.
Then, the formula of $\tilde{H_{j}^{*}}$ for CBS-P in Equation (\ref{eq:10.1})
is obtained by simply combining Equation (\ref{eq:10.2}) with the
formula $\tilde{H_{j}^{*}}$ for CBS in Equation (\ref{eq:10.1}).
Equation (\ref{eq:8}) thus holds for CBS-P as well, with only $\tilde{H_{j}^{*}}$
being different from CBS. 

Solving the system (\ref{eq:8}) is quite involved, as shown in \cite{chung1994performance}.
However, the expected value of $W_{i}$ can be obtained by solving
the system (\ref{eq:8}) with $z_{j}=0\mbox{ or }1$ for $j=1,...,k$,
where only $k(2^{k}-1)$ linear equations need to be solved. The expected
time $W_{i}$ is then given as $\boldsymbol{E}[W_{i}]=\frac{\gamma_{i}}{\lambda_{i}\alpha_{i}}-\frac{1}{\lambda_{i}},$
where $\alpha_{i}=1-F_{i}(1,...,\overset{i}{0},...,1)$ (the notation
$F_{i}(1,...,\overset{i}{0},...,1)$ means that $z_{i}=0$ and $z_{l\neq i}=1$
in $F_{i}(z_{1},...,z_{k})$) and $\gamma_{i}$ is given in Equation
(\ref{eq:10.3}).

To obtain $\boldsymbol{E}[G_{i}]$, we need to get $W_{i}^{*}(\lambda_{i})$.
From \cite{chung1994performance,takine1988exact} we have $W_{i}^{*}(s)=\frac{1}{\alpha_{i}}\frac{\lambda_{i}}{s-\lambda_{i}}\left\{ 1-\alpha_{i}-f_{i}(1-\frac{s}{\lambda_{i}})\right\} ,$
where $f_{i}(z)=F_{i}(1,...,\overset{i}{z},...,1)$. Using L'Hospital
rule, we have $W_{i}^{*}(\lambda_{i})=\frac{f_{i}^{(1)}(0)}{\alpha_{i}}=\frac{1}{\alpha_{i}}\frac{\partial F_{i}(1,...,z,...,1)}{\partial z}|_{z=0},$
in which the derivative of $F_{i}(1,...,z,...,1)$ is needed. Therefore
we need to compute the partial derivative of Equation (\ref{eq:8})
with respect to $z_{l}$ for $l=1,...,k$, which is to solve Equation
(\ref{eq:9}). Note here we only need to solve system (\ref{eq:9})
for $z_{j}=0\mbox{ or }1$ for $j=1,...,k$ to obtain $W_{i}^{*}(\lambda_{i})$,
so that $k^{2}2^{k}$ number of equations need to be solved. After
solving system (\ref{eq:8}) and (\ref{eq:9}), the closed-form expression
of PAoI can be obtained from the Equation (\ref{eq:10}), and we can
also have the following theorem.

\begin{strip}
\begin{eqnarray}
\gamma_{i} & = & \begin{cases}
\frac{\lambda_{i}}{\pi_{i}}\sum_{j=1}^{k}\pi_{j}(\alpha_{j}h_{j}+\sum_{l=1}^{k}p_{jl}u_{jl})-\lambda_{i}\alpha_{i}h_{i} & \mbox{for CBS,}\\
\frac{\lambda_{i}}{\pi_{i}}\sum_{j=1}^{k}\pi_{j}(\alpha_{j}h_{j}+\sum_{l=1}^{k}p_{jl}u_{jl}) & \mbox{for BRS, and}\\
\frac{\lambda_{i}}{\pi_{i}}\sum_{j=1}^{k}\pi_{j}(\alpha_{j}\frac{1-H_{j}^{*}(\lambda_{j})}{\lambda H_{j}^{*}(\lambda_{j})}+\sum_{l=1}^{k}p_{jl}u_{jl})-\lambda_{i}\alpha_{i}\frac{1-H_{i}^{*}(\lambda_{i})}{\lambda_{i}H^{*}(\lambda_{i})} & \mbox{for CBS-P.}
\end{cases}\label{eq:10.3}
\end{eqnarray}
\begin{eqnarray}
\frac{\partial F_{i}(z_{1},...,z_{k})}{\partial z_{l}} & = & \frac{\partial}{\partial z_{l}}\left\{ \sum_{j=1}^{k}\frac{\pi_{j}}{\pi_{i}}p_{ji}\tilde{U}_{ij}^{*}\left((1-\tilde{H_{j}^{*}})F_{j}(z_{1},...,z_{k})_{z_{j}=0}+\tilde{H_{j}^{*}}F_{j}(z_{1},...,z_{k})_{z_{j}=1}\right)\right\} \nonumber \\
 &  & \mbox{for }i=1,...,k\mbox{ and }l=1,...,k.\label{eq:9}
\end{eqnarray}
\begin{eqnarray}
\boldsymbol{E}[A_{i}] & = & \begin{cases}
-\frac{1}{\lambda_{i}}W_{i}^{*}(\lambda_{i})+\frac{2}{\lambda_{i}}+\boldsymbol{E}[W_{i}]+2\boldsymbol{E}[H_{i}] & \mbox{for CBS,}\\
-\frac{1}{\lambda_{i}}W_{i}^{*}(\lambda_{i})+\frac{2}{\lambda_{i}}+\boldsymbol{E}[W_{i}]+\boldsymbol{E}[H_{i}] & \mbox{for BRS, and}\\
-\frac{H_{i}^{*(1)}(\lambda_{i})}{H_{i}^{*}(\lambda_{i})}+H_{i}^{*}(\lambda_{i})\frac{1}{\lambda_{i}}(1-W_{i}^{*}(\lambda_{i}))+\boldsymbol{E}[W_{i}]+\frac{1}{\lambda_{i}H_{i}^{*}(\lambda_{i})} & \mbox{for CBS-P.}
\end{cases}\label{eq:10}
\end{eqnarray}
\end{strip}

\begin{thm}
\label{thm:If-the-service-1}If the service time for each queue is
exponentially distributed in a polling system, then CBS-P will always
have a PAoI than that in CBS.
\end{thm}
\begin{IEEEproof}
See Appendix \ref{sec:Proof-for-Theorem-2} of the supplementary material. 
\end{IEEEproof}
However, when the service time is not exponential, CBS-P does not
always have a smaller PAoI than CBS. We will show more computational
results in Section \ref{sec:Numerical-Study}.

\section{Numerical Study: Verification, Findings, and Explanations\label{sec:Numerical-Study}}

In this section, we first perform a set of numerical experiments for
systems with i.i.d. vacations, and then provide the numerical results
to verify the exact solution of PAoI for polling systems. We then
provide the results for the polling system under different Markovian
polling schemes and develop insights.

\subsection{CBS, BRS and CBS-P with i.i.d. Vacations\label{subsec:Study-for-CBS,}}

\begin{figure*}[t]
\subfloat[AoI Comparison, $H\sim exp(1),$$V\sim exp(\frac{1}{2})$]{\includegraphics[scale=0.35]{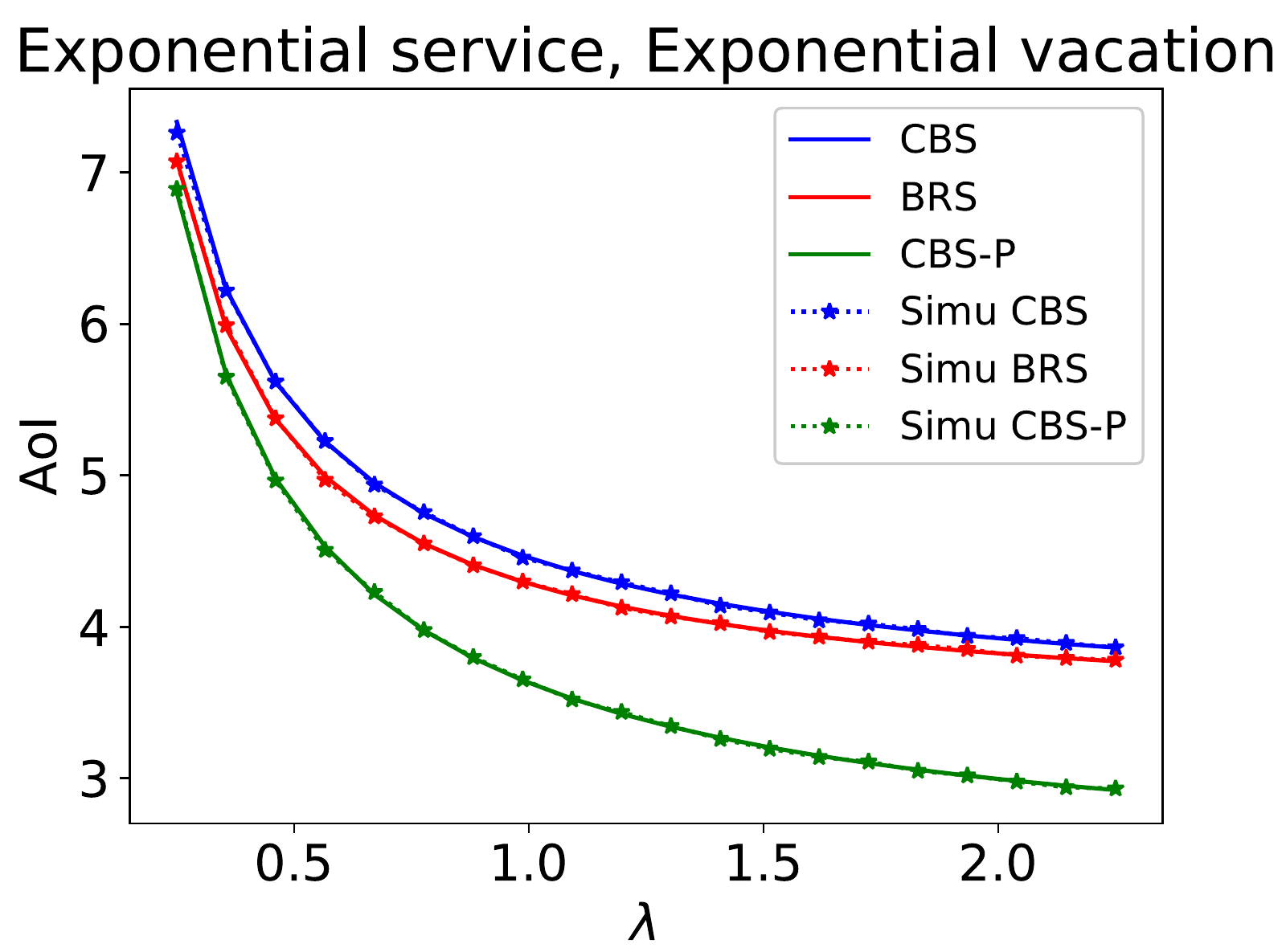}

\captionsetup{width=.3\linewidth}}\subfloat[PAoI Comparison, $H\sim exp(1),$$V\sim exp(\frac{1}{2})$]{\includegraphics[scale=0.35]{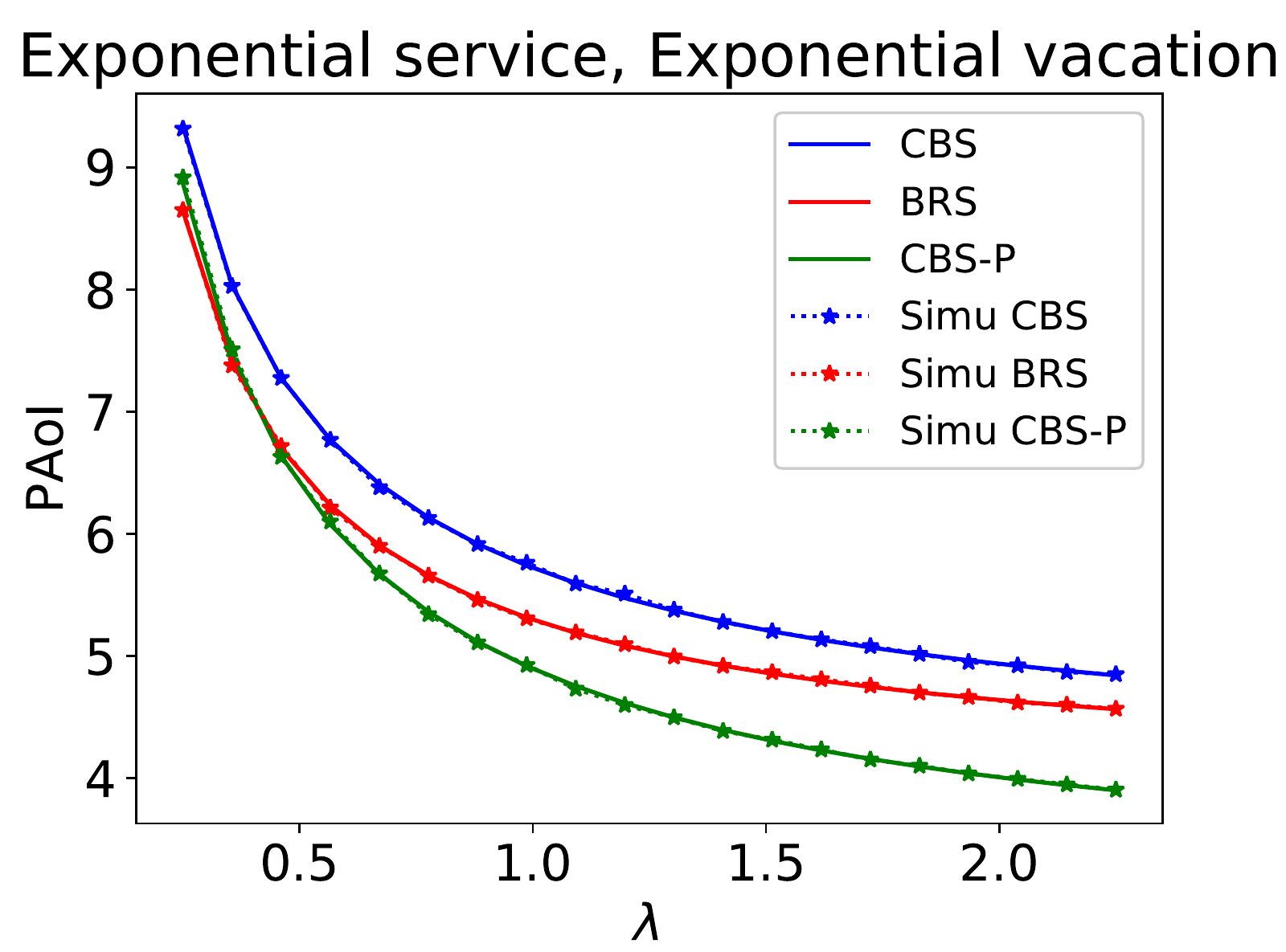}\captionsetup{width=.3\linewidth}

}\subfloat[Variance of Peak Age Comparison, $H=exp(1)$, $V\sim exp(\frac{1}{2})$]{\includegraphics[scale=0.35]{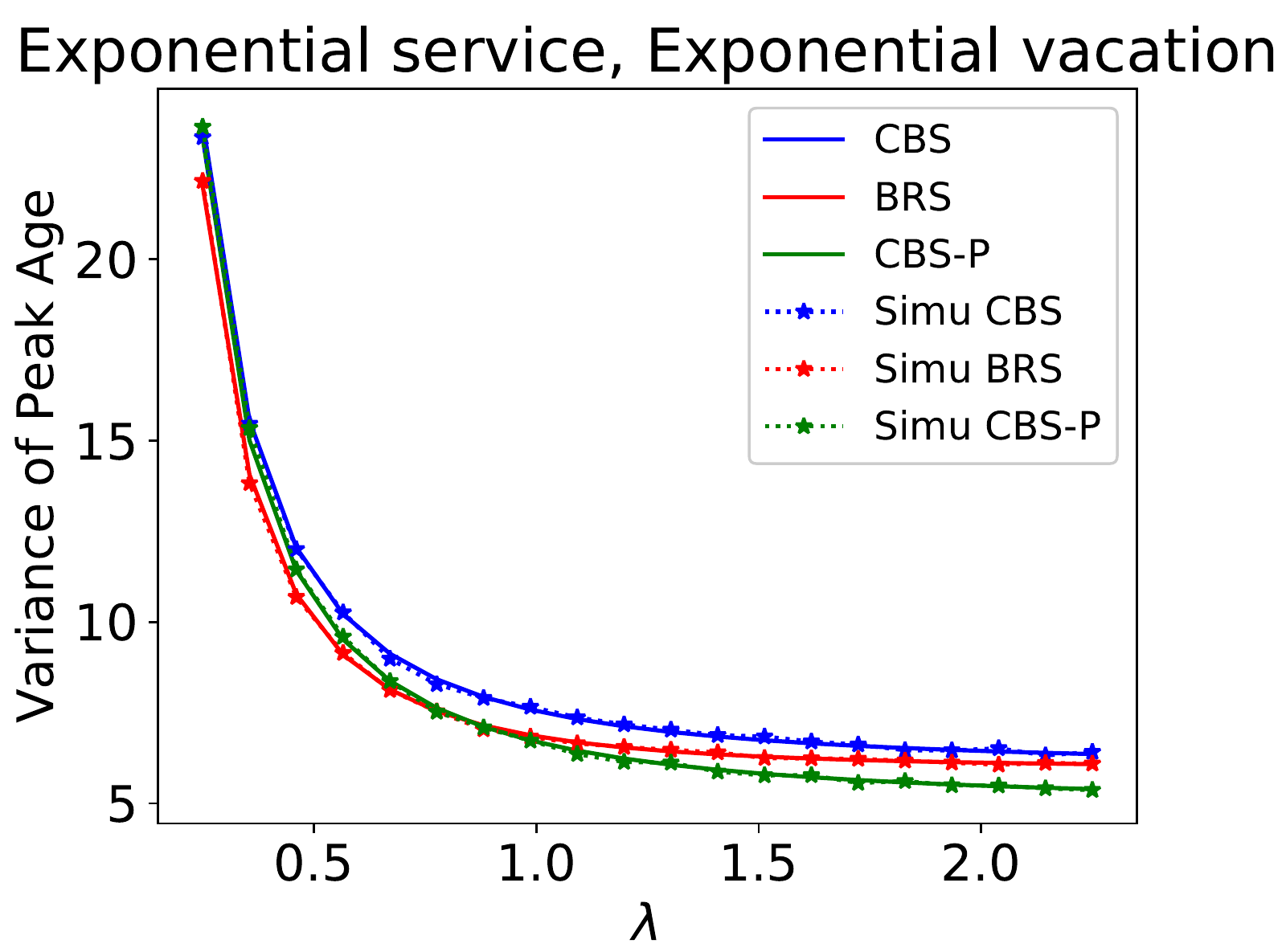}

\captionsetup{width=.2\linewidth}}

\subfloat[AoI Comparison, $H=1$, $V\sim gamma(2,1)$]{\includegraphics[scale=0.35]{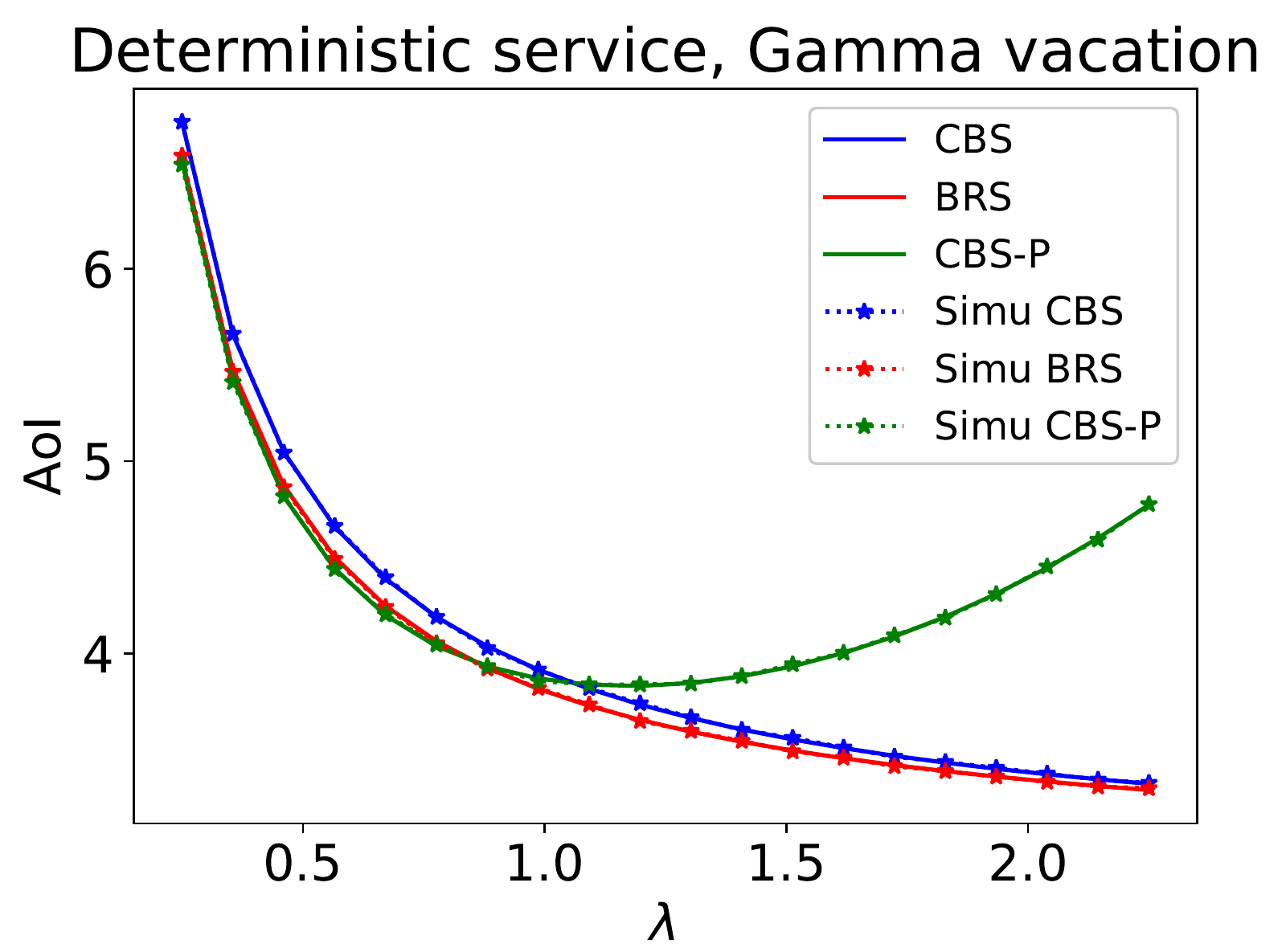}\captionsetup{width=.3\linewidth}}\subfloat[PAoI Comparison, $H=1$, $V\sim gamma(2,1)$]{\includegraphics[scale=0.35]{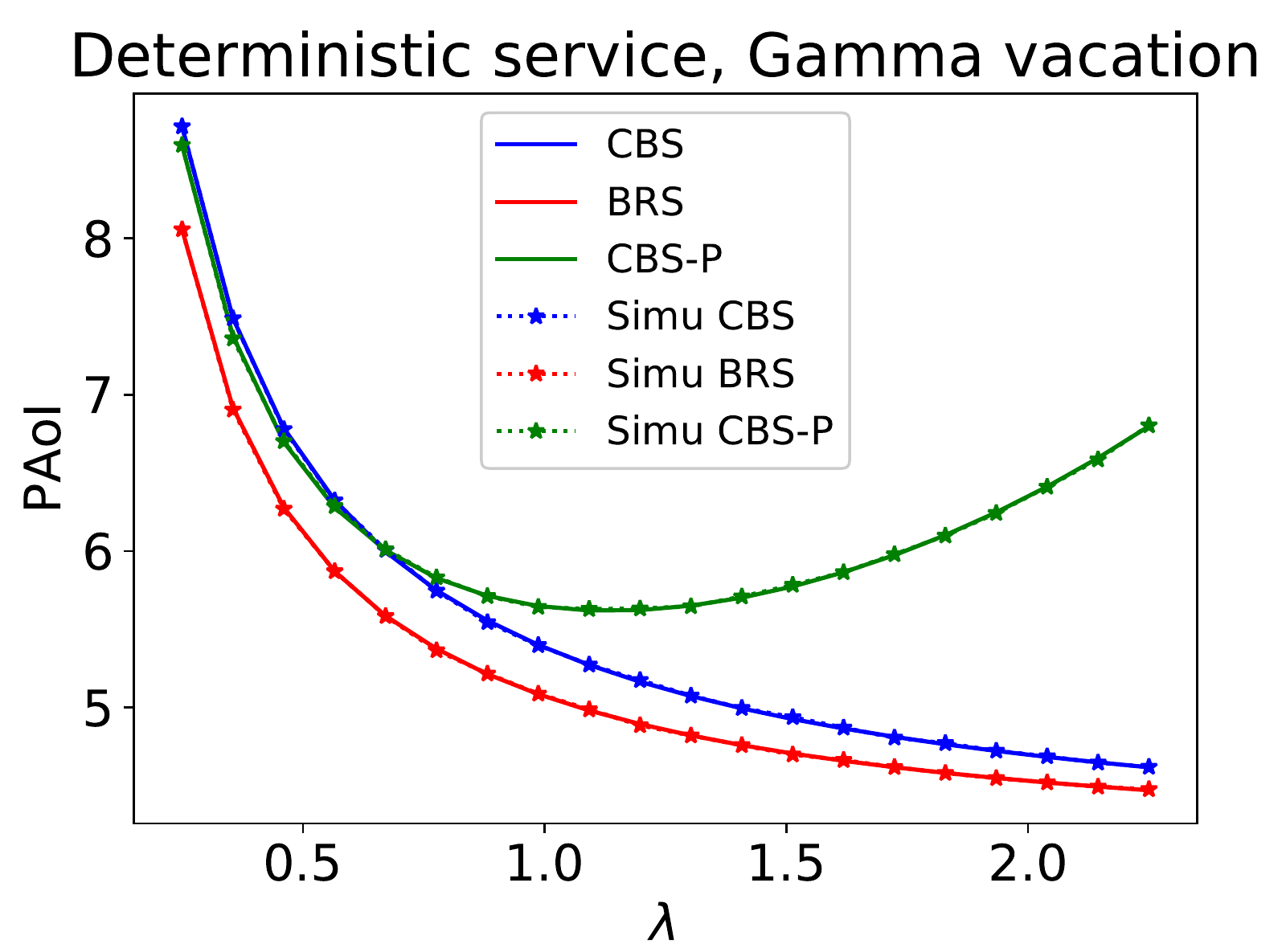}\captionsetup{width=.3\linewidth}

}\subfloat[Variance of Peak Age Comparison, $H=1$, $V\sim gamma(2,1)$]{\includegraphics[scale=0.35]{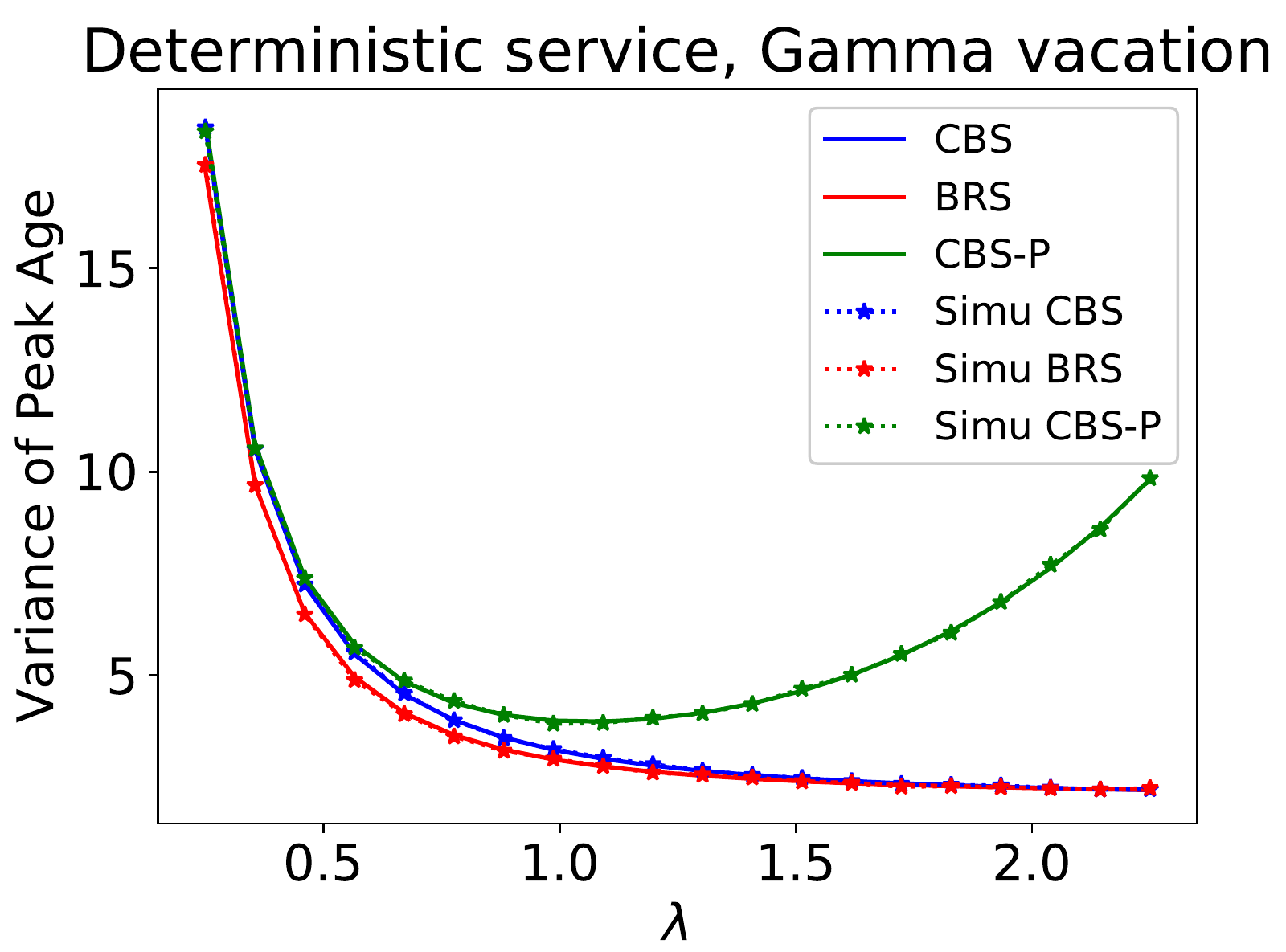}\captionsetup{width=.2\linewidth}

}

\caption{Vacation Server Systems with $\boldsymbol{E}[H]=1$ and $\boldsymbol{E}[V]=2$\label{fig:Vacation-Server}}
\end{figure*}

We begin our discussion by comparing the AoI, PAoI, and variance of
peak age for CBS, BRS, and CBS-P, as shown in Fig. \ref{fig:Vacation-Server}.
In each subfigure of Fig. \ref{fig:Vacation-Server}, the simulation
results match the exact results, which verifies our analysis.

Fig. \ref{fig:Vacation-Server}(a) and Fig. \ref{fig:Vacation-Server}(d)
compare the AoI for these three systems under different service and
vacation times. It is shown in Fig. \ref{fig:Vacation-Server}(a)
that CBS-P has the advantage over the other two systems in minimizing
AoI, when service time is exponentially distributed. When the arrival
rate is large, this advantage becomes more significant. However, in
Fig. \ref{fig:Vacation-Server}(d) where service time is deterministic,
AoI in CBS-P is greater than that in the other two systems when the
arrival rate is large. In CBS-P, the server would process the new
packet when an arrival preempts the service. The server will continuously
serve only until an inter-arrival time is smaller than the constant
service time. If the arrival rate is large (which means the expected
inter-arrival time is small), then the probability of the inter-arrival
time being smaller than the constant service time is small. Thus the
AoI of CBS-P becomes large when the arrival rate is large for deterministic
service time cases. In Section \ref{subsec:Study-for-Systems_2} we
will observe a similar phenomenon when the server does not take vacations.

In Fig. \ref{fig:Vacation-Server}(b) and Fig. \ref{fig:Vacation-Server}(e)
we compare the PAoI of these three systems. We find that CBS always
has a larger PAoI than BRS for both exponential and deterministic
service times, which matches Theorem \ref{thm:BRS-always-has}. It
can be observed from Fig. \ref{fig:Vacation-Server}(a) and Fig. \ref{fig:Vacation-Server}(b)
that CBS-P has smaller AoI and PAoI than CBS for the exponential service
cases, which matches the results in Theorem \ref{thm:If-the-service}.
In Fig. \ref{fig:Vacation-Server}(c) and Fig. \ref{fig:Vacation-Server}(f),
we compare the variance of peak age for these three systems. When
service time is exponential, CBS has a larger variance of peak age
than the other two systems when $\lambda$ is large. From all the
subfigures in Fig. \ref{fig:Vacation-Server}, we find that for both
CBS and BRS, increasing the arrival rate would reduce AoI, PAoI, and
variance of peak age for the given service time and vacation time
distributions. 

\begin{figure*}[h!]
\subfloat[AoI Comparison]{\includegraphics[scale=0.35]{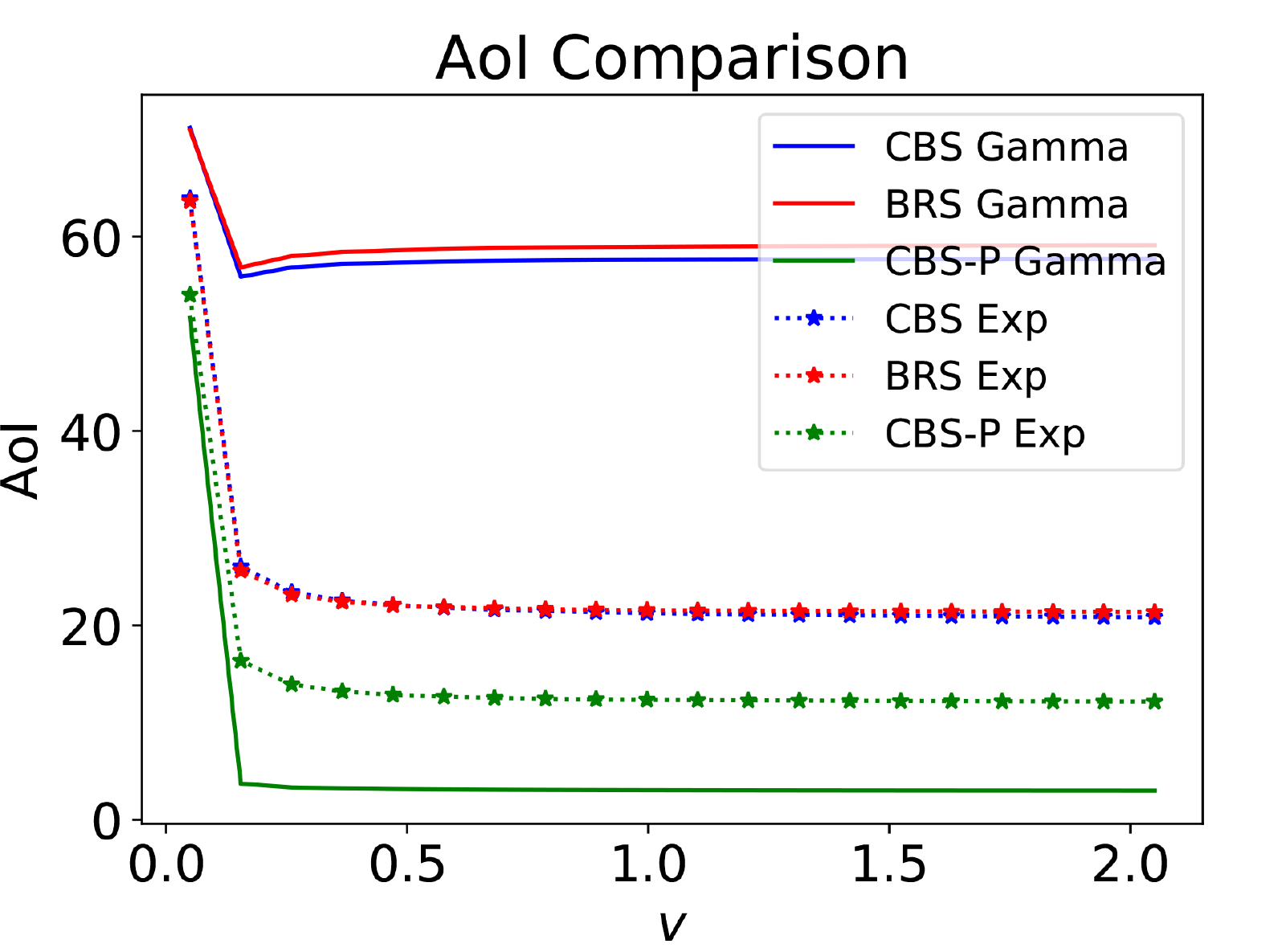}

\captionsetup{width=.3\linewidth}}\subfloat[PAoI Comparison]{\includegraphics[scale=0.35]{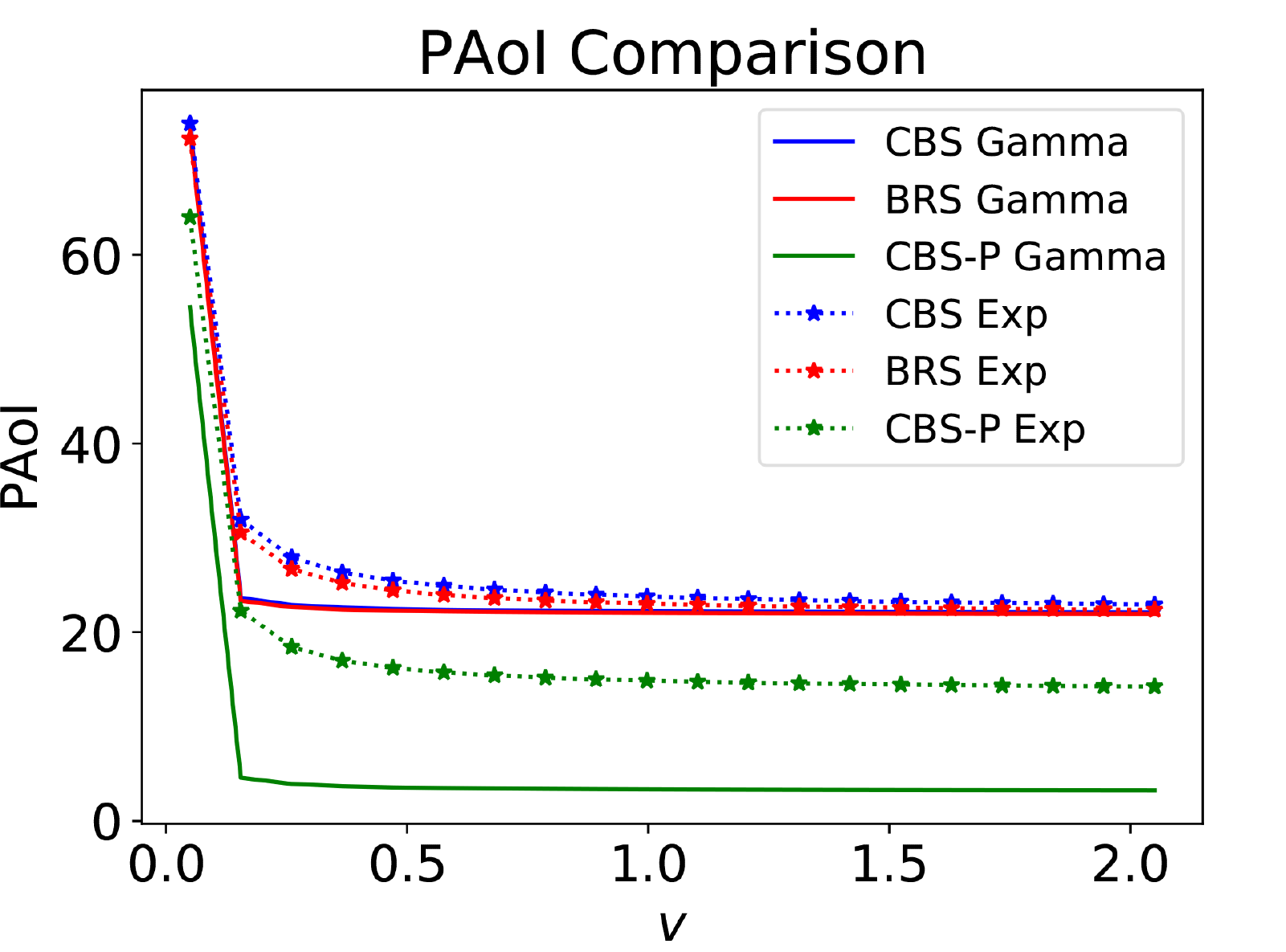}\captionsetup{width=.3\linewidth}}\subfloat[Variance of Peak Age Comparison]{\includegraphics[scale=0.35]{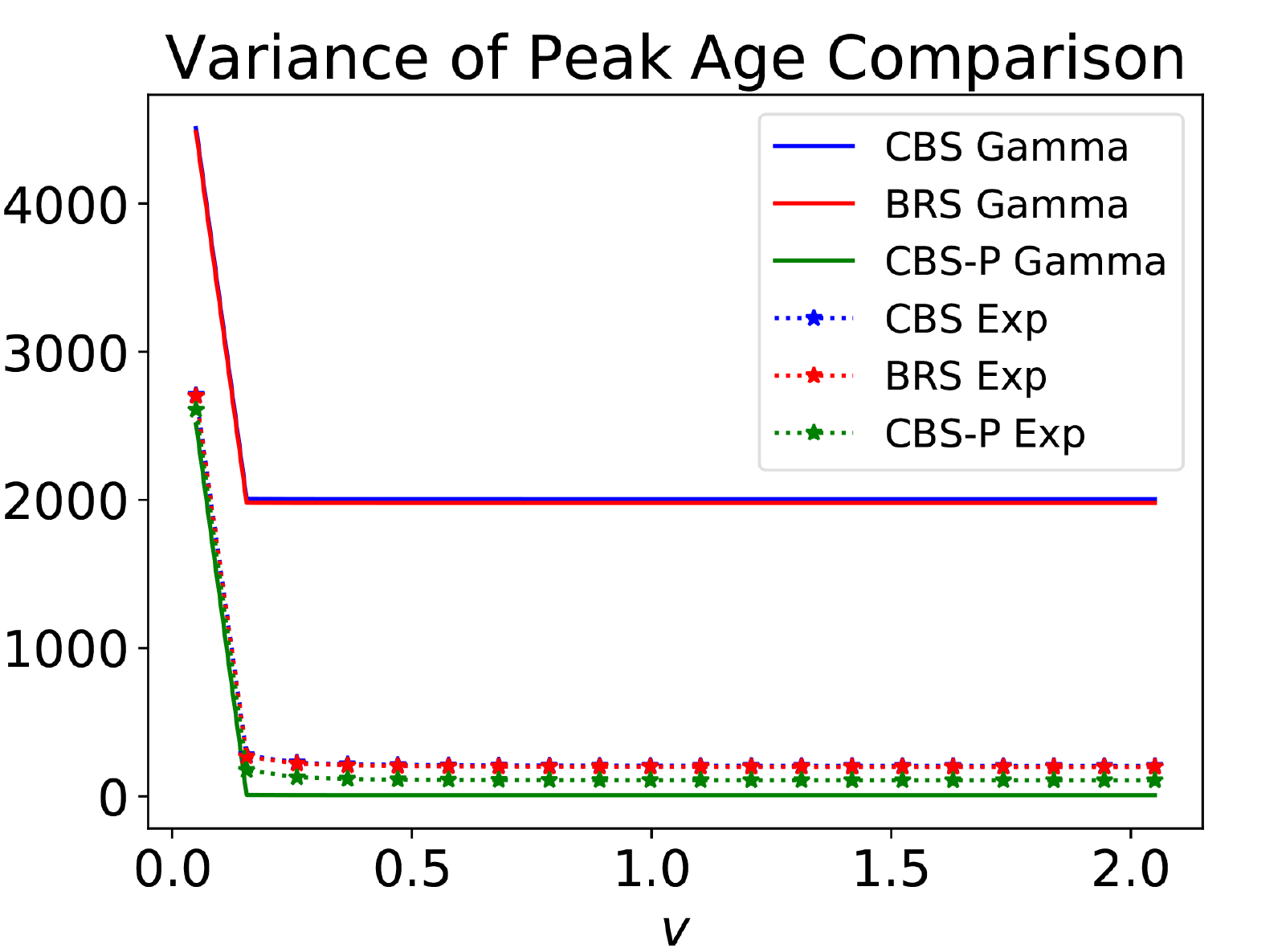}

\captionsetup{width=.2\linewidth}}

\caption{Vacation Server Systems with $\boldsymbol{E}[H]=10$, $\lambda=0.5$,
$V\sim exp(v)$\label{fig:Vacation-Server-1}}
\end{figure*}

{In Fig. \ref{fig:Vacation-Server-1} we plot the
age metrics as functions of the vacation rate $v$. We compare the
metrics with $H\sim Gamma(0.1,100)$ and $H\sim exp(0.1)$. Interestingly,
we find that reducing the vacation time decreases the PAoI and $Var(A)$
for CBS, BRS, and CBS-P, but it does not always reduce the AoI, as
shown in Fig. \ref{fig:Vacation-Server-1}(a). When $H\sim Gamma(0.1,100)$,
the AoI under CBS and BRS does not always decrease as $v$ increases.
The reason is that as shown in Theorems \ref{thm:The-AoI-of-CBS}
and \ref{thm:The-PAoI-for}, $\boldsymbol{E}[\Delta_{CBS}]$ and $\boldsymbol{E}[\Delta_{BRS}]$
depend on the term $\frac{\boldsymbol{E}[I^{2}]}{2\boldsymbol{E}[I]}$.
While reducing vacation time would reduce $\boldsymbol{E}[I]$, it
does not always reduce $\frac{\boldsymbol{E}[I^{2}]}{2\boldsymbol{E}[I]}$,
especially when $\boldsymbol{E}[H^{2}]$ is large. We also find from
Fig. \ref{fig:Vacation-Server-1} that the AoI, PAoI, and variance
of peak age under CBS-P is significantly smaller than those under
CBS and BRS when $H\sim Gamma(0.1,100)$, which shows the advantage
of having preemption in processing when $H$ is Gamma distributed
with a small scale parameter. }

\subsection{Systems with No Vacations\label{subsec:Study-for-Systems_2}}

\begin{figure*}[h!]
\subfloat[AoI Comparison with $H\sim exp(1)$]{\includegraphics[scale=0.35]{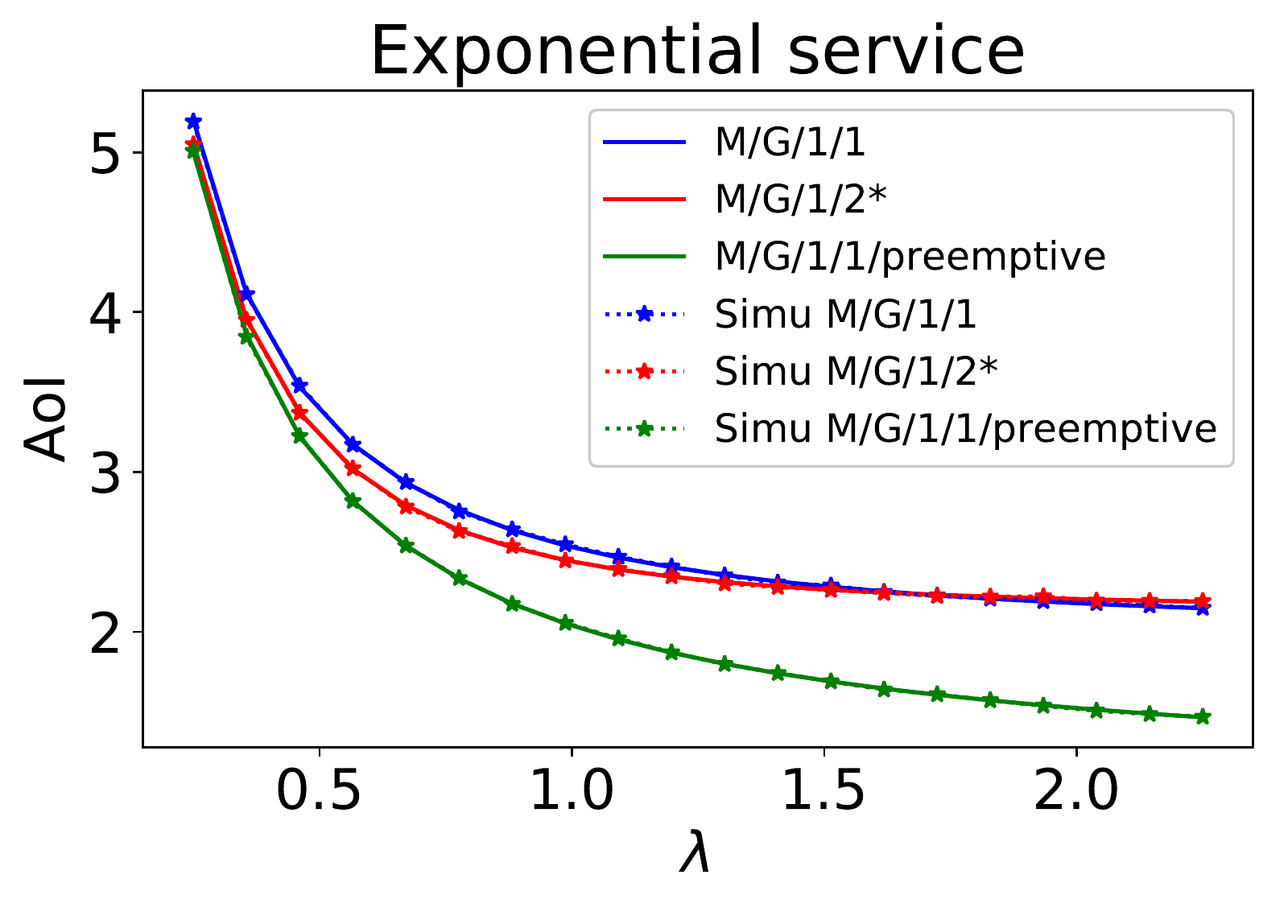}

\captionsetup{width=.3\linewidth}

}\subfloat[PAoI Comparison with $H\sim exp(1)$]{\includegraphics[scale=0.35]{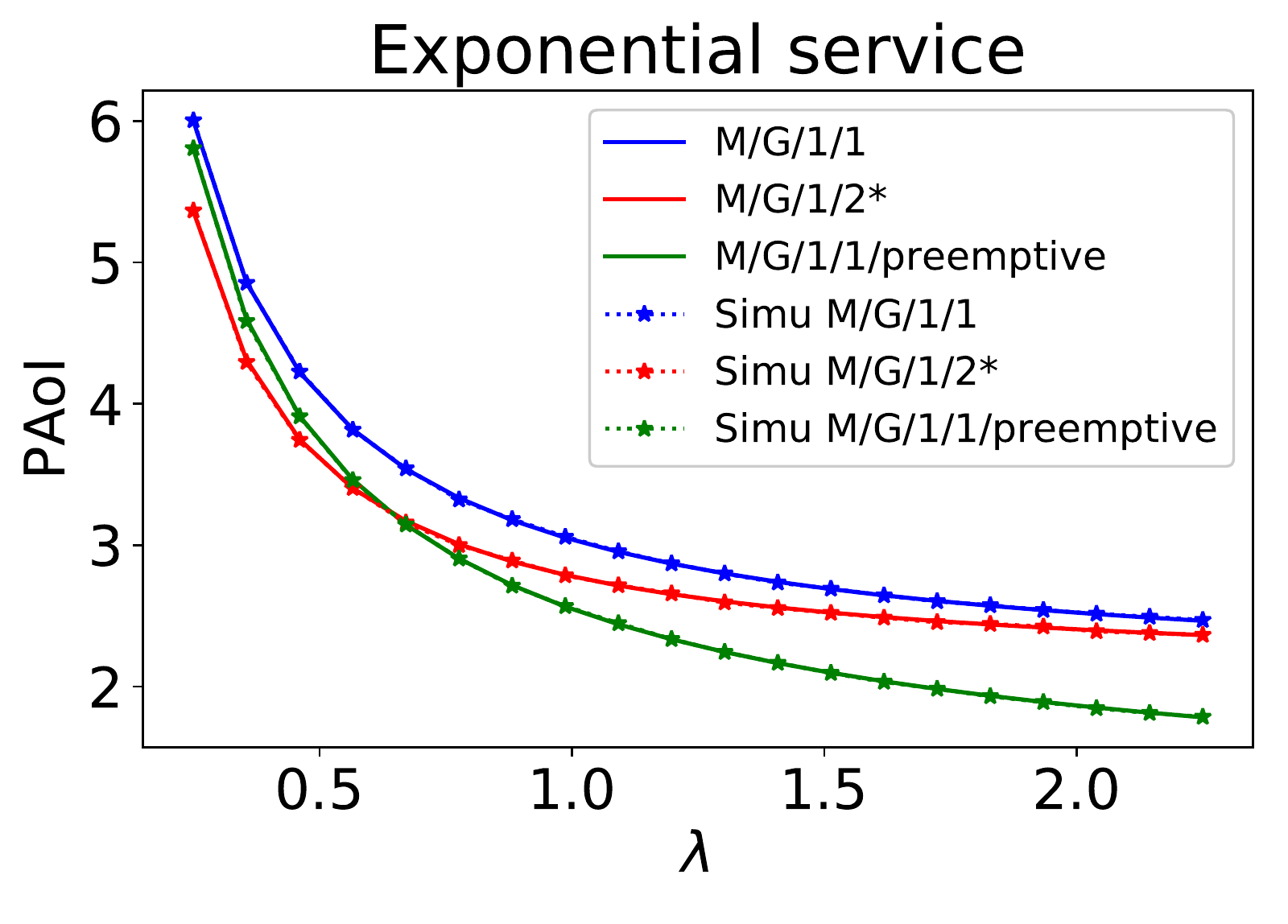}\captionsetup{width=.3\linewidth}

}\subfloat[Variance of Peak Age Comparison with $H\sim exp(1)$]{\includegraphics[scale=0.35]{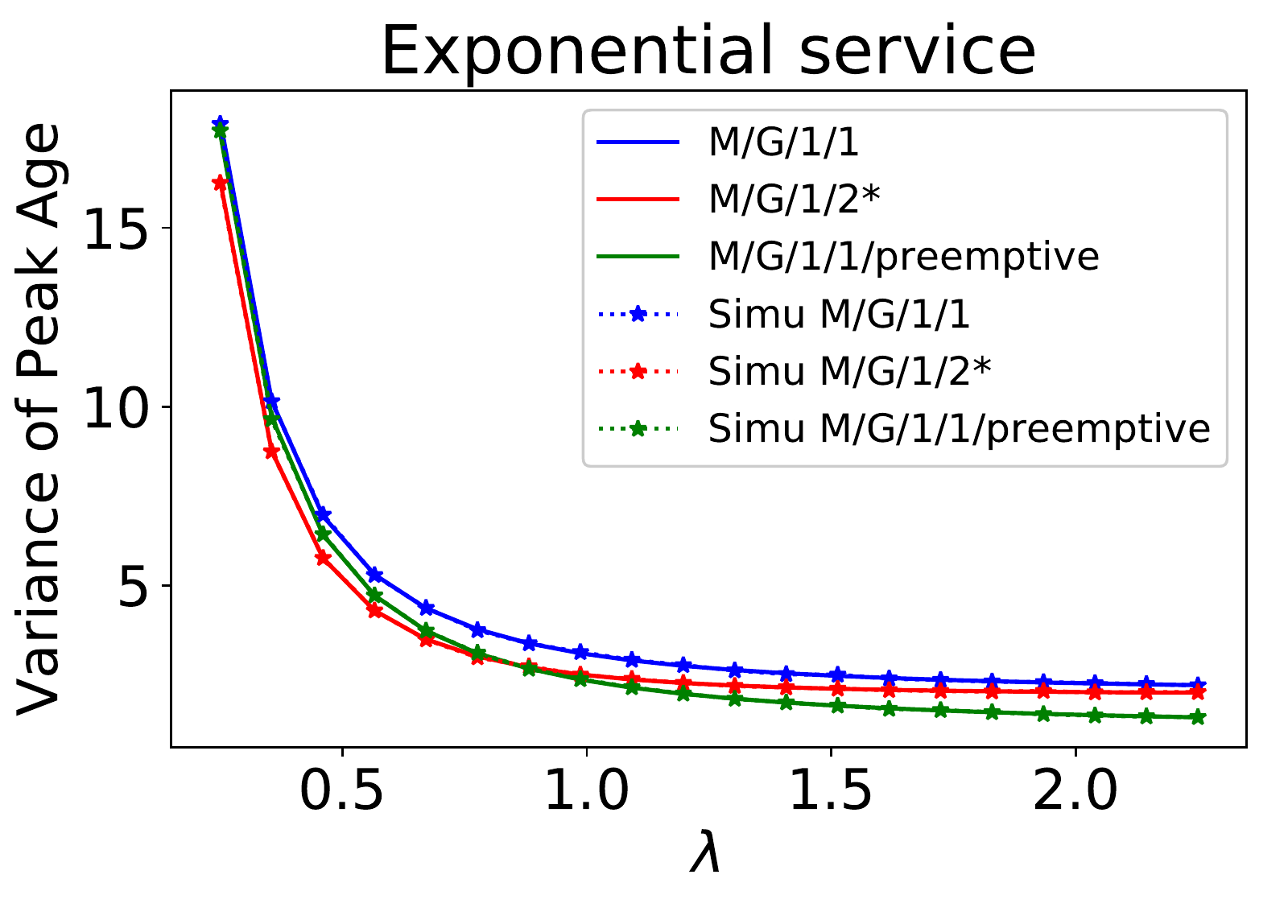}

\captionsetup{width=.25\linewidth}}

\subfloat[AoI Comparison with $H=1$]{\includegraphics[scale=0.35]{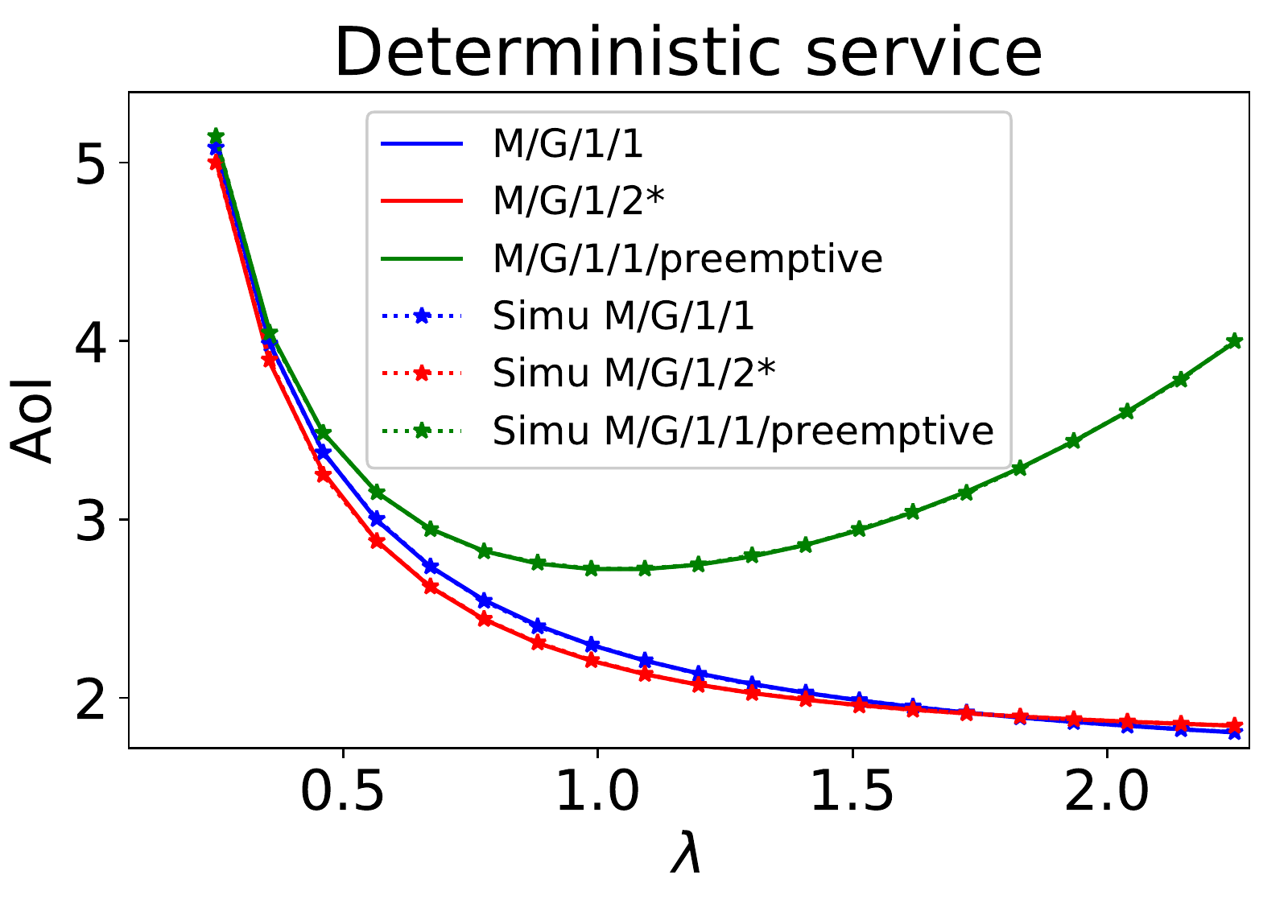}\captionsetup{width=.3\linewidth}}\subfloat[PAoI Comparison with $H=1$]{\includegraphics[scale=0.35]{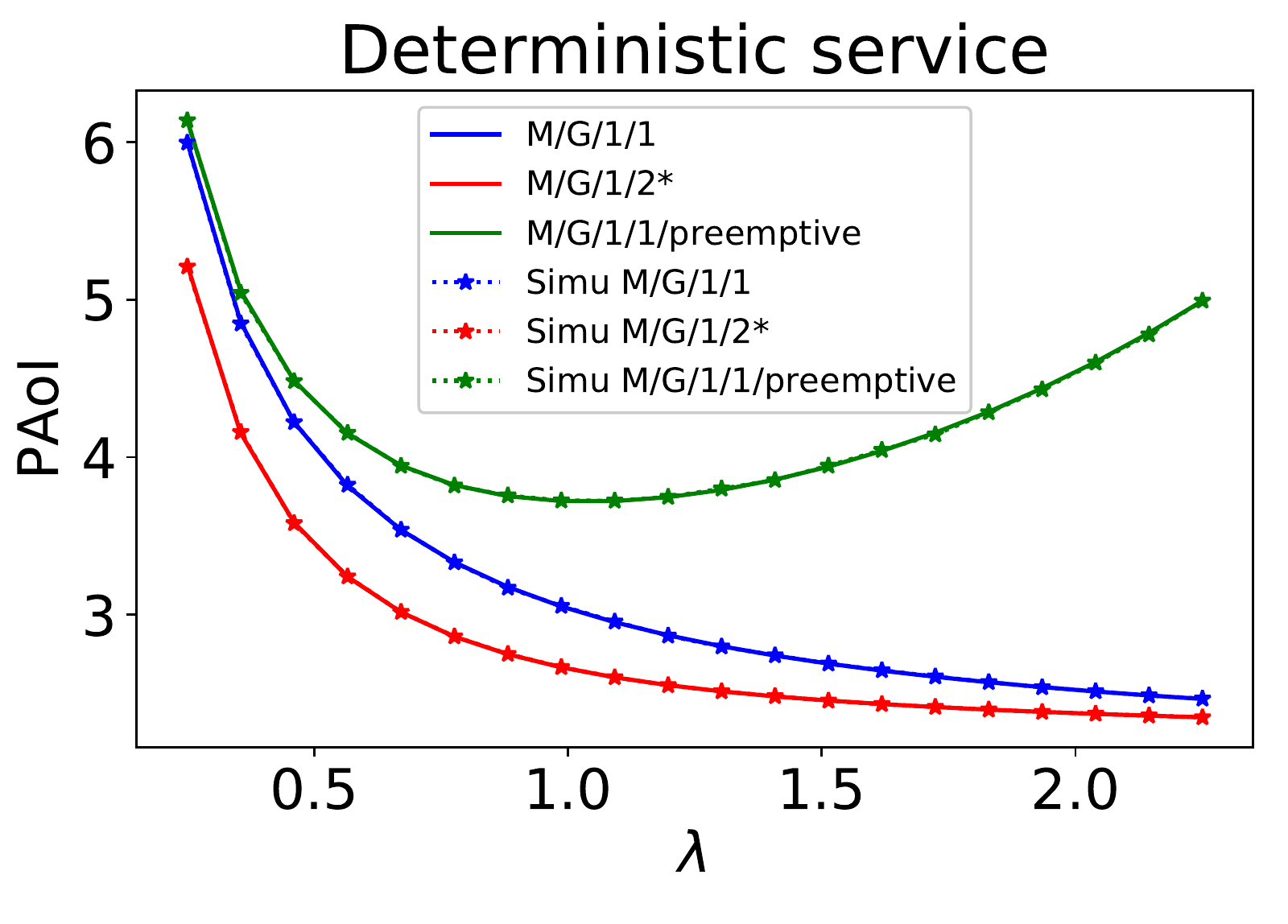}

\captionsetup{width=.3\linewidth}}\subfloat[Variance of Peak Age Comparison with $H=1$]{\includegraphics[scale=0.35]{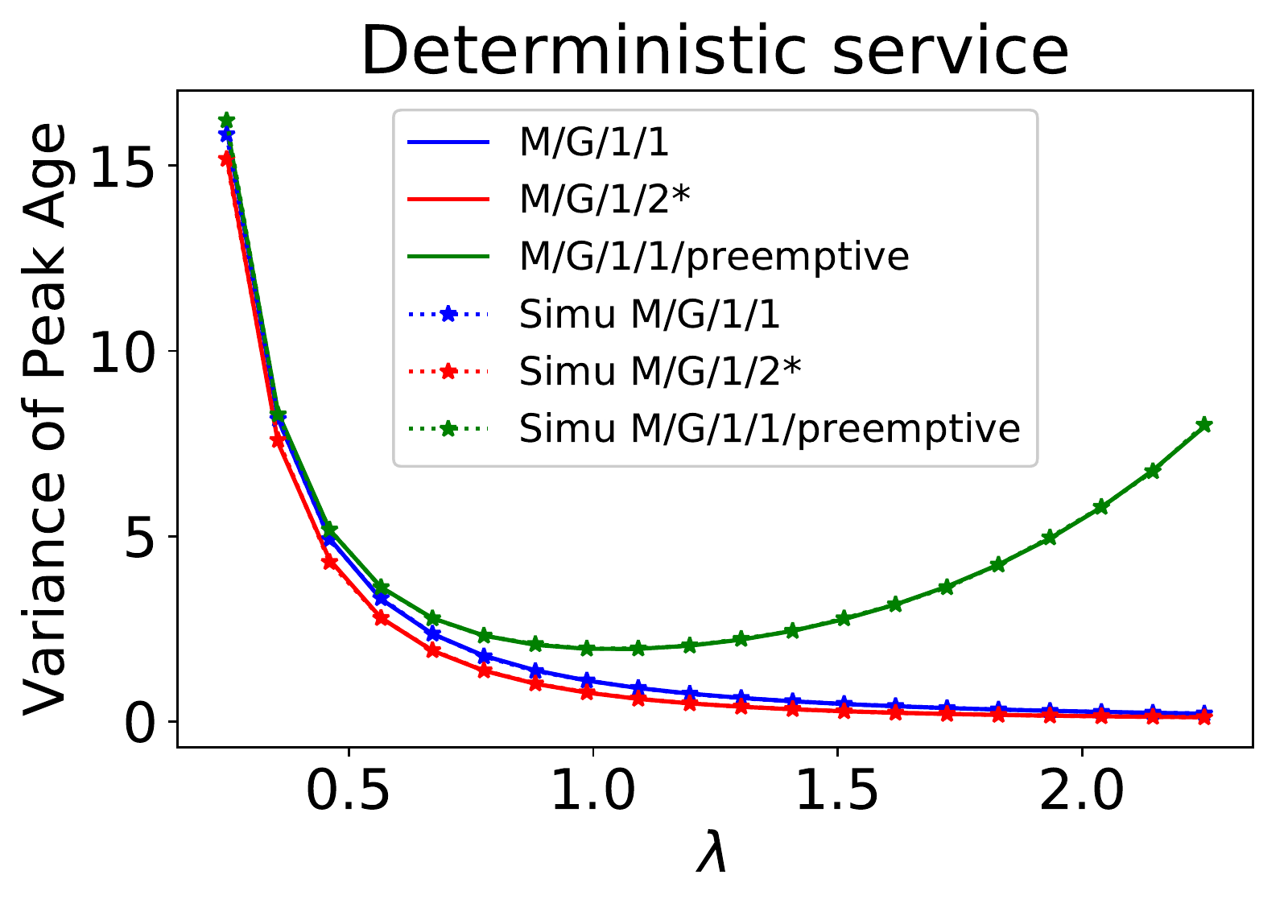}

\captionsetup{width=.25\linewidth}

}

\caption{\label{fig:Single-Queue-System}Single Queue System with $\boldsymbol{E}[H]=1$}

\end{figure*}

We next compare the AoI, PAoI, and variance of peak age for M/G/1/1,
M/G/1/2{*}, and M/G/1/1/preemptive systems under exponential and deterministic
service cases. The simulation results match the exact results in Section
\ref{subsec:Discussions-for-Systems} for each system, as shown in
Fig. \ref{fig:Single-Queue-System}. We find that although the AoI
in M/G/1/2{*} system is not always smaller than that in M/G/1/1 system,
the PAoI and variance of peak age in M/G/1/2{*} system are smaller
than those in M/G/1/1 system. Especially when the arrival rate is
low, the advantage that M/G/1/2{*} system has over M/G/1/1 system
in minimizing PAoI and variance of peak age becomes significant. For
M/G/1/1/preemptive system, the AoI, PAoI, and variance of peak age
will increase dramatically when the arrival rate becomes large when
the service time is deterministic. We also find that when the service
time is exponential, M/M/1/1 system has the largest variance of peak
age among all the three systems, which verifies our discussion in
Section \ref{subsec:Discussions-for-Systems}. When the service time
is deterministic, M/D/1/2{*} system has a lower variance of peak age
than the other two systems. 

\subsection{Polling Systems\label{subsec:Study-for-Polling}}

\begin{figure*}[h]
\subfloat[PAoI of Queue 1]{\includegraphics[scale=0.35]{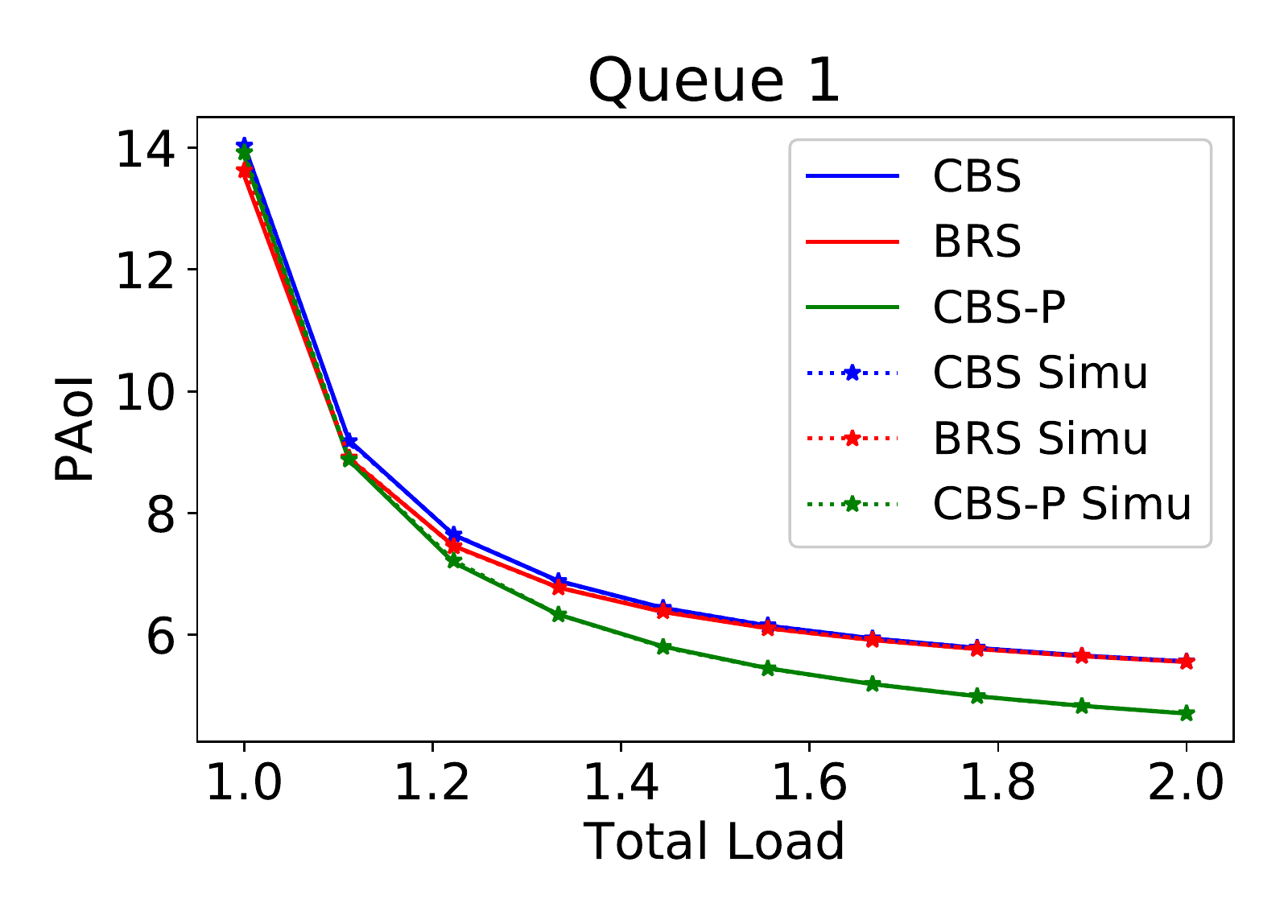}

}\subfloat[PAoI of Queue 2]{\includegraphics[scale=0.35]{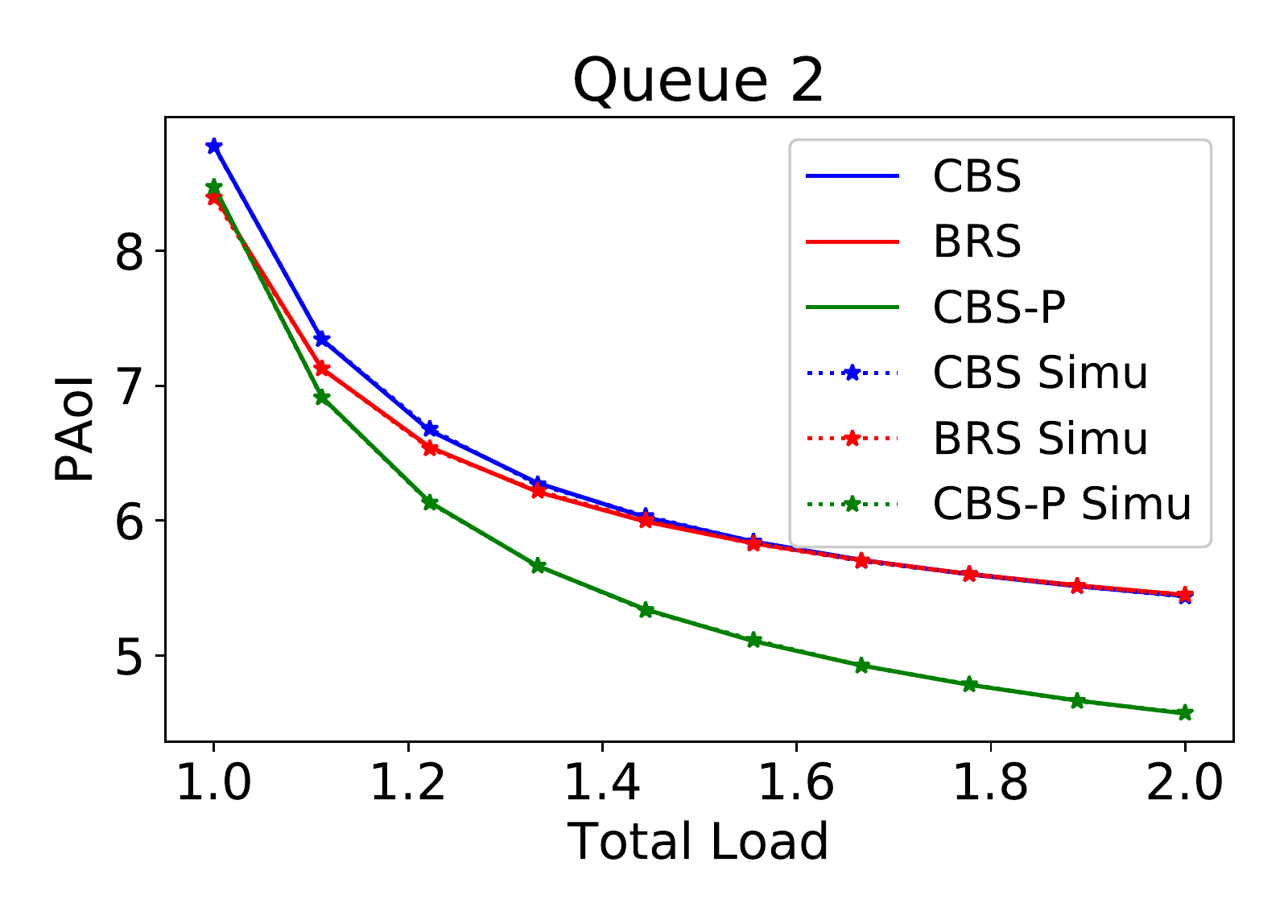}

}\subfloat[PAoI of Queue 3]{\includegraphics[scale=0.35]{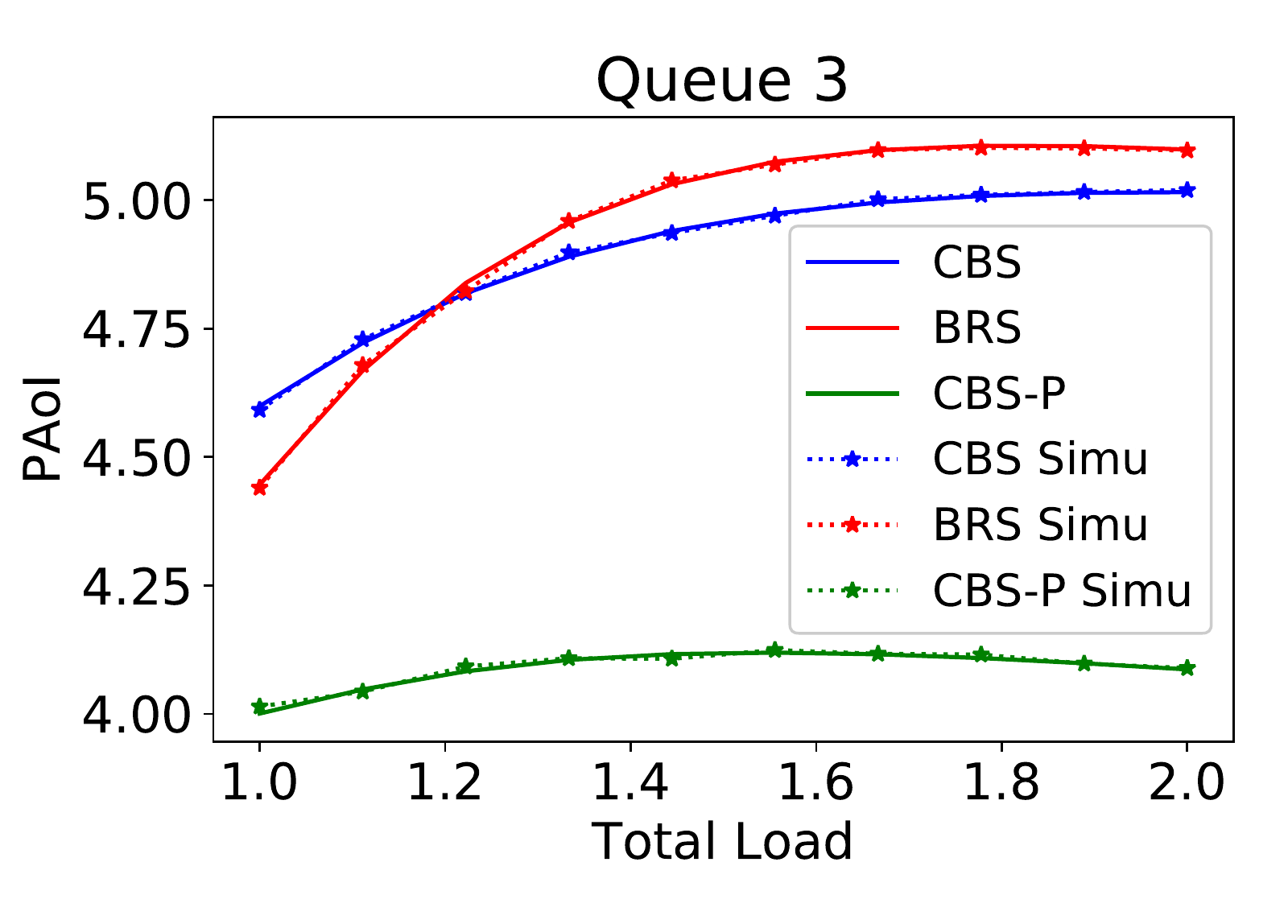}

}

\caption{PAoI of Polling Systems with Cyclic Scheme, $\boldsymbol{\lambda}=(0.1,0.2,0.7)*Total\,Load$,
$H_{i}=H\sim exp(1),U_{ij}=U=0.2$\label{fig:PAoI-of-Polling}}

\end{figure*}

We now perform numerical studies for different polling systems. In
Fig. \ref{fig:PAoI-of-Polling} we compare the exact solutions of
PAoI that we provided in Section \ref{sec:AoI-Polling} with the simulation
results for the polling system with $k=3$ and cyclic polling scheme.
We find that the exact results match the simulation results from Fig.
\ref{fig:PAoI-of-Polling}. Interestingly, we find that increasing
the traffic load will not always reduce the PAoI for CBS, BRS, and
CBS-P. This observation is different from that for i.i.d. vacation
systems, where increasing the traffic rate can reduce the PAoI when
service time is exponential. As we observed from Fig. \ref{fig:PAoI-of-Polling}(c),
the PAoI of queue 3 in all three systems will increase when the traffic
load increases. This phenomenon is because the numerical test of Fig.
\ref{fig:PAoI-of-Polling} is based on the cyclic polling scheme.
For queue 3, the vacation time increases since the other queues are
more likely to be served during the server's vacation. Although increasing
the traffic load will reduce the waiting time of an informative packet
(i.e., the server is more likely to find a fresh packet when a vacation
is over), the increase in vacation time for queue 3 would overshadow
the reduction in $G$, so that the PAoI is increasing for queue 3
as the total traffic load increases. 

The numerical study for a polling system with $k=8$ and cyclic scheme
is provided in Table \ref{tab:Exact-PAoI-and-1}. We choose the same
system parameters as the numerical study in \cite{takine1988exact}
by heavily loading two queues (each queue takes 45\% of the total
load). We proved in Theorem \ref{thm:BRS-always-has} that BRS always
has a no greater PAoI than CBS when the server's vacations are i.i.d.
However, Table \ref{tab:Exact-PAoI-and-1} shows that the PAoI of
BRS is not always smaller than PAoI of CBS in the polling system.
\cite{kofman1993blocking}. The non-i.i.d. vacations in the polling
system thus prevent Theorem \ref{thm:BRS-always-has} from holding
true. However, we can see that when the arrival rate is low, BRS still
has a smaller PAoI than CBS. Table \ref{tab:Exact-PAoI-and-1} also
shows that PAoI in CBS is larger than that in CBS-P when the service
time is exponential, as we proved in Theorem \ref{thm:If-the-service-1}.

\begin{table}[h]
\begin{center}\subfloat[Total load = 0.85]{\scalebox{0.8}{%
\begin{tabular}{|c|c|c|c|c|c|c|}
\hline 
\multirow{2}{*}{Queue} & \multicolumn{2}{c|}{CBS} & \multicolumn{2}{c|}{BRS} & \multicolumn{2}{c|}{CBS-P}\tabularnewline
\cline{2-7} \cline{3-7} \cline{4-7} \cline{5-7} \cline{6-7} \cline{7-7} 
 & PAoI & Simu & PAoI & Simu & PAoI & Simu\tabularnewline
\hline 
1 & 5.4396 & 5.4235 & 5.0996 & 5.1078 & 5.0688 & 5.0567\tabularnewline
\hline 
2 & 74.2941 & 75.7875 & 73.6306 & 73.9982 & 74.2684 & 74.1001\tabularnewline
\hline 
3 & 74.2984 & 74.6491 & 73.6372 & 74.9442 & 74.2726 & 72.9671\tabularnewline
\hline 
4 & 5.4386 & 5.4292 & 5.0985 & 5.1076 & 5.0677 & 5.0804\tabularnewline
\hline 
5 & 74.2897 & 73.3433 & 73.6236 & 75.2181 & 74.2639 & 74.6225\tabularnewline
\hline 
6 & 74.2938 & 73.2033 & 73.6300 & 74.3852 & 74.2680 & 75.7437\tabularnewline
\hline 
7 & 74.2980 & 75.8521 & 73.6366 & 74.2756 & 74.2723 & 75.8249\tabularnewline
\hline 
8 & 74.3024 & 75.7529 & 73.6433 & 73.2163 & 74.2766 & 73.6263\tabularnewline
\hline 
\end{tabular}}

}\end{center}

\begin{center}\subfloat[Total load = 30]{\scalebox{0.8}{%
\begin{tabular}{|c|c|c|c|c|c|c|}
\hline 
\multirow{2}{*}{Queue} & \multicolumn{2}{c|}{CBS} & \multicolumn{2}{c|}{BRS} & \multicolumn{2}{c|}{CBS-P}\tabularnewline
\cline{2-7} \cline{3-7} \cline{4-7} \cline{5-7} \cline{6-7} \cline{7-7} 
 & PAoI & Simu & PAoI & Simu & PAoI & Simu\tabularnewline
\hline 
1 & 8.7298 & 8.7368 & 8.8934 & 8.8892 & 7.7298 & 7.7360\tabularnewline
\hline 
2 & 10.9433 & 10.9366 & 10.9663 & 10.9606 & 10.0502 & 10.0833\tabularnewline
\hline 
3 & 10.9513 & 10.9366 & 10.9697 & 10.9589 & 10.0584 & 10.0699\tabularnewline
\hline 
4 & 8.7296 & 8.7433 & 8.8935 & 8.8942 & 7.72963 & 7.7357\tabularnewline
\hline 
5 & 10.9352 & 10.9290 & 10.9630 & 10.9432 & 10.0419 & 10.0419\tabularnewline
\hline 
6 & 10.9426 & 10.9026 & 10.9662 & 10.9835 & 10.0494 & 10.0817\tabularnewline
\hline 
7 & 10.9509 & 10.9874 & 10.9698 & 10.9990 & 10.0578 & 10.0799\tabularnewline
\hline 
8 & 10.9601 & 10.9653 & 10.9735 & 10.9509 & 10.0672 & 10.0768\tabularnewline
\hline 
\end{tabular}}

}\end{center}

\caption{Exact PAoI for the system with $k=8$ and cyclic scheme. Queue 1 and
4 are heavily loaded: each with 45\% total load. $H_{i}=H\sim exp(1),U_{ij}=U=\frac{1}{80}.$\label{tab:Exact-PAoI-and-1}}

\end{table}

Now we consider the PAoI of the polling system under different polling
schemes described in Section \ref{sec:AoI-Polling}. We keep the same
set of parameters for service and switching time for Table \ref{tab:Exact-PAoI-for-2}
and \ref{tab:Exact-PAoI-for-2-1}, and provide the computational results
for cyclic, LOP, and symmetric random polling schemes with different
total traffic loads. From both Tables \ref{tab:Exact-PAoI-for-2}
and \ref{tab:Exact-PAoI-for-2-1}, we find that cyclic and symmetric
random schemes perform similarly when the total traffic load is low.
When the traffic load is high, the symmetric scheme provides a lower
PAoI for those queues with high arrival rates than the cyclic scheme,
but provides higher PAoI for other queues than the cyclic scheme.
LOP has a lower PAoI than the other two schemes for queues with high
arrival rates, especially when the total traffic load is high. However,
LOP causes very large PAoI for queues with low arrival rates. This
is because the server under LOP would serve queues with high arrival
rates more frequently. Note that Theorem \ref{thm:If-the-service-1}
does not specify the polling scheme for CBS or CBS-P. So when service
time is exponential, CBS-P will always have a PAoI no greater than
that in CBS regardless of the polling scheme, as shown in Tables \ref{tab:Exact-PAoI-for-2}
and \ref{tab:Exact-PAoI-for-2-1}.

\begin{table*}
\begin{center}\scalebox{0.8}{%
\begin{tabular}{|c|c|c|c|c|c|c|c|c|c|}
\hline 
\multirow{2}{*}{Queue} & \multicolumn{3}{c|}{CBS} & \multicolumn{3}{c|}{BRS} & \multicolumn{3}{c|}{CBS-P}\tabularnewline
\cline{2-10} \cline{3-10} \cline{4-10} \cline{5-10} \cline{6-10} \cline{7-10} \cline{8-10} \cline{9-10} \cline{10-10} 
 & Cyclic & LOP & Symmetric & Cyclic & LOP & Symmetric & Cyclic & LOP & Symmetric\tabularnewline
\hline 
1 & 7.0216 & 6.9340 & 7.1243 & 6.4694 & 6.3262 & 6.5428 & 6.7901 & 6.7137 & 6.8840\tabularnewline
\hline 
2 & 123.1109 & 125.6743 & 123.2638 & 122.2918 & 126.2261 & 122.6218 & 123.0980 & 125.5646 & 123.2504\tabularnewline
\hline 
3 & 123.1121 & 125.6743 & 123.2638 & 122.2935 & 126.2261 & 122.6218 & 123.0992 & 125.5646 & 123.2504\tabularnewline
\hline 
4 & 7.0212 & 6.9340 & 7.1243 & 6.4690 & 6.3262 & 6.5428 & 6.7897 & 6.7137 & 6.8840\tabularnewline
\hline 
5 & 123.1097 & 125.6743 & 123.2638 & 122.2900 & 126.2261 & 122.6218 & 123.0969 & 125.5646 & 123.2504\tabularnewline
\hline 
6 & 123.1108 & 125.6743 & 123.2638 & 122.2917 & 126.2261 & 122.6218 & 123.0980 & 125.5646 & 123.2504\tabularnewline
\hline 
7 & 123.1120 & 125.6743 & 123.2638 & 122.2933 & 126.2261 & 122.6218 & 123.0991 & 125.5646 & 123.2504\tabularnewline
\hline 
8 & 123.1131 & 125.6743 & 123.2638 & 122.2951 & 126.2261 & 122.6218 & 123.1003 & 125.5646 & 123.2504\tabularnewline
\hline 
\end{tabular}}\end{center}

\caption{Exact PAoI for the system with $k=8$ and different polling schemes.
Queue 1 and 4 are heavily loaded: each with 45\% total load. Total
load = 0.5. $H_{i}=H\sim exp(1),U_{ij}=U=\frac{1}{80}.$\label{tab:Exact-PAoI-for-2}}

\end{table*}

\begin{table*}
\begin{center}\scalebox{0.8}{%
\begin{tabular}{|c|c|c|c|c|c|c|c|c|c|}
\hline 
\multirow{2}{*}{Queue} & \multicolumn{3}{c|}{CBS} & \multicolumn{3}{c|}{BRS} & \multicolumn{3}{c|}{CBS-P}\tabularnewline
\cline{2-10} \cline{3-10} \cline{4-10} \cline{5-10} \cline{6-10} \cline{7-10} \cline{8-10} \cline{9-10} \cline{10-10} 
 & Cyclic & LOP & Symmetric & Cyclic & LOP & Symmetric & Cyclic & LOP & Symmetric\tabularnewline
\hline 
1 & 8.0632 & 3.5189 & 6.9849 & 8.3780 & 3.3630 & 7.0477 & 7.0635 & 2.5353 & 5.9902\tabularnewline
\hline 
2 & 11.6450 & 42.6585 & 12.2968 & 11.6605 & 63.3207 & 12.3081 & 10.8810 & 41.7180 & 11.5688\tabularnewline
\hline 
3 & 11.6663 & 42.6585 & 12.2968 & 11.6715 & 63.3207 & 12.3081 & 10.9019 & 41.7180 & 11.5688\tabularnewline
\hline 
4 & 8.0620 & 3.5189 & 6.9849 & 8.3778 & 3.3630 & 7.0477 & 7.0622 & 2.5353 & 5.9902\tabularnewline
\hline 
5 & 11.6232 & 42.6585 & 12.2968 & 11.6493 & 63.3207 & 12.3081 & 10.8596 & 41.7180 & 11.5688\tabularnewline
\hline 
6 & 11.6413 & 42.6585 & 12.2968 & 11.6590 & 63.3207 & 12.3081 & 10.8773 & 41.7180 & 11.5688\tabularnewline
\hline 
7 & 11.6624 & 42.6585 & 12.2968 & 11.6700 & 63.3207 & 12.3081 & 10.8980 & 41.7180 & 11.5688\tabularnewline
\hline 
8 & 11.6870 & 42.6585 & 12.2968 & 11.6825 & 63.3207 & 12.3081 & 10.9221 & 41.7180 & 11.5688\tabularnewline
\hline 
\end{tabular}}\end{center}

\caption{Exact PAoI for the system with $k=8$ and different polling schemes.
Queue 1 and 4 are heavily loaded: each with 45\% total load. Total
load = 20. $H_{i}=H\sim exp(1),U_{ij}=U=\frac{1}{80}.$\label{tab:Exact-PAoI-for-2-1}}
\end{table*}

\begin{figure*}[t]
\subfloat[CBS]{\includegraphics[scale=0.35]{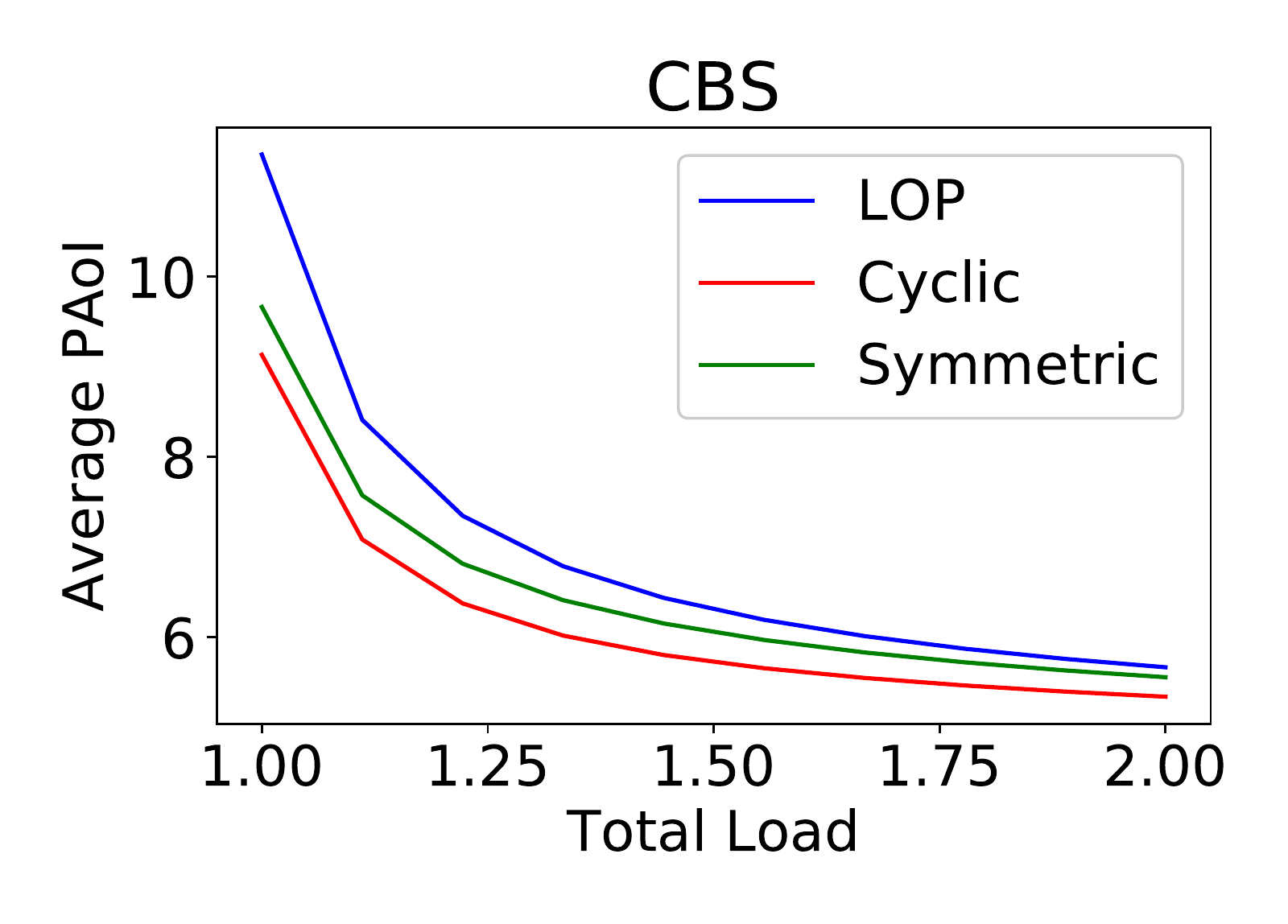}

}\subfloat[BRS]{\includegraphics[scale=0.35]{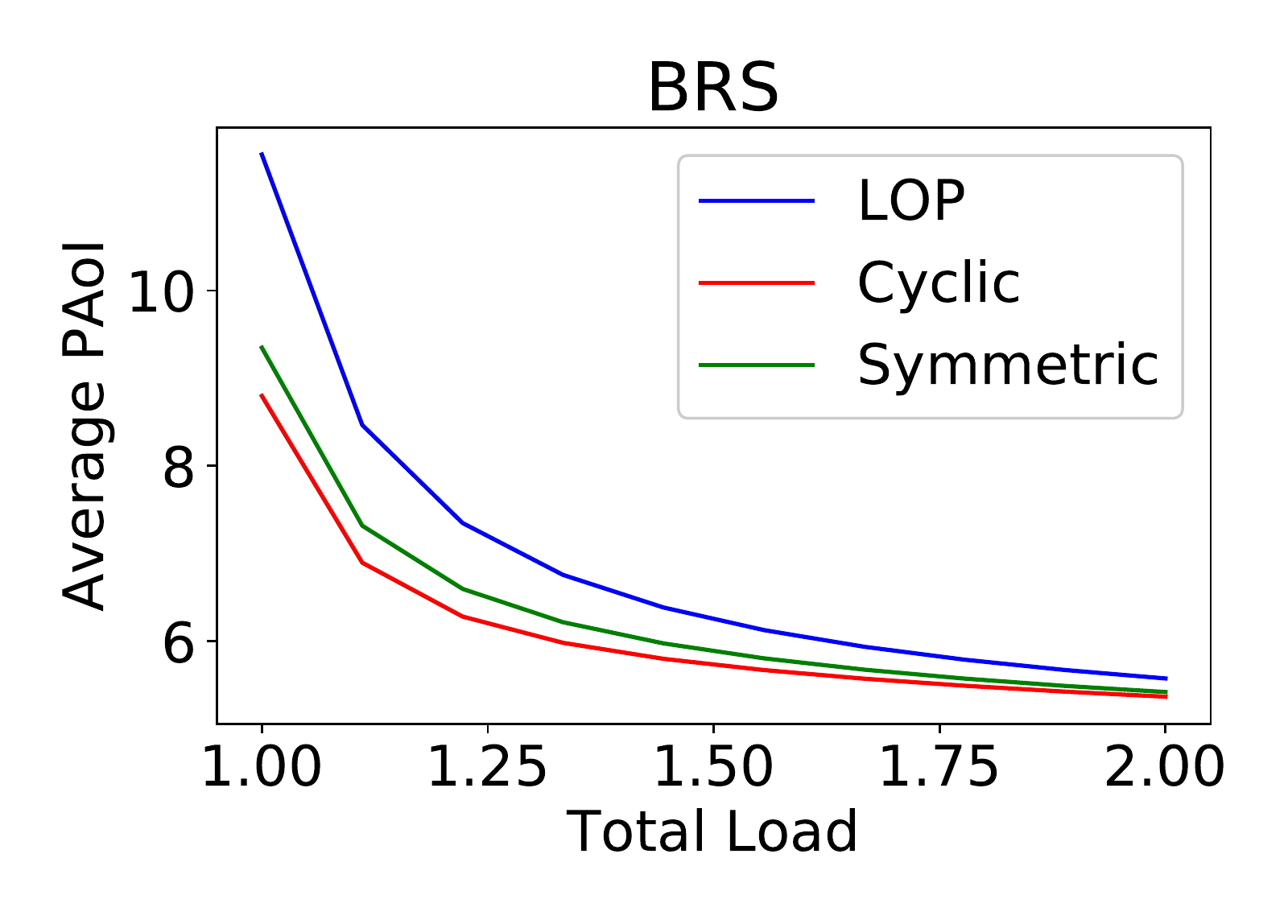}

}\subfloat[CBS-P]{\includegraphics[scale=0.35]{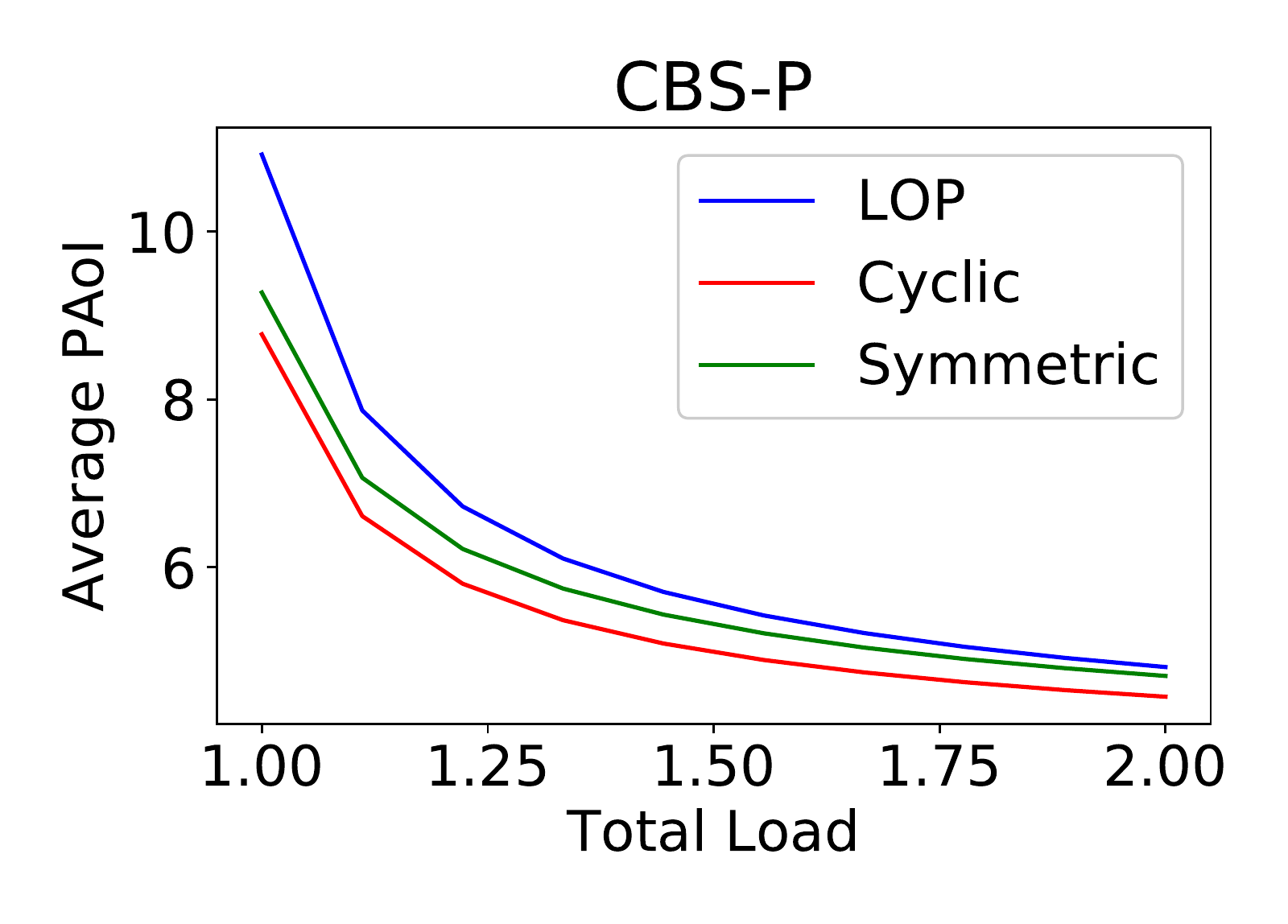}

}

\caption{Average PAoI Across Queues in Polling Systems, $\boldsymbol{\lambda}=(0.1,0.2,0.7)*Total\,Load$,
$H_{i}=H\sim exp(1),U_{ij}=U=0.2$ \label{fig:Sum-PAoI-for}}
\end{figure*}

Next, we consider the average PAoI across queues (i.e., $\frac{1}{k}\sum_{i=1}^{k}\boldsymbol{E}[A_{i}]$)
under those three different Markovian polling schemes, as shown in
Fig. \ref{fig:Sum-PAoI-for}. The average PAoI across queues was also
considered in \cite{bedewy2018age,xu2019towards}. In Fig. \ref{fig:Sum-PAoI-for}
we find that the cyclic scheme achieves the lowest average PAoI under
different traffic loads for CBS, BRS, and CBS-P. LOP has the highest
average PAoI among these three polling schemes. This is because, under
LOP, the server would likely serve the queues with high arrival rates,
and queues with low arrival rates would be polled infrequently. Since
PAoI is more sensitive to the arrival rate change when the arrival
rate is low (which we can observe from Fig. \ref{fig:Vacation-Server}
and \ref{fig:Single-Queue-System}), the PAoI reduction in queues
with high arrival rates would be overshadowed by the PAoI increase
in queues with low arrival rates, when LOP is applied. This observation
implies that if one wants to reduce the average PAoI for the entire
system, a potential strategy is to avoid polling specific queues too
frequently. Therefore, policies with even polling frequency for queues,
such as the cyclic scheme, are recommended for achieving a small average
PAoI. 

\section{Concluding Remarks\label{sec:Concluding-Remarks}}

In this paper, we investigated the information freshness on queueing
systems with server vacations. We evaluated the performance of three
scheduling policies, i.e., CBS, BRS, and CBS-P, in systems with both
i.i.d. vacations and non-i.i.d. vacations. For i.i.d. vacation systems,
we provided a general decomposition approach that decomposes the system
age into independent components. We further used the decomposition
approach to derive information freshness metrics such as AoI, PAoI,
and the variance of peak age for these three policies. We showed that
BRS always achieves a PAoI no greater than CBS regardless of the service
time and vacation time distributions, and BRS has the advantage over
CBS in minimizing information freshness metrics when the arrival rate
is low. We also proved that the AoI and PAoI in CBS-P are always no
greater than those in CBS when the service time is exponential, and
we showed that CBS-P has the advantage over CBS in minimizing information
freshness metrics when the service time is Gamma distributed with
a small scale parameter. However, no system always performs better
than the other two in terms of AoI, PAoI, and variance of peak age
altogether. We also found that reducing vacation time does not always
reduce AoI, due to the special definition of AoI. 

For systems with non-i.i.d. vacations, we investigated the polling
system as an example. We provided an approach to calculate the PAoI
for the three policies in the polling system and proved that when
service times are exponential, CBS-P has a PAoI no greater than that
in CBS, under any Markovian polling schemes. Our numerical studies
showed that BRS no longer has a smaller PAoI than CBS in the polling
system. However, when the arrival rate is low, the PAoI in BRS can
still be much smaller than that in CBS. We also found that the cyclic
polling scheme performs better than the symmetric scheme and LOP in
reducing the average PAoI across queues in polling systems. In our
future work, we will consider the closed-form expressions of AoI for
systems with non-i.i.d. vacations, such as polling systems. We will
also consider the optimal switching scheme and scheduling scheme for
polling systems in the future.

%\section*{Acknowledgment}
%The authors are grateful to the editorial board and anonymous reviewers. Their comments and suggestions
%have significantly improved the content and presentation of this work.

\bibliographystyle{IEEEtran}
\bibliography{IEEEabrv,Jin_AoI_1}

% Generated by IEEEtran.bst, version: 1.14 (2015/08/26)
\begin{thebibliography}{10}
\providecommand{\url}[1]{#1}
\csname url@samestyle\endcsname
\providecommand{\newblock}{\relax}
\providecommand{\bibinfo}[2]{#2}
\providecommand{\BIBentrySTDinterwordspacing}{\spaceskip=0pt\relax}
\providecommand{\BIBentryALTinterwordstretchfactor}{4}
\providecommand{\BIBentryALTinterwordspacing}{\spaceskip=\fontdimen2\font plus
\BIBentryALTinterwordstretchfactor\fontdimen3\font minus
  \fontdimen4\font\relax}
\providecommand{\BIBforeignlanguage}[2]{{%
\expandafter\ifx\csname l@#1\endcsname\relax
\typeout{** WARNING: IEEEtran.bst: No hyphenation pattern has been}%
\typeout{** loaded for the language `#1'. Using the pattern for}%
\typeout{** the default language instead.}%
\else
\language=\csname l@#1\endcsname
\fi
#2}}
\providecommand{\BIBdecl}{\relax}
\BIBdecl

\bibitem{kaul2012real}
S.~Kaul, R.~Yates, and M.~Gruteser, ``Real-time status: How often should one
  update?'' in \emph{INFOCOM, 2012 Proceedings IEEE}.\hskip 1em plus 0.5em
  minus 0.4em\relax IEEE, 2012, pp. 2731--2735.

\bibitem{costa2016age}
M.~Costa, M.~Codreanu, and A.~Ephremides, ``On the age of information in status
  update systems with packet management,'' \emph{IEEE Transactions on
  Information Theory}, vol.~62, no.~4, pp. 1897--1910, 2016.

\bibitem{zhong2018two}
J.~Zhong, R.~D. Yates, and E.~Soljanin, ``Two freshness metrics for local cache
  refresh,'' in \emph{2018 IEEE International Symposium on Information Theory
  (ISIT)}.\hskip 1em plus 0.5em minus 0.4em\relax IEEE, 2018, pp. 1924--1928.

\bibitem{huang2015optimizing}
L.~Huang and E.~Modiano, ``Optimizing age-of-information in a multi-class
  queueing system,'' in \emph{2015 IEEE International Symposium on Information
  Theory (ISIT)}.\hskip 1em plus 0.5em minus 0.4em\relax IEEE, 2015, pp.
  1681--1685.

\bibitem{bedewy2019age}
A.~M. Bedewy, Y.~Sun, and N.~B. Shroff, ``The age of information in multihop
  networks,'' \emph{IEEE/ACM Transactions on Networking}, vol.~27, no.~3, pp.
  1248--1257, 2019.

\bibitem{song2019generic}
\BIBentryALTinterwordspacing
C.~Song, K.~Liu, and X.~Zhang, ``A generic framework for multisensor
  degradation modeling based on supervised classification and failure
  surface,'' \emph{IISE Transactions}, vol.~51, no.~11, pp. 1288--1302, 2019.
  [Online]. Available: \url{https://doi.org/10.1080/24725854.2018.1555384}
\BIBentrySTDinterwordspacing

\bibitem{cheng2008vision}
Y.~Cheng and M.~A. Jafari, ``Vision-based online process control in
  manufacturing applications,'' \emph{IEEE Transactions on Automation Science
  and Engineering}, vol.~5, no.~1, pp. 140--153, 2008.

\bibitem{yao2018constrained}
B.~Yao and H.~Yang, ``Constrained markov decision process modeling for
  sequential optimization of additive manufacturing build quality,'' \emph{IEEE
  Access}, vol.~6, pp. 54\,786--54\,794, 2018.

\bibitem{guo2016delay}
X.~Guo, Z.~Niu, S.~Zhou, and P.~Kumar, ``Delay-constrained energy-optimal base
  station sleeping control,'' \emph{IEEE Journal on Selected Areas in
  Communications}, vol.~34, no.~5, pp. 1073--1085, 2016.

\bibitem{heidemann2012underwater}
J.~Heidemann, M.~Stojanovic, and M.~Zorzi, ``Underwater sensor networks:
  applications, advances and challenges,'' \emph{Philosophical Transactions of
  the Royal Society A: Mathematical, Physical and Engineering Sciences}, vol.
  370, no. 1958, pp. 158--175, 2012.

\bibitem{vasilescu2005data}
I.~Vasilescu, K.~Kotay, D.~Rus, M.~Dunbabin, and P.~Corke, ``Data collection,
  storage, and retrieval with an underwater sensor network,'' in
  \emph{Proceedings of the 3rd international conference on Embedded networked
  sensor systems}, 2005, pp. 154--165.

\bibitem{mohapatra2012combined}
A.~K. Mohapatra, N.~Gautam, and R.~L. Gibson, ``Combined routing and node
  replacement in energy-efficient underwater sensor networks for seismic
  monitoring,'' \emph{IEEE Journal of Oceanic Engineering}, vol.~38, no.~1, pp.
  80--90, 2012.

\bibitem{heidemann2006research}
J.~Heidemann, W.~Ye, J.~Wills, A.~Syed, and Y.~Li, ``Research challenges and
  applications for underwater sensor networking,'' in \emph{IEEE Wireless
  Communications and Networking Conference, 2006. WCNC 2006.}, vol.~1.\hskip
  1em plus 0.5em minus 0.4em\relax IEEE, 2006, pp. 228--235.

\bibitem{doukas2011managing}
C.~Doukas and I.~Maglogiannis, ``Managing wearable sensor data through cloud
  computing,'' in \emph{2011 IEEE Third International Conference on Cloud
  Computing Technology and Science}.\hskip 1em plus 0.5em minus 0.4em\relax
  IEEE, 2011, pp. 440--445.

\bibitem{majumder2017wearable}
S.~Majumder, T.~Mondal, and M.~J. Deen, ``Wearable sensors for remote health
  monitoring,'' \emph{Sensors}, vol.~17, no.~1, p. 130, 2017.

\bibitem{takagi1991priority}
H.~Takagi and Y.~Takahashi, ``Priority queues with batch poisson arrivals,''
  \emph{Operations Research Letters}, vol.~10, no.~4, pp. 225--232, 1991.

\bibitem{xu2019towards}
J.~{Xu} and N.~{Gautam}, ``Peak age of information in priority queuing
  systems,'' \emph{IEEE Transactions on Information Theory}, vol.~67, no.~1,
  pp. 373--390, 2021.

\bibitem{kella1988priorities}
O.~Kella and U.~Yechiali, ``Priorities in {M/G/1} queue with server
  vacations,'' \emph{Naval Research Logistics (NRL)}, vol.~35, no.~1, pp.
  23--34, 1988.

\bibitem{boon2011applications}
M.~A. Boon, R.~Van~der Mei, and E.~M. Winands, ``Applications of polling
  systems,'' \emph{Surveys in Operations Research and Management Science},
  vol.~16, no.~2, pp. 67--82, 2011.

\bibitem{fuhrmann1985stochastic}
S.~W. Fuhrmann and R.~B. Cooper, ``Stochastic decompositions in the {M/G/1}
  queue with generalized vacations,'' \emph{Operations research}, vol.~33,
  no.~5, pp. 1117--1129, 1985.

\bibitem{maccio2015optimal}
V.~J. Maccio and D.~G. Down, ``On optimal policies for energy-aware servers,''
  \emph{Performance Evaluation}, vol.~90, pp. 36--52, 2015.

\bibitem{mankar2021spatial}
P.~D. Mankar, M.~A. Abd-Elmagid, and H.~S. Dhillon, ``Spatial distribution of
  the mean peak age of information in wireless networks,'' \emph{IEEE
  Transactions on Wireless Communications}, 2021.

\bibitem{maatouk2018age}
A.~Maatouk, M.~Assaad, and A.~Ephremides, ``The age of updates in a simple
  relay network,'' in \emph{2018 IEEE Information Theory Workshop (ITW)}.\hskip
  1em plus 0.5em minus 0.4em\relax IEEE, 2018, pp. 1--5.

\bibitem{yates2019age}
R.~D. Yates and S.~K. Kaul, ``The age of information: Real-time status updating
  by multiple sources,'' \emph{IEEE Transactions on Information Theory},
  vol.~65, no.~3, pp. 1807--1827, 2019.

\bibitem{kosta2019age}
A.~Kosta, N.~Pappas, A.~Ephremides, and V.~Angelakis, ``Age of information
  performance of multiaccess strategies with packet management,'' \emph{Journal
  of Communications and Networks}, vol.~21, no.~3, pp. 244--255, 2019.

\bibitem{fuhrmann1984note}
S.~Fuhrmann, ``A note on the {M/G/1} queue with server vacations,''
  \emph{Operations research}, vol.~32, no.~6, pp. 1368--1373, 1984.

\bibitem{lee1984m}
T.~T. Lee, ``{M/G/1/N} queue with vacation time and exhaustive service
  discipline,'' \emph{Operations Research}, vol.~32, no.~4, pp. 774--784, 1984.

\bibitem{lee1989m}
------, ``{M/G/1/N} queue with vacation time and limited service discipline,''
  \emph{Performance Evaluation}, vol.~9, no.~3, pp. 181--190, 1989.

\bibitem{frey1997note}
A.~Frey and Y.~Takahashi, ``A note on an {M/GI/1/N} queue with vacation time
  and exhaustive service discipline,'' \emph{Operations Research lLetters},
  vol.~21, no.~2, pp. 95--100, 1997.

\bibitem{takagi1991analysis}
H.~Takagi, ``Analysis of finite-capacity polling systems,'' \emph{Advances in
  Applied Probability}, vol.~23, no.~2, pp. 373--387, 1991.

\bibitem{hur2003analysis}
S.~Hur, J.~Kim, and C.~Kang, ``An analysis of the m/g/1 system with n and t
  policy,'' \emph{Applied Mathematical Modelling}, vol.~27, no.~8, pp.
  665--675, 2003.

\bibitem{artalejo2001m}
J.~R. Artalejo, ``On the m/g/1 queue with d-policy,'' \emph{Applied
  Mathematical Modelling}, vol.~25, no.~12, pp. 1055--1069, 2001.

\bibitem{najm2018status}
E.~Najm and E.~Telatar, ``Status updates in a multi-stream {M/G/1/1} preemptive
  queue,'' in \emph{IEEE Infocom 2018-IEEE Conference On Computer
  Communications Workshops (Infocom Wkshps)}.\hskip 1em plus 0.5em minus
  0.4em\relax IEEE, 2018, pp. 124--129.

\bibitem{zou2019benefis}
P.~{Zou}, O.~{Ozel}, and S.~{Subramaniam}, ``Waiting before serving: A
  companion to packet management in status update systems,'' \emph{IEEE
  Transactions on Information Theory}, vol.~66, no.~6, pp. 3864--3877, 2020.

\bibitem{kaul2018age}
S.~K. Kaul and R.~D. Yates, ``Age of information: Updates with priority,'' in
  \emph{2018 IEEE International Symposium on Information Theory (ISIT)}.\hskip
  1em plus 0.5em minus 0.4em\relax IEEE, 2018, pp. 2644--2648.

\bibitem{maatouk2019age}
A.~Maatouk, M.~Assaad, and A.~Ephremides, ``Age of information with prioritized
  streams: When to buffer preempted packets?'' in \emph{2019 IEEE International
  Symposium on Information Theory (ISIT)}.\hskip 1em plus 0.5em minus
  0.4em\relax IEEE, 2019, pp. 325--329.

\bibitem{moltafet2019age}
M.~Moltafet, M.~Leinonen, and M.~Codreanu, ``On the age of information in
  multi-source queueing models,'' \emph{arXiv preprint arXiv:1911.07029}, 2019.

\bibitem{kam2018age}
C.~Kam, S.~Kompella, G.~D. Nguyen, J.~E. Wieselthier, and A.~Ephremides, ``On
  the age of information with packet deadlines,'' \emph{IEEE Transactions on
  Information Theory}, vol.~64, no.~9, pp. 6419--6428, 2018.

\bibitem{inoue2019general}
Y.~Inoue, H.~Masuyama, T.~Takine, and T.~Tanaka, ``A general formula for the
  stationary distribution of the age of information and its application to
  single-server queues,'' \emph{IEEE Transactions on Information Theory},
  vol.~65, no.~12, pp. 8305--8324, 2019.

\bibitem{soysal2018age}
A.~Soysal and S.~Ulukus, ``Age of information in {G/G/1/1} systems,''
  \emph{arXiv preprint arXiv:1805.12586}, 2018.

\bibitem{najm2016age}
E.~Najm and R.~Nasser, ``Age of information: The gamma awakening,'' in
  \emph{2016 IEEE International Symposium on Information Theory (ISIT)}.\hskip
  1em plus 0.5em minus 0.4em\relax IEEE, 2016, pp. 2574--2578.

\bibitem{najm2017status}
E.~Najm, R.~Yates, and E.~Soljanin, ``Status updates through {M/G/1/1} queues
  with {HARQ},'' in \emph{2017 IEEE International Symposium on Information
  Theory (ISIT)}.\hskip 1em plus 0.5em minus 0.4em\relax IEEE, 2017, pp.
  131--135.

\bibitem{chen2016age}
K.~Chen and L.~Huang, ``Age-of-information in the presence of error,'' in
  \emph{2016 IEEE International Symposium on Information Theory (ISIT)}.\hskip
  1em plus 0.5em minus 0.4em\relax IEEE, 2016, pp. 2579--2583.

\bibitem{jiang2019timely}
Z.~Jiang, B.~Krishnamachari, X.~Zheng, S.~Zhou, and Z.~Niu, ``Timely status
  update in wireless uplinks: Analytical solutions with asymptotic
  optimality,'' \emph{IEEE Internet of Things Journal}, vol.~6, no.~2, pp.
  3885--3898, 2019.

\bibitem{he2017optimal}
Q.~He, D.~Yuan, and A.~Ephremides, ``Optimal link scheduling for age
  minimization in wireless systems,'' \emph{IEEE Transactions on Information
  Theory}, vol.~64, no.~7, pp. 5381--5394, 2017.

\bibitem{hsu2017age}
Y.-P. Hsu, E.~Modiano, and L.~Duan, ``Age of information: Design and analysis
  of optimal scheduling algorithms,'' in \emph{2017 IEEE International
  Symposium on Information Theory (ISIT)}.\hskip 1em plus 0.5em minus
  0.4em\relax IEEE, 2017, pp. 561--565.

\bibitem{talak2019optimizing}
R.~{Talak}, S.~{Karaman}, and E.~{Modiano}, ``Optimizing information freshness
  in wireless networks under general interference constraints,'' \emph{IEEE/ACM
  Transactions on Networking}, vol.~28, no.~1, pp. 15--28, 2020.

\bibitem{kadota2018scheduling}
I.~Kadota, A.~Sinha, E.~Uysal-Biyikoglu, R.~Singh, and E.~Modiano, ``Scheduling
  policies for minimizing age of information in broadcast wireless networks,''
  \emph{IEEE/ACM Transactions on Networking}, vol.~26, no.~6, pp. 2637--2650,
  2018.

\bibitem{najm2019content}
E.~Najm, R.~Nasser, and E.~Telatar, ``Content based status updates,''
  \emph{IEEE Transactions on Information Theory}, 2019.

\bibitem{tripathi2019age1}
V.~Tripathi, R.~Talak, and E.~Modiano, ``Age of information for discrete time
  queues,'' \emph{arXiv preprint arXiv:1901.10463}, 2019.

\bibitem{tripathi2019age}
------, ``Age optimal information gathering and dissemination on graphs,'' in
  \emph{IEEE INFOCOM 2019-IEEE Conference on Computer Communications}.\hskip
  1em plus 0.5em minus 0.4em\relax IEEE, 2019, pp. 2422--2430.

\bibitem{niu2003vacation}
Z.~Niu, T.~Shu, and Y.~Takahashi, ``A vacation queue with setup and close-down
  times and batch markovian arrival processes,'' \emph{Performance Evaluation},
  vol.~54, no.~3, pp. 225--248, 2003.

\bibitem{gupta2006computing}
U.~Gupta and K.~Sikdar, ``Computing queue length distributions in map/g/1/n
  queue under single and multiple vacation,'' \emph{Applied mathematics and
  computation}, vol. 174, no.~2, pp. 1498--1525, 2006.

\bibitem{lee1996exact}
T.~Y.~S. Lee and J.~Sunjaya, ``Exact analysis of asymmetric random polling
  systems with single buffers and correlated input process,'' \emph{Queueing
  Systems}, vol.~23, no. 1-4, pp. 131--156, 1996.

\bibitem{takine1988exact}
T.~Takine, Y.~Takahashi, and T.~Hasegawa, ``Exact analysis of asymmetric
  polling systems with single buffers,'' \emph{IEEE Transactions on
  Communications}, vol.~36, no.~10, pp. 1119--1127, 1988.

\bibitem{chung1994performance}
H.~Chung, C.~K. Un, and W.~Y. Jung, ``Performance analysis of markovian polling
  systems with single buffers,'' \emph{Performance Evaluation}, vol.~19, no.~4,
  pp. 303--315, 1994.

\bibitem{mukherjee1990comments}
B.~Mukherjee, C.~K. Kwok, A.~C. Lantz, and W.-H. Moh, ``Comments on" exact
  analysis of asymmetric polling systems with single buffers,'' \emph{IEEE
  Transactions on Communications}, vol.~38, no.~7, pp. 944--946, 1990.

\bibitem{kulkarni2016modeling}
V.~G. Kulkarni, \emph{Modeling and analysis of stochastic systems}.\hskip 1em
  plus 0.5em minus 0.4em\relax Chapman and Hall/CRC, 2016.

\bibitem{takagi1988queuing}
H.~Takagi, ``Queuing analysis of polling models,'' \emph{ACM Computing Surveys
  (CSUR)}, vol.~20, no.~1, pp. 5--28, 1988.

\bibitem{gautam2012analysis}
N.~Gautam, \emph{Analysis of queues: methods and applications}.\hskip 1em plus
  0.5em minus 0.4em\relax CRC Press, 2012.

\bibitem{kofman1993blocking}
D.~Kofman, ``Blocking probability, throughput and waiting time in finite
  capacity polling systems,'' \emph{Queueing Systems}, vol.~14, no. 3-4, pp.
  385--411, 1993.

\bibitem{wierman2007scheduling}
A.~Wierman, E.~M. Winands, and O.~J. Boxma, ``Scheduling in polling systems,''
  \emph{Performance Evaluation}, vol.~64, no.~9, pp. 1009--1028, 2007.

\bibitem{2020arXiv200102530X}
J.~{Xu} and N.~{Gautam}, ``On competitive analysis for polling systems,''
  \emph{Naval Research Logistics (NRL)}, vol.~67, no.~6, pp. 404--419, 2020.

\bibitem{takagi2000analysis}
H.~Takagi, ``Analysis and application of polling models,'' in \emph{Performance
  Evaluation: Origins and Directions}.\hskip 1em plus 0.5em minus 0.4em\relax
  Springer, 2000, pp. 423--442.

\bibitem{ferguson1985exact}
M.~Ferguson and Y.~Aminetzah, ``Exact results for nonsymmetric token ring
  systems,'' \emph{IEEE Transactions on Communications}, vol.~33, no.~3, pp.
  223--231, 1985.

\bibitem{boxma1989waiting}
O.~J. Boxma and J.~A. Weststrate, ``Waiting times in polling systems with
  markovian server routing,'' in \emph{Messung, Modellierung und Bewertung von
  Rechensystemen und Netzen}.\hskip 1em plus 0.5em minus 0.4em\relax Springer,
  1989, pp. 89--104.

\bibitem{bedewy2018age}
A.~M. Bedewy, Y.~Sun, S.~Kompella, and N.~B. Shroff, ``Age-optimal sampling and
  transmission scheduling in multi-source systems,'' \emph{arXiv preprint
  arXiv:1812.09463}, 2018.

\bibitem{lehmann1966some}
E.~L. Lehmann, ``Some concepts of dependence,'' \emph{The Annals of
  Mathematical Statistics}, pp. 1137--1153, 1966.

\end{thebibliography}

\newpage
\pagebreak
$ $
\newpage
\appendices
\begin{strip}
\begin{center}
\huge{Supplementary Material for the paper "Age of Information for Single Buffer Systems with Vacation Server"}
\end{center}
\end{strip}

\section{\label{sec:Proof-for-Theorem}Proof for Theorem \ref{thm:The-PAoI-for}}
\begin{IEEEproof}
Different from the case of non-preemptive service, in the case where
service is preempted by new arrivals, we decompose the age peak into
three pieces 
\begin{eqnarray}
\boldsymbol{E}[A_{\{l\}}] & = & \boldsymbol{E}[D_{\{l-1\}}]+\boldsymbol{E}[B_{\{l\}}]+\boldsymbol{E}[L_{\{l+1\}}],\label{eq:A0}
\end{eqnarray}
where $D_{\{l-1\}}$ is the delay (time in the system) of an informative
packet, \textbf{$B_{\{l\}}$ }is the time period when the server is
on vacation during the $l^{th}$ regenerative cycle (the same as we
defined in Theorem \ref{thm:The-AoI-of-CBS}), and $L_{\{l+1\}}$
is the time when the server is processing during the $l^{th}$ regenerative
cycle. We let $r_{j}$, $S_{j}$, and $C_{j}$ be the arrival time,
time to start service, and completion time of the $j^{th}$ packet
that arrives in the system from time 0. Note that not all the packets
have $S_{j}$ and $C_{j}$, as some packets are preempted and discarded.
A demonstrative graph is given in Fig. \ref{fig:Age-of-Information-1},
and the three decomposed components are mutually independent. This
is because $B_{\{l\}}$ is the time when the server is on vacation,
which is independent of delay $D_{\{l-1\}}$ and processing time $L_{\{l+1\}}$.
$L_{\{l+1\}}$ is independent of $D_{\{l-1\}}.$ Therefore the AoI
of this system can be given as 
\begin{eqnarray}
\boldsymbol{E}[\Delta] & = & \lim_{l\rightarrow\infty}\frac{\boldsymbol{E}[(D_{\{l-1\}}+B_{\{l\}}+L_{\{l+1\}})^{2}]-\boldsymbol{E}[D_{\{l\}}^{2}]}{2(\boldsymbol{E}[L_{\{l+1\}}]+\boldsymbol{E}[B_{\{l\}}])}\nonumber \\
 & = & \frac{\boldsymbol{E}[L^{2}]+2\boldsymbol{E}[L]\boldsymbol{E}[B]+\boldsymbol{E}[B^{2}]}{2(\boldsymbol{E}[L]+\boldsymbol{E}[B])}+\boldsymbol{E}[D].\label{eq:A1}
\end{eqnarray}

\begin{figure}
\begin{center}\includegraphics[scale=0.3]{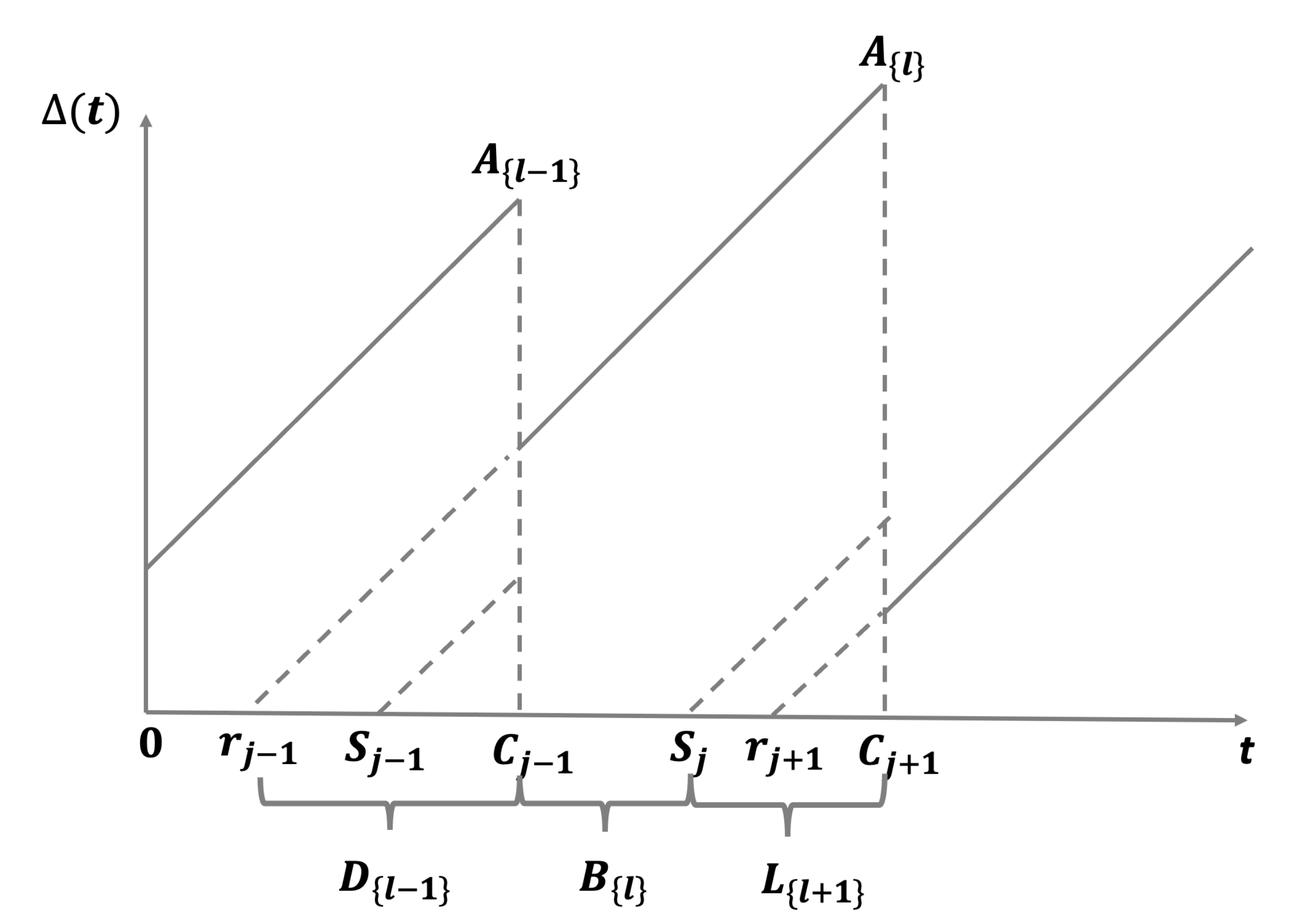}\end{center}

\caption{\label{fig:Age-of-Information-1}Age of Information Decomposition
for Preemptive Service Systems. The $l^{th}$ age peak is decomposed
into three components: $A_{\{l\}}=D_{\{l-1\}}+B_{\{l\}}+L_{\{l+1\}}$,
where $D_{\{l-1\}}$ is the delay of the $(l-1)^{th}$ informative
packet, $B_{\{l\}}$ is period when the server is on vacation, and
$L_{\{l+1\}}$ is the time period when the server is serving. In CBS-P,
packet indices may differ from the indices for age peaks. In this
example, packet $j$ is preempted by packet $j+1$ at time $r_{\{j+1\}},$
and packet $j+1$ is not preempted by any packet.}
\end{figure}

We now derive the LST of $D$, denoted as $D^{*}(s).$ We first notice
that if the service time of a packet $H$ is smaller than the inter-arrival
time $T$, then the packet is served without being preempted. Therefore,
all the packets that are eventually processed must have the service
time smaller than the inter-arrival time. If the packet that we serve
arrives during the last vacation, then its delay $D$ is its waiting
time $G$ plus its service time. If it arrives during service (it
preempts the previous packet in service), then the delay is its service
time only. Thus we have $\boldsymbol{E}[e^{-sD}|H<T]=G^{*}(s)\hat{H}(s)$
and $\boldsymbol{E}[e^{-sD}|H\geq T]=\hat{H}(s),$ where $\hat{H}(s)=\boldsymbol{E}[e^{-sH}|H<T].$

Since the inter-arrival time is exponential, we have $\hat{H}(s)=\frac{\int_{u=0}^{\infty}\int_{x=u}^{\infty}\lambda e^{-\lambda x}e^{-su}dF_{H}(u)dx}{\boldsymbol{P}(H<I)}=\frac{\int_{u=0}^{\infty}e^{-(s+\lambda)u}dF_{H}(u)}{\int_{u=0}^{\infty}\int_{x=u}^{\infty}\lambda e^{-\lambda x}dF_{H}(u)dx}=\frac{H^{*}(\lambda+s)}{H^{*}(\lambda)}.$
Then we have 
\begin{eqnarray*}
D^{*}(s) & = & G^{*}(s)\frac{H^{*}(\lambda+s)}{H^{*}(\lambda)}H^{*}(\lambda)\\
 &  & +\frac{H^{*}(\lambda+s)}{H^{*}(\lambda)}(1-H^{*}(\lambda))\\
 & = & H^{*}(\lambda+s)\left(G^{*}(s)+\frac{1}{H^{*}(\lambda)}-1\right).
\end{eqnarray*}
Using the expression for $\boldsymbol{E}[G]$ in Theorem \ref{thm:The-AoI-of-CBS},
we have 
\begin{eqnarray}
\boldsymbol{E}[D] & = & -\frac{H^{*(1)}(\lambda)}{H^{*}(\lambda)}-H^{*}(\lambda)G^{*(1)}(0)\nonumber \\
 & = & -\frac{H^{*(1)}(\lambda)}{H^{*}(\lambda)}+H^{*}(\lambda)\big(\frac{1}{\lambda}+\frac{V^{*(1)}(\lambda)}{1-V^{*}(\lambda)}\big).\label{eq:A2}
\end{eqnarray}
The LST of $B$ is given in Theorem \ref{thm:The-AoI-of-CBS} as $B^{*}(s)=\frac{V^{*}(s)-V^{*}(s+\lambda)}{1-V^{*}(s+\lambda)},$
with $\boldsymbol{E}[B]=-\frac{V^{*(1)}(0)}{1-V^{*}(\lambda)}$ and
$\boldsymbol{E}[B^{2}]=\frac{V^{*(2)}(0)}{1-V^{*}(\lambda)}+2\frac{V^{*(1)}(0)V^{*(1)}(\lambda)}{(1-V^{*}(\lambda))^{2}}$.
Now we derive the LST for $L$, i.e., $L^{*}(s)$. Notice that if
the inter-arrival time $T$ is greater than service time $H$, then
the packet is processed without being preempted. If the inter-arrival
time $T$ is smaller than $H$, then a new period $L$ is started
after $T$. We then have $\boldsymbol{E}[e^{-sL}|H<T]=\hat{H}(s)$
and $\boldsymbol{E}[e^{-sL}|H\geq T]=\boldsymbol{E}[e^{-sT}L(s)|H\geq T].$
Thus 
\begin{eqnarray*}
L^{*}(s) & = & \int_{u=0}^{\infty}\int_{x=u}^{\infty}\lambda e^{-\lambda x}e^{-su}dF_{H}(u)dx\\
 &  & +L(s)\int_{u=0}^{\infty}\int_{x=0}^{u}e^{-sx}\lambda e^{-\lambda x}dF_{H}(u)dx\\
 & = & H^{*}(s+\lambda)+L^{*}(s)\frac{\lambda}{s+\lambda}(1-H^{*}(s+\lambda)).
\end{eqnarray*}
We can then get 
\begin{eqnarray}
L^{*}(s) & = & \frac{H^{*}(s+\lambda)}{\frac{s}{s+\lambda}+\frac{\lambda}{s+\lambda}H^{*}(s+\lambda)},\label{eq:A2.1}
\end{eqnarray}
 
\begin{eqnarray}
\boldsymbol{E}[L] & = & \frac{1-H^{*}(\lambda)}{\lambda H^{*}(\lambda)},\label{eq:A3}
\end{eqnarray}
and 
\begin{eqnarray*}
\boldsymbol{E}[L^{2}] & = & \frac{2}{\lambda H^{*}(\lambda)^{2}}\left(\frac{1}{\lambda}-\frac{H^{*}(\lambda)}{\lambda}+H^{*(1)}(\lambda)\right).
\end{eqnarray*}
The PAoI for the system can now be given as 
\begin{eqnarray*}
\boldsymbol{E}[A] & = & \boldsymbol{E}[D]+\boldsymbol{E}[B]+\boldsymbol{E}[L]\\
 & = & \frac{1-H^{*}(\lambda)-\lambda H^{*(1)}(\lambda)+H^{*}(\lambda)^{2}}{\lambda H^{*}(\lambda)}\\
 &  & +\frac{H^{*}(\lambda)V^{*(1)}(\lambda)-V^{*(1)}(0)}{1-V^{*}(\lambda)}.
\end{eqnarray*}
The variance of peak age can be given as $Var(A)=Var(L)+Var(B)+Var(D),$
where the variance of each component can be computed using corresponding
LST. The expression for $Var(A)$ is involved, and we do not present
here.
\end{IEEEproof}
\begin{comment}
\newpage\clearpage\newpage
\end{comment}

\section{\label{sec:Proof-for-Theorem-4-1}Proof for Theorem \ref{thm:The-AoI-of}.}
\begin{IEEEproof}
We first derive $I^{*}(s)$ in BRS. Since each $I$ starts with processing
a packet with processing time $H,$ if there is more than one arrival
during the processing time $H$, then the server only takes one vacation
after processing the current packet. If there is no arrival during
this processing time, the server takes vacations until a packet is
observed in buffer when a vacation is over. By conditioning on scenarios
during $H$, we have $\boldsymbol{E}[e^{-sI}|H=h,m(H)\geq1]=e^{-sh}V^{*}(s),$
and $\boldsymbol{E}[e^{-sI}|H=h,m(H)=0]=e^{-sh}B^{*}(s)$. We thus
have $\boldsymbol{E}[e^{-sI}|H=h]=e^{-sh}V^{*}(s)(1-e^{-\lambda h})+e^{-sh}B^{*}(s)e^{-\lambda h}.$
Therefore $\boldsymbol{E}[e^{-sI}]=H^{*}(s)V^{*}(s)-H^{*}(\lambda+s)V^{*}(s)+H^{*}(\lambda+s)B^{*}(s),$
where $B^{*}(s)=\frac{V^{*}(s)-V^{*}(s+\lambda)}{1-V^{*}(s+\lambda)}.$

We next derive $\boldsymbol{E}[G]$ for BRS. From Equation (\ref{eq:1.2})
we know that $\boldsymbol{E}[G]$ can be written as a formula of the
LST of $W$. So in the following we first derive the LST of $W$.
If there is more than one arrival before the server returns from the
first vacation, then $\boldsymbol{E}[e^{-sW}|m(V_{1}+H)\geq1]=\frac{V^{*}(\lambda)H^{*}(\lambda)-V^{*}(s)H^{*}(s)}{(s-\lambda)(1-V^{*}(\lambda)H^{*}(\lambda))}\lambda.$
If there is no arrival before the server returns from the first vacation,
we have $\boldsymbol{E}[e^{-sW}|m(V_{1}+H)=0]=\frac{V^{*}(\lambda)-V^{*}(s)}{(s-\lambda)(1-V^{*}(\lambda))}\lambda.$
We thus have 
\begin{eqnarray*}
\boldsymbol{E}[e^{-sW}] & = & \frac{\lambda[1-V^{*}(\lambda)H^{*}(\lambda)]}{(s-\lambda)(1-V^{*}(\lambda)H^{*}(\lambda))}\bigg\{ V^{*}(\lambda)H^{*}(\lambda)\\
 &  & -V^{*}(s)H^{*}(s)\bigg\}\\
 &  & +\frac{V^{*}(\lambda)-V^{*}(s)}{(s-\lambda)(1-V^{*}(\lambda))}\lambda V^{*}(\lambda)H^{*}(\lambda).
\end{eqnarray*}
Using L'Hospital rule at $s=\lambda$, we have 

\begin{eqnarray*}
\boldsymbol{E}[e^{-\lambda W}] & = & -\lambda V^{*(1)}(\lambda)H^{*}(\lambda)-\lambda V^{*}(\lambda)H^{*(1)}(\lambda)\\
 &  & -\frac{V^{*(1)}(\lambda)}{1-V^{*}(\lambda)}\lambda V^{*}(\lambda)H^{*}(\lambda).
\end{eqnarray*}
Therefore $\boldsymbol{E}[G]=\frac{1}{\lambda}+V^{*(1)}(\lambda)H^{*}(\lambda)+V^{*}(\lambda)H^{*(1)}(\lambda)+\frac{V^{*(1)}(\lambda)}{1-V^{*}(\lambda)}V^{*}(\lambda)H^{*}(\lambda)$.
Using Equation (\ref{eq:0}) and (\ref{eq:1}) we can then obtain
the PAoI and AoI of BRS.
\end{IEEEproof}

\section{\label{sec:Proof-for-Theorem-3}Proof for Theorem \ref{thm:BRS-always-has}}
\begin{IEEEproof}
From Theorems \ref{thm:The-AoI-of-CBS} and \ref{thm:The-AoI-of}
we have 
\begin{eqnarray*}
 &  & \boldsymbol{E}[A_{CBS}]-\boldsymbol{E}[A_{BRS}]\\
 & = & \frac{1}{1-V^{*}(\lambda)}\bigg\{[V^{*(1)}(\lambda)-V^{*(1)}(0)V^{*}(\lambda)][1-H^{*}(\lambda)]\\
 &  & +V^{*}(\lambda)H^{*(1)}(\lambda)(V^{*}(\lambda)-1)\bigg\}.
\end{eqnarray*}
Notice that $H^{*(1)}(\lambda)\leq0$ and $0\leq V^{*}(\lambda)\leq1$,
we have $V^{*}(\lambda)H^{*(1)}(\lambda)(V^{*}(\lambda)-1)\geq0.$
Since $0\leq H^{*}(\lambda)\leq1$, to show that $\boldsymbol{E}[A_{CBS}]-\boldsymbol{E}[A_{BRS}]\geq0$,
we only need to show $V^{*(1)}(\lambda)-V^{*(1)}(0)V^{*}(\lambda)\geq0.$
Since $V^{*(1)}(\lambda)-V^{*(1)}(0)V^{*}(\lambda)=-\boldsymbol{E}[Ve^{-\lambda V}]+\boldsymbol{E}[V]\boldsymbol{E}[e^{-\lambda V}],$
we let $X=V$, $Y=e^{-\lambda V}$ with CDF $F_{X}(x),$ $F_{Y}(x)$
and joint CDF $F(x,y)$. We now show that $\boldsymbol{P}(X\leq x,Y\leq y)\leq\boldsymbol{P}(X\leq x)\boldsymbol{P}(Y\leq y).$
Notice that 
\begin{eqnarray*}
 &  & F(x,y)=\boldsymbol{P}(X\leq x,Y\leq y)\\
 & = & \boldsymbol{P}(V\leq x,e^{-\lambda V}\leq y)=\boldsymbol{P}(-\frac{\ln y}{\lambda}\leq V\leq x)\\
 & = & \boldsymbol{P}(V\leq x)-\boldsymbol{P}(V\leq-\frac{\ln y}{\lambda})\\
 & \leq & \boldsymbol{P}(V\leq x)-\boldsymbol{P}(V\leq-\frac{\ln y}{\lambda})\boldsymbol{P}(V\leq x)\\
 & = & F_{X}(x)F_{Y}(y).
\end{eqnarray*}
From \cite{lehmann1966some} we know $\boldsymbol{E}[XY]-\boldsymbol{E}[X]\boldsymbol{E}[Y]=\int_{-\infty}^{\infty}\int_{-\infty}^{\infty}\left[F(x,y)-F_{X}(x)F_{Y}(y)\right]dxdy.$
Therefore, $V^{*(1)}(\lambda)-V^{*(1)}(0)V^{*}(\lambda)=\boldsymbol{E}[X]\boldsymbol{E}[Y]-\boldsymbol{E}[XY]\geq0$
and $\boldsymbol{E}[A_{CBS}]-\boldsymbol{E}[A_{BRS}]\geq0$. 
\end{IEEEproof}

\section{\label{sec:Proof-for-Theorem-1}Proof for Theorem \ref{thm:If-the-service}}

We first provide a lemma that will be useful in Proof of Theorem \ref{thm:If-the-service}.
\begin{lem}
\label{lem:It-holds-true}It holds true for any LST function $V^{*}(s)$
that $\frac{V^{*(1)}(\lambda)}{1-V^{*}(\lambda)}\geq-\frac{1}{\lambda}$
for any positive $\lambda$.
\end{lem}
\begin{IEEEproof}
Since $\frac{1}{\lambda}+\frac{V^{*(1)}(\lambda)}{1-V^{*}(\lambda)}=\frac{1-\boldsymbol{E}[e^{-\lambda V}]-\boldsymbol{E}[\lambda Ve^{-\lambda V}]}{\lambda(1-\boldsymbol{E}[e^{-\lambda V}])},$
we only need to show that $\boldsymbol{E}[1-e^{-\lambda V}-\lambda Ve^{-\lambda V}]\geq0$.
Let $\beta(v)=1-e^{-\lambda v}-\lambda ve^{-\lambda v}$, then $\beta(0)=0$
and $\frac{\partial\beta(v)}{\partial v}=\lambda^{2}ve^{-\lambda v}\geq0$
for $v\geq0$. Therefore $\frac{V^{*(1)}(\lambda)}{1-V^{*}(\lambda)}\geq-\frac{1}{\lambda}.$
\end{IEEEproof}
\textbf{Proof for Theorem \ref{thm:If-the-service}:}
\begin{IEEEproof}
We assume that the service time is exponentially distributed with
parameter $\mu.$ We first show the conclusion holds for AoI. When
the service time is exponentially distributed, by Lemma \ref{lem:It-holds-true},
we have
\begin{eqnarray*}
 &  & \boldsymbol{E}[\Delta_{CBS}]-\boldsymbol{E}[\Delta_{CBS-P}]\\
 & = & \frac{\lambda}{\mu(\mu+\lambda)}+\frac{\lambda}{\mu+\lambda}(\frac{1}{\lambda}+\frac{V^{*(1)}(\lambda)}{1-V^{*}(\lambda)})\geq0
\end{eqnarray*}
Now we show the result holds true for PAoI. Since we have $\boldsymbol{E}[A_{CBS}]=\frac{1}{\lambda}+\frac{V^{*(1)}(\lambda)-V^{*(1)}(0)}{1-V^{*}(\lambda)}+\frac{2}{\mu}$
and $\boldsymbol{E}[A_{CBS-P}]=\frac{1-\frac{\mu}{\mu+\lambda}+\frac{\lambda\mu}{(\mu+\lambda)^{2}}+(\frac{\mu}{\mu+\lambda})^{2}}{\frac{\lambda\mu}{\mu+\lambda}}+\frac{\frac{\mu}{\mu+\lambda}V^{*(1)}(\lambda)-V^{*(1)}(0)}{1-V^{*}(\lambda)},$
then 
\begin{eqnarray*}
 &  & \boldsymbol{E}[A_{CBS}]-\boldsymbol{E}[A_{CBS-P}]\\
 & = & \frac{1}{\lambda}+\frac{1}{\mu}-\frac{1}{\lambda}+\frac{\lambda}{\mu+\lambda}\frac{V^{*(1)}(\lambda)}{1-V^{*}(\lambda)}\\
 & \geq & \frac{1}{\mu}-\frac{1}{\mu+\lambda}\geq0.
\end{eqnarray*}
\end{IEEEproof}

\section{\label{sec:Proof-for-Theorem-4}Proof for Theorem \ref{thm:If-the-service-2}}
\begin{IEEEproof}
We first have 
\begin{eqnarray*}
 &  & \boldsymbol{E}[A_{CBS}]-\boldsymbol{E}[A_{CBS-P}]\\
 & = & \frac{1}{\lambda}+\frac{V^{*(1)}(\lambda)-V^{*(1)}(0)}{1-V^{*}(\lambda)}+2\boldsymbol{E}[H]\\
 &  & -\frac{1-H^{*}(\lambda)-\lambda H^{*(1)}(\lambda)+H^{*}(\lambda)^{2}}{\lambda H^{*}(\lambda)}\\
 &  & -\frac{H^{*}(\lambda)V^{*(1)}(\lambda)-V^{*(1)}(0)}{1-V^{*}(\lambda)}\\
 & = & (1-H^{*}(\lambda))\left(\frac{1}{\lambda}+\frac{V^{*(1)}(\lambda)}{1-V^{*}(\lambda)}\right)\\
 &  & +2\boldsymbol{E}[H]-\frac{1-H^{*}(\lambda)-\lambda H^{*(1)}(\lambda)}{\lambda H^{*}(\lambda)}.
\end{eqnarray*}

From Lemma \ref{lem:It-holds-true} we know that $\frac{1}{\lambda}+\frac{V^{*(1)}(\lambda)}{1-V^{*}(\lambda)}\geq0$
and $1-H^{*}(\lambda)\geq-\lambda H^{*(1)}(\lambda)$, then we have
$\boldsymbol{E}[A_{CBS}]-\boldsymbol{E}[A_{CBS-P}]\geq2\boldsymbol{E}[H]-2\frac{1-H^{*}(\lambda)}{\lambda H^{*}(\lambda)}\geq0.$
\end{IEEEproof}

\section{Derivations for Systems without Vacations\label{sec:Derivations-for-Systems}.}

We first derive the variance of peak age in M/G/1/1. Realizing that
in M/G/1/1 system, once a packet arrives, the server will start processing
it immediately. Thus there is no waiting time for all packets. Then
the LST of peak age in M/G/1/1 can be given as $A^{*}(s)=I^{*}(s)H^{*}(s).$
The inter-service time $I$ can be further decomposed into the idling
time $T$ (exponentially distributed) and service time $H$, i.e.,
$I=T+H$. We thus have $A^{*}(s)=T^{*}(s)H^{*}(s)^{2}.$ By some simple
algebra, we can obtain the results for M/G/1/1 system.

Similarly, for M/G/1/1/preemptive system, there is no waiting time
for packets. Thus by the argument in Appendix \ref{sec:Proof-for-Theorem},
we have $D^{*}(s)=\frac{H^{*}(s+\lambda)}{H^{*}(\lambda)}$. Then
the LST of peak age can be given as $A^{*}(s)=D^{*}(s)T^{*}(s)L^{*}(s),$
where $L^{*}(s)$ is given by Equation (\ref{eq:A2.1}). And the results
for M/G/1/1/preemptive can be obtained.

For M/G/1/2{*} system, the inter-service time is $H$ if there is
an arrival during processing time. If there is no arrival during processing
time, the next service starts when the next arrival occurs. By memoryless
property of Poisson arrivals, we have $I=T$ in this case. Therefore
$I=\max\{H,T\}.$ To calculate the LST of $I,$ we have 
\begin{eqnarray*}
I^{*}(s) & = & \int_{h=0}^{\infty}\int_{t=h}^{\infty}\lambda e^{-\lambda t}e^{-st}dF_{H}(h)dt\\
 &  & +\int_{h=0}^{\infty}\int_{t=0}^{h}e^{-sh}\lambda e^{-\lambda t}dF_{H}(h)dt\\
 & = & \frac{\lambda}{\lambda+s}H^{*}(s+\lambda)+H^{*}(s)-H^{*}(s+\lambda)\\
 & = & H^{*}(s)-\frac{s}{\lambda+s}H^{*}(s+\lambda).
\end{eqnarray*}
We can then have $I^{*(1)}(0)=H^{*(1)}(0)-\frac{H^{*}(\lambda)}{\lambda},$
and $I^{*(2)}(0)=H^{*(2)}(0)+\frac{2}{\lambda^{2}}H^{*}(\lambda)-\frac{2}{\lambda}H^{*(1)}(\lambda).$
The waiting time only occurs when there is an arrival during processing
time $H$, so that $W=\max\{H-T,0\}.$ The LST of $W$ is thus be
given as 
\begin{eqnarray*}
W^{*}(s) & = & \int_{h=0}^{\infty}\int_{t=0}^{h}e^{-s(h-t)}dF_{H}(h)\lambda e^{-\lambda t}dt\\
 &  & +\int_{h=0}^{\infty}dF_{H}(h)\int_{t=h}^{\infty}\lambda e^{-\lambda t}dt\\
 & = & \frac{\lambda}{\lambda-s}H^{*}(s)-\frac{s}{\lambda-s}H^{*}(\lambda).
\end{eqnarray*}
From Lemma \ref{lem:1} we have 
\begin{align*}
G^{*}(s) & =\frac{\lambda}{\lambda+s}+\frac{s}{\lambda+s}W^{*}(\lambda+s)\\
 & =\frac{\lambda}{\lambda+s}-\frac{\lambda}{\lambda+s}H^{*}(\lambda+s)+H^{*}(\lambda).
\end{align*}
By taking the first and second derivative of $G^{*}(s)$, we have
$G^{*(1)}(0)=-\frac{1}{\lambda}+\frac{1}{\lambda}H^{*}(\lambda)-H^{*(1)}(\lambda)$
and $G^{*(2)}(0)=\frac{2}{\lambda^{2}}-\frac{2}{\lambda^{2}}H^{*}(\lambda)+\frac{2}{\lambda}H^{*(1)}(\lambda)-H^{*(2)}(\lambda).$
By Equation (\ref{eq:0}) and (\ref{eq:1}), we can directly obtain
$\boldsymbol{E}[A_{M/G/1/2^{*}}]$ and $\boldsymbol{E}[\Delta_{M/G/1/2^{*}}]$.
Using Equation (\ref{eq:1.1}), we can directly derive the variance
of peak age.

\section{Exact Solution for PAoI in CBS-P with Dependent Vacation\label{sec:Bounds-for-AoI}}

Notice that in CBS-P, the server's vacation time $B$ can be divided
into $B=T+W$, where $T$ is the inter-arrival time of packets, which
is exponentially distributed, and $W$ is the time when the buffer
is occupied. Because of the memoryless property of exponential distribution,
we have $\boldsymbol{E}[B]=\frac{1}{\lambda}+\boldsymbol{E}[W]$.
From Equation (\ref{eq:A2}) we have $\boldsymbol{E}[D]=-\frac{H^{*(1)}(\lambda)}{H^{*}(\lambda)}-H^{*}(\lambda)G^{*(1)}(0)$.
By combining it with Equations (\ref{eq:1.2}), (\ref{eq:A0}), and
(\ref{eq:A3}), the PAoI for CBS-P can be written as 
\begin{eqnarray*}
\boldsymbol{E}[A] & = & \boldsymbol{E}[D]+\boldsymbol{E}[B]+\boldsymbol{E}[L]\\
 & = & -\frac{H^{*(1)}(\lambda)}{H^{*}(\lambda)}+H^{*}(\lambda)\frac{1}{\lambda}(1-W^{*}(\lambda))\\
 &  & +\boldsymbol{E}[W]+\frac{1}{\lambda H^{*}(\lambda)}.
\end{eqnarray*}

\section{\label{sec:Proof-for-Theorem-2}Proof for Theorem \ref{thm:If-the-service-1}}
\begin{IEEEproof}
When the service time is exponentially distributed, from Equation
(\ref{eq:10.2}), we have 
\begin{eqnarray*}
L_{j}^{*}(s) & = & \frac{H_{j}^{*}(s+\lambda_{j})}{\frac{s}{s+\lambda_{j}}+\frac{\lambda_{j}}{s+\lambda_{j}}H_{j}^{*}(s+\lambda_{j})}=\frac{\frac{1}{h_{j}}}{s+\frac{1}{h_{j}}}.
\end{eqnarray*}
So that the expressions for $\tilde{H_{j}^{*}}$ in Equation (\ref{eq:10.1})
are identical for CBS and CBS-P. Both systems will have the same $F_{j}(z_{1},...,z_{k})$
for all $j$ after solving for Equation (\ref{eq:8}). Similarly,
since $\frac{1-H_{j}^{*}(\lambda_{j})}{\lambda_{j}H^{*}(\lambda_{j})}=h_{j},$
both CBS and CBS-P will have the same expression for $\gamma_{j}$
in Equation (\ref{eq:10.3}) for all $j$. Therefore, CBS and CBS-P
have the same expressions for $W_{j}^{*}(\lambda_{j})$ and $\boldsymbol{E}[W_{j}]$
for all queue $j$. We then have 
\begin{eqnarray*}
 &  & \boldsymbol{E}[A_{j}^{CBS}]-\boldsymbol{E}[A_{j}^{CBS-P}]\\
 & = & -\frac{1}{\lambda_{j}}W_{j}^{*}(\lambda_{j})+\frac{2}{\lambda_{j}}+\boldsymbol{E}[W_{j}]+2\boldsymbol{E}[H_{j}]\\
 &  & -\bigg\{-\frac{H_{j}^{*(1)}(\lambda_{j})}{H_{j}^{*}(\lambda_{j})}+H_{j}^{*}(\lambda_{j})\frac{1}{\lambda_{j}}(1-W_{j}^{*}(\lambda_{j}))\\
 &  & +\frac{1}{\lambda_{j}}+\boldsymbol{E}[W_{j}]+\frac{1-H_{j}^{*}(\lambda_{j})}{\lambda_{j}H_{j}^{*}(\lambda_{j})}\bigg\}\\
 & = & \left(1-H_{j}^{*}(\lambda_{j})\right)\frac{1}{\lambda_{j}}\left(1-W_{j}^{*}(\lambda_{j})\right)+h_{j}-\frac{1}{\frac{1}{h_{j}}+\lambda_{j}}\geq0.
\end{eqnarray*}
\end{IEEEproof}

\end{document}